\documentclass[usenatbib, fleqn, a4paper, useAMS]{mnras}

\usepackage{times}
\usepackage[T1]{fontenc}
\usepackage{ae, aecompl}

\usepackage{graphicx}
\usepackage{amsmath}
\usepackage{amssymb}
\usepackage[math]{cellspace}

\usepackage{aas_macros}
\usepackage{bm}

\title[The Kelvin-Helmholtz instability and SPH]{The Kelvin-Helmholtz instability and smoothed particle hydrodynamics}

\author[Tricco]{Terrence S. Tricco$^{1}$\thanks{E-mail: \href{mailto:ttricco@cita.utoronto.ca}{ttricco@cita.utoronto.ca}} \\
$^1$ Canadian Institute for Theoretical Astrophysics, University of Toronto, 60 St. George Street, Toronto, ON M5S 3H8, Canada}

\date{\today}

\pubyear{2018}

\begin{document}
\label{firstpage}
\pagerange{\pageref{firstpage}--\pageref{lastpage}}

\maketitle

\begin{abstract}
We perform simulations of the Kelvin-Helmholtz instability using smoothed particle hydrodynamics (SPH). The instability is studied both in the linear and strongly non-linear regimes. The smooth, well-posed initial conditions of Lecoanet et al. (2016) are used, along with an explicit Navier-Stokes viscosity and thermal conductivity to enforce the evolution in the non-linear regime. We demonstrate convergence to the reference solution using SPH. The evolution of the vortex structures and the degree of mixing, as measured by a passive scalar `colour' field, match the reference solution. Tests with an initial density contrast produce the correct qualitative behaviour. The $\mathcal{L}_2$ error of the SPH calculations decreases as the resolution is increased. The primary source of error is numerical dissipation arising from artificial viscosity, and tests with reduced artificial viscosity have reduced $\mathcal{L}_2$ error. A high-order smoothing kernel is needed in order to resolve the initial velocity amplitude of the seeded mode and inhibit excitation of spurious modes. We find that standard SPH with an artificial viscosity has no difficulty in correctly modelling the Kelvin-Helmholtz instability and yields convergent solutions.
\end{abstract}


\section{Introduction}
There exists doubt about the ability of smoothed particle hydrodynamics (SPH) to correctly model the Kelvin-Helmholtz instability and hydrodynamical mixing instabilities in general \citep[e.g.,][]{arepo, hs10, bs12, sijackietal12, hopkins13, hopkins15}. This contention should be taken seriously, for if true, it would imply that SPH is unable to solve the fundamental equations of hydrodynamics. This disbelief is borne primarily from the work of \citet{agertzetal07}, who, after comparing results between grid-based methods and SPH, concluded that SPH has a fundamental shortcoming in the mixing of multiphase fluids. 

\citet*{wvc08} and \citet{price08} were quick to point out that the results of \citet{agertzetal07} were not due to a fundamental flaw of SPH, but are explainable by the combined effect of discontinuous initial conditions with the lack of numerical treatment for contact discontinuities. These are solvable issues. 

The primary issue in \citet{agertzetal07}'s interpretation of their results arises from the presence of a contact discontinuity in the initial conditions. Discontinuities of any form lead to a multivalued fluid, and obtaining a solution numerically may lead to an unphysical solution if the fluid is assumed to be differentiable. What is required is either to make no assumption about differentiability (e.g., by solving the equations in integral form), or to include a numerical method to keep the fluid single valued. In the case of shocks, for example, the fluid is kept single valued by the addition of a Riemann solver (for grid codes) or artificial viscosity (SPH). Contact discontinuities require similar treatment. For grid methods, some measure of numerical dissipation is inherently present due to the truncation error involved in advection of fluid between grid cells \citep[e.g.,][]{robertsonetal10}. For SPH, there is no such corresponding numerical dissipation since the equations of motion for the particles are derived from the Lagrangian. Advection of entropy is exact \citep{sh02, price08}. What results is a spurious surface tension at contact discontinuities that keeps the two phases apart (the `pressure gap'; \citealt{okamotoetal03, agertzetal07}; \citet*{rha10}; \citet{springel10, abel11}). In some sense, the problem is that SPH is too good, with the only dissipation present that which is explicitly added. 

Solutions exist to handle contact discontinuities in SPH. One option is to explicitly add dissipation in the form of an artificial conductivity, similar to artificial viscosity, which keeps the fluid single valued by smoothing the contact discontinuity to the order of the resolution length \citep{price08, wvc08}. This has been shown to be effective in capturing contact discontinuities, removing the pressure gap, and promoting mixing \citep{valckeetal10, kawataetal13}. One challenge lies in the detection of contact discontinuities so that the artificial dissipation does not unnecessarily diffuse pressure gradients, for example, when pressure gradients are balanced by external forces (such as gravity). Options have been suggested for such cases \citep{valdarnini12, phantom}. 

Other solutions have focused on changing the base SPH formulation. If a kernel summation is used for the internal energy or entropy, then the assumption of differentiability is not implied in the resulting equations. These methods, often referred to as `density-independent' or `pressure-entropy' formulations \citep{rt01, sm13, hopkins13}, perform well in promoting mixing, though no formal convergence study of the Kelvin-Helmholtz instability has been conducted. That they can be used for general astrophysical simulations without introducing other issues remains to be determined. \citet{hopkins13} found that `pressure-entropy' formulations do not provide as good an estimate of the mass density, recommending that a hybrid scheme be used of the density summation and entropy summation when the mass density is required (i.e., additional physics), and \citet{rh12} found that this formulation performs poorly when dealing with shocks. Alternative approaches to handle contact discontinuities include kernel correction techniques \citep*[e.g.,][]{gce12, fro17}. 

Beyond the context of SPH, it is a problem in general to use discontinuous initial conditions to study the Kelvin-Helmholtz instability. Discontinuities present in the initial conditions, such as in the velocity or density, prohibits formal convergence since the initial conditions are ill-defined, remaining unresolved irrespective of the spatial resolution \citep{robertsonetal10, mlp12}. Convergence cannot be achieved because increasing the resolution of the calculation only permits access to higher wavenumber modes. Even if a chosen mode is seeded initially, higher wavenumber modes can be excited by the discontinuity and become dominant since they are the fastest growing modes \citep{chandrasekhar61}. 

This has not prevented authors from using discontinuous initial conditions to assert the accuracy of their simulations. \citet{arepo} and \citet{hopkins15}, for example, argued that since their voronoi tessallation and meshless finite mass (MFM) methods produce more small-scale structure than grid-based methods, they are closer to a `true' physical solution. This may be untrue. Since these small-scale structures are generated by high wavenumber modes seeded by the numerics, there is no guarantee that they represent physical structures \citep{mlp12,dm15}. It is inappropriate to attribute structure generated from numerical noise as corresponding to resolved physical structure.

Smooth, well-defined initial conditions of the Kelvin-Helmholtz instability produce converged solutions, avoiding the issues inherent to discontinuous initial conditions. \citet{robertsonetal10}, \citet{mlp12} and \citet{lecoanetetal16} have put forward well-defined initial conditions. They each demonstrate convergence in the linear regime, and, for \citet{lecoanetetal16}, additionally in the non-linear regime by the inclusion of a physical Navier-Stokes viscosity and thermal conductivity. Convergence is demonstrated by these authors for a variety of codes and methods. Collectively, they use the Eulerian grid-based codes {\sc Art} \citep*{kkk97, kkh02}, {\sc Enzo} \citep{osheaetal05, enzo}, {\sc Athena} \citep{athena}, the {\sc Pencil code} \citep{bd02}, the pseudo-spectral code {\sc Dedalus}, and the Lagrangian mesh-free code {\sc Phurbas} \citep*{phurbas1, phurbas2}. Convergence of the \citet{mlp12} test has also been demonstrated for the implicit large eddy simulation code {\sc Music} \citep{goffreyetal17}. 

%\footnote{\url{http://enzo-project.org}}
%\footnote{\url{https://trac.princeton.edu/Athena/}}
%\footnote{\url{http://pencil-code.nordita.org}}
%\footnote{\url{http://dedalus-project.org}}

What is missing is a comprehensive study demonstrating convergence with SPH. A number of studies have been conducted using discontinuous initial conditions \citep{price08, cin10, valckeetal10, junketal10, sy11, hfg13}. Studies with smooth initial conditions have been limited. While \citet{mlp12} included results from the SPH code {\sc ndspmhd} \citep{price12} in their work, their analyses were not focused on SPH, with a total of five codes compared simultaneously. Other studies have either been limited in scope \citep{phantom}, for incompressible fluids \citep{yueetal15}, or have used higher-order corrections \citep{cgf17, fro17}.

In this paper, a convergence study of the Kelvin-Helmholtz instability in SPH is undertaken using the smooth, well-defined initial conditions of \citet{lecoanetetal16}. Importantly, a physical Navier-Stokes viscosity and thermal conductivity are included to set the dissipation. The aim is to demonstrate convergence in both the linear and non-linear regime. 

The standard formulation of SPH is used, rather than alternative `pressure-entropy' formulations or higher-order corrections. This is motivated by the many public astrophysical SPH codes implemented using standard formulations of SPH with artificial dissipation terms, e.g., {\sc Gasoline 2} \citep{gasoline2}, {\sc Gandalf} \citep{gandalf}, and {\sc Phantom} \citep{phantom}. 

The paper is organised as follows. The setup, initial conditions and numerical method used for the Kelvin-Helmholtz test is described Section~\ref{sec:problem}. Results are presented and analysed in Section~\ref{sec:results}. A summary and conclusion is given in Section~\ref{sec:summary}.

%\footnote{\url{http://gasoline-code.com}}
%\footnote{\url{https://gandalfcode.github.io}}
%\footnote{\url{https://phantomsph.bitbucket.io}}

\section{Problem setup}
\label{sec:problem}

\subsection{Equations}

We solve the Navier-Stokes equations of hydrodynamics that include a thermal conductivity, given by
\begin{gather}
\frac{{\rm d}\rho}{{\rm d}t} = - \rho \nabla \cdot {\bm v} , \label{eq:cty} \\
\frac{{\rm d}{\bm v}}{{\rm d}t} = - \frac{\nabla P}{\rho} - \frac{1}{\rho} \nabla \cdot {\bm \Pi} , \label{eq:momentum} \\
\frac{{\rm d}u}{{\rm d}t} = - \frac{P}{\rho} \nabla \cdot {\bm v} +  \nabla \cdot ({\bm v} \cdot {\bm \Pi}) + \frac{1}{\rho} \nabla \cdot (\chi \rho \nabla T) ,\label{eq:energy}
\end{gather}
where $\rho$ is the density, ${\bm v}$ is the velocity, $P$ is the thermal pressure, $u$ is the internal energy, and $T$ is the temperature. The equations are written in Lagrangian form, using the material derivative ${\rm d}/{\rm d}t \equiv \partial / \partial t + {\bm v} \cdot \nabla$. The Navier-Stokes stress tensor is
\begin{equation}
\Pi^{ij} = \nu \rho \left( \frac{\partial v^i}{\partial x^j} + \frac{\partial v^j}{\partial x^i} - \frac{2}{3} \frac{\partial v^k}{\partial x^k} \delta^{ij} \right) ,
\label{eq:stresstensor}
\end{equation}
where $\nu$ is the shear viscosity. Heating from the viscosity is included as the second term in Equation~\ref{eq:energy}. The third term in Equation~\ref{eq:energy} is a physical thermal conductivity, with $\chi$ the thermal diffusivity. An ideal gas equation of state is used, $P = \rho T$, with the ratio of specific heats $\gamma = 5/3$. The temperature is thus related to the internal energy according to $T = (\gamma - 1) u$.

A scalar field is added to the simulation to quantify mixing. This colour field, or dye, is evolved according to
\begin{equation}
\frac{{\rm d}c}{{\rm d}t} = \frac{1}{\rho} \nabla \cdot (\nu_{\rm c} \rho \nabla c) . \label{eq:colour}
\end{equation}
The colour field is passively advected with the flow and does not influence the dynamics. A `physical' diffusion term is present to set the diffusion rate. This is important to achieve convergence in grid codes since numerical diffusion is introduced due to advection. For SPH, advection is dissipationless, but the explicitly added diffusion is retained for these calculations in order to converge to the results of \citet{lecoanetetal16}.

\subsection{Initial Conditions}
\label{sec:ics}

The initial conditions are given by
\begin{gather}
\rho = 1 + \frac{\Delta \rho}{\rho_0} \left[ \tanh \left( \frac{y - y_1}{a} \right) - \tanh \left( \frac{y - y_2}{a} \right) \right] , \label{eq:ic-rho} \\
v_x = v_0 \left[ \tanh\left( \frac{y - y_1}{a} \right) - \tanh \left( \frac{y - y_2}{a} \right) - 1 \right] , \\
v_y = A \sin(2 \pi x) \left[ \exp\left( - \frac{(y - y_1)^2}{\sigma^2} \right) + \exp \left( - \frac{(y - y_2)^2}{\sigma^2} \right) \right] , \label{eq:icsvy} \\
P = 10 , \\
c = \frac{1}{2} \left[ \tanh \left( \frac{y - y_1}{a} \right) - \tanh \left( \frac{y - y_2}{a} \right) + 2 \right] ,
\end{gather}
where $a = 0.05$ and $\sigma = 0.2$ dictate the width of the interface regions. The simulations are run in a periodic box of size $x \in [0, L]$ and $y \in [0, 2L]$ with $L=1$. The interfaces are at $y_1 = 0.5$ and $y_2 = 1.5$.  For the chosen initial conditions, the top half mirrors the bottom half. 

We take $v_0 = 1$, with $x$-velocity that smoothly transitions between the two regions. The flow remains subsonic with Mach number $\mathcal{M} \approx 0.25$. The amplitude of the $v_y$ velocity perturbation used to seed the instability is $A = 0.01$. These initial conditions are well-resolved even at modest resolutions.

The Reynolds number is Re=$10^5$, where the Reynolds number is defined according to
\begin{equation}
\text{Re} = \frac{L \Delta v}{\nu} .
\end{equation}
The box size is $L=1$ and the difference in fluid velocity is $\Delta v = 2 v_0$. This yields $\nu = 2 \times 10^{-5}$, with $\chi = \nu_{\rm c} = \nu$.

\citet{lecoanetetal16} considered cases where $\Delta \rho / \rho_0 = 0$ and 1. That is, for uniform density with $\rho=1$ throughout, and with a density contrast between densities of $\rho=1$ and $\rho=2$. The stratified test introduces secondary instabilities that produce complex, small-scale structure in the flow. This a far more computationally demanding test case, with \citet{lecoanetetal16} only achieving convergence with the grid-code {\sc Athena} in the non-linear regime when $16~384 \times 32~768$ grid cells are used, requiring $\sim 1$ million cpu-hours. Since we do not have a million spare cpu-hours available, our tests focus on the unstratified, uniform density test case. This is sufficient to quantify the convergence properties of the Kelvin-Helmholtz instability in SPH. Tests including a density contrast are studied in Section~\ref{sec:densitycontrast}.

\subsection{Smoothed particle hydrodynamics}
\label{sec:sph}

The equations are solved using smoothed particle hydrodynamics (SPH). The continuum equations are discretised into a set of particles that are smoothed over their local volume to recover continuum-like behaviour. The implementation of SPH used is similar to that of the {\sc Phantom} SPH code \citep{phantom}. No modifications have been made to the SPH formalism in order to simulate the Kelvin-Helmholtz instability. The calculations use the standard form of SPH with an artificial viscosity, of the type that has been employed for decades \citep{monaghan92, monaghan05}. 

The discrete set of equations for Euler fluid flow, equivalent to Equations~\ref{eq:cty}--\ref{eq:energy} absent the Navier-Stokes viscous terms and thermal conductivity, are given by
\begin{gather}
\rho_a = \sum_b m_b W_{ab}(h_a) , \label{eq:denssum} \\
\frac{{\rm d} {\bm v}_a}{{\rm d}t} = -  \sum_b m_b \left[ \frac{P_a}{\Omega_a \rho_a^2} \nabla_a W_{ab}(h_a) + \frac{P_b}{\Omega_b \rho_b^2} \nabla_a W_{ab}(h_b) \right]  \nonumber \\
\hspace{11mm} + \sum_b \frac{m_b}{\overline{\rho}_{ab}} v^{\rm sig}_{ab} {\bm v}_{ab} \cdot \hat{\bm r}_{ab} \overline{\nabla_a W}_{ab}  , \label{eq:sphmom} \\
\frac{{\rm d}u_a}{{\rm d}t} = \frac{P_a}{\Omega_a \rho_a^2} \sum_b m_b {\bm v}_{ab} \cdot \nabla_a W_{ab}(h_a) \nonumber \\
\hspace{11mm} - \sum_b \frac{m_b}{\overline{\rho}_{ab}} v^{\rm sig}_{ab} \left( {\bm v}_{ab} \cdot \hat{\bm r}_{ab} \right)^2 \hat{\bm r}_{ab} \cdot \overline{\nabla_a W}_{ab} . \label{eq:sphenergy}
\end{gather}
Here, $a$ and $b$ refer to particle indices and the summations are over neighbouring particles. The shorthand notation, ${\bm v}_{ab} \equiv {\bm v}_a - {\bm v}_b$, is used, along with $\hat{\bm r}_{ab} \equiv ({\bm r}_a - {\bm r}_b) / \vert {\bm r}_a - {\bm r}_b \vert$, which is the unit vector along the line of the sight between particles. The smoothing kernel is $W_{ab}(h_a) \equiv W(\vert {\bm r}_a - {\bm r}_b \vert, h_a)$, with $\overline{W}_{ab} \equiv \tfrac{1}{2} [W_{ab}(h_a)/\Omega_a + W_{ab}(h_b)/\Omega_b]$ the averaged smoothing kernel. The particle mass is $m$, and these set of calculations use particles of equal mass. 

The smoothing length, $h$, is obtained through iteration with the density, $\rho$, according to the relation
\begin{equation}
h_a = \eta \left( \frac{m_a}{\rho_a} \right)^p
\label{eq:sml}
\end{equation}
where $p$ is the number of dimensions (two for these calculations). We use $\eta = 1.2$ for the ratio of smoothing length to interparticle spacing, which is ideal for the bell-shaped spline kernels \citep{da12}. Gradients of the smoothing length are accounted for by the factor
\begin{equation}
\Omega_a = 1 + \frac{h_a}{p \rho_a} \sum_b m_b \frac{\partial W_{ab}(h_a)}{\partial h_a} ,
\end{equation}
which is important to ensure exact energy conservation in the spatial discretisation \citep{sh02, monaghan02}. The discretised momentum equation exactly conserves linear and angular momentum.

Artificial viscosity is included by the second terms of the momentum and energy equations (Equations~\ref{eq:sphmom}--\ref{eq:sphenergy}). This implementation closely matches that of \citet{monaghan97}. The signal velocity used is 
\begin{equation}
v^{\rm sig}_{ab} =  \frac{1}{2} \overline{\alpha}_{ab} ( c_{{\rm s},a} + c_{{\rm s},b}) - \beta {\bm v}_{ab} \cdot \hat{\bm r}_{ab} ,
\end{equation}
where $c_{\rm s}$ is the sound speed, $\alpha$ and $\beta$ are dimensionless quantities, and $\overline{\alpha}_{ab} = \tfrac{1}{2} (\alpha_a + \alpha_b)$. The \citet{mm97} switch is used to reduce numerical dissipation. Each particle has an individual $\alpha$ which is time integrated according to 
\begin{equation}
\frac{{\rm d}\alpha_a}{{\rm d}t} = \max( \nabla \cdot {\bm v}_a, 0) - \frac{\alpha_a - \alpha_0}{\tau_a} ,
\label{eq:mm97}
\end{equation}
such that $\alpha$ is increased in regions of compression. The range of values is enforced to be $\alpha \in [\alpha_0, 1]$, with $\alpha_0 = 0.1$ the minimum value used in these calculations. The time-scale over which $\alpha$ decays is $\tau_a = 0.1 h_a / \max(v^{\rm sig}_{ab})$, where $\max(v^{\rm sig}_{ab})$ is the maximum signal velocity as calculated over all neighbouring particles. This decay time-scale roughly corresponds to the time it takes for a sound wave to cross five smoothing lengths \citep{mm97}. The $\beta$ term represents a von Neumann-Richtmyer type viscosity and is used to prevent particle interpenetration \citep{monaghan89}. In following of the {\sc Phantom} and {\sc Gasoline2} codes, this term is not multiplied by $\alpha$. In general, the kinetic energy dissipated by the $\beta$ term tends to lower than the $\alpha$ term since it scales as $\propto h^2$.

The one important ingredient required to achieve convergence in these simulations is a high-order smoothing kernel. The amplitude of the initial velocity perturbation for the seeded Kelvin-Helmholtz mode has Mach number on the order of $\mathcal{M} \approx 0.0025$, notably smaller than in the initial conditions used by \citet{arepo}, \citet{robertsonetal10}, \citet{mlp12}, and \citet{hopkins15}. It is required that velocity noise caused by errors in the pressure gradient be below this amplitude in order to resolve the linear growth phase. Additionally, the excitation of other, non-seeded modes by velocity noise must be inhibited in order to correctly model the non-linear phase. We found that the cubic spline is unable to do this for this problem, and that while the quintic spline produces all of the correct features of the reference solution, the kernel bias prevents convergence. The calculations presented here use the septic spline, which exhibits convergence, as will be demonstrated. Though the septic spline has a smoothing volume greater than the cubic or quintic splines, thus includes more neighbour particles, the resolution length of the calculation is the same. The ratio of smoothing length to interparticle spacing, as obtained via Equation~\ref{eq:sml}, is unchanged. The bias of the smoothing kernel on our calculations is discussed further in Section~\ref{sec:kernelbias}, and a list of the spline kernels, up to the nonic spline, is given in Appendix~\ref{sec:kernels}.

\subsection{Physical dissipation in SPH}

Navier-Stokes viscosity is implemented using two first derivatives, in a manner similar to \citet{flebbeetal94}, \citet{watkinsetal96}, and \citet{ss06}. Each component of the stress tensor (Equation~\ref{eq:stresstensor}) is calculated using a difference derivative estimate \citep{price12} according to
\begin{equation}
\frac{\partial v_a^i}{\partial x_a^j} = - \frac{1}{\Omega_a \rho_a} \sum_b m_b (v_a^i - v_b^i) \frac{\partial W_{ab}(h_a)}{\partial x_a^j} .
\label{eq:sphdiffderiv}
\end{equation}
The corresponding term in the momentum equation can be derived from the Lagrangian to be
\begin{equation}
\frac{{\rm d} v_a^i}{{\rm d}t} = - \sum_b m_b \left[ \frac{\Pi^{ij}_a}{\Omega_a \rho_a^2} \nabla_a^j W_{ab}(h_a) + \frac{\Pi^{ij}_b}{\Omega_b \rho_b^2} \nabla_a^j W_{ab}(h_b) \right] .
\label{eq:sphNS}
\end{equation}
The derivative of the stress tensor is a symmetric derivative operator. The conjugate nature of difference and symmetric derivative operators has been noted in other contexts \citep{cr99, price10, tp12}. 

Heating from the Navier-Stokes viscosity is given by
\begin{equation}
\frac{{\rm d}u_a}{{\rm d}t} = \nu \rho_a \left( \frac{\partial v_a^i}{\partial x_a^j} + \frac{\partial v_a^j}{\partial x_a^i} \right)^2 - \frac{2}{3} \nu \rho_a \left( \frac{\partial v_a^k}{\partial x_a^k} \right)^2 ,
\end{equation}
where by the square of the first term we mean the tensor summation $\sum_i \sum_j T_{ij} T_{ij}$ with $T_{ij} = \partial v_a^i / \partial x_a^j + \partial v_a^j / \partial x_a^i$. The heating term is positive definite provided the terms of the stress tensor are calculated using Equation~\ref{eq:sphdiffderiv}. The derivation of the viscous heating term is given in Appendix~\ref{sec:NSheating}. This spatial discretisation of the Navier-Stokes viscosity exactly conserves energy and linear momentum.

Thermal conductivity is implemented in SPH according to \citet{cm99}, given by
\begin{equation}
\frac{{\rm d}u_a}{{\rm d}t} = \chi \sum_b m_b (\rho_a + \rho_b) (u_a - u_b) \frac{{\bm r}_{ab} \cdot \overline{\nabla_a W}_{ab}(h_a)}{r_{ab}^2} .
\end{equation}
This uses a direct second derivative in the manner of \citet{brookshaw85}. \citet{cm99} proposed that the harmonic mean of the thermal diffusivity coefficients, $4 \chi \rho_a \rho_b / (\rho_a + \rho_b)$, is better for discontinuous thermal conductivities. For these calculations, the arithmetic mean, $\tfrac{1}{2} \chi (\rho_a + \rho_b)$, is used instead since the thermal conductivity for this problem varies smoothly.

Colour diffusion is implemented in an analogous fashion to the thermal conductivity, given by
\begin{equation}
\frac{{\rm d}c_a}{{\rm d}t} =  \nu_{\rm c} \sum_b m_b (\rho_a + \rho_b)  (c_a - c_b)  \frac{{\bm r}_{ab} \cdot \overline{ \nabla_a W}_{ab} }{r_{ab}^2} .
\label{eq:sphcolourdiffusion}
\end{equation}
This formulation exactly conserves the volume integral of colour, that is $\sum_a m_a c_a$, since it is equal and opposite for each particle pair.

\subsection{Calculations}

Calculations are performed using $256 \times 592$, $512 \times 1184$, $1024 \times 2364$, and $2048 \times 4728$ particles in the $x$ and $y$ directions, respectively, labelled as $n_{\rm x} = 256$, 512, 1024, and 2048. The particles are arranged on triangular lattices, with the largest calculation consisting of $\sim 10$ million particles ($n_{\rm x} = 2048$), requiring $\sim 70~000$ cpu-hours of computational time. 

\subsection{Reference solution}

There is no analytic solution to the \citet{lecoanetetal16} Kelvin-Helmholtz test. In its absence, we turn to a solution obtained using high-resolution calculations, which acts as an approximation to the true solution. The SPH calculations performed in this work are compared to the data from the D2048 calculation of \citet{lecoanetetal16}. This solution is obtained using the pseudo-spectral code {\sc Dedalus}\footnote{\url{http://dedalus-project.org}}, on a grid of $2048 \times 4096$ points. The data consists of the density and colour fields at times of $t=2,4,6$, and $8$.

\section{Results}
\label{sec:results}

The SPH calculations are qualitatively and quantitatively compared with the solution of the D2048 {\sc Dedalus} calculation from \citet{lecoanetetal16}. Our focus is to study the convergence properties of SPH with respect to this solution, particularly in the strongly non-linear regime.

\subsection{Linear regime; $t < 2$}
\label{sec:linear}

\begin{figure*}
\centering
\hspace{-1.85mm}
\includegraphics[width=0.245\linewidth]{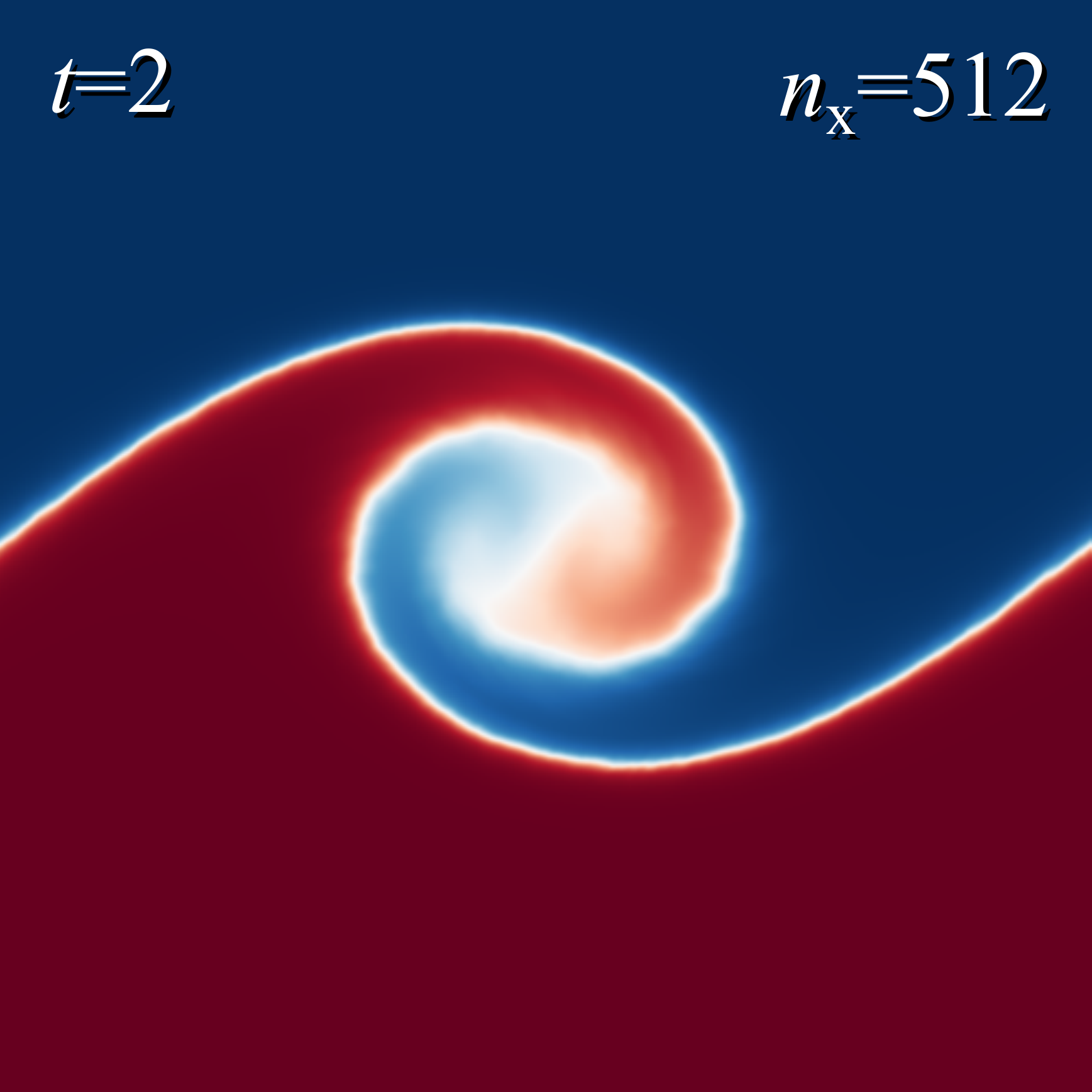}  \hspace{-1.85mm}
\includegraphics[width=0.245\linewidth]{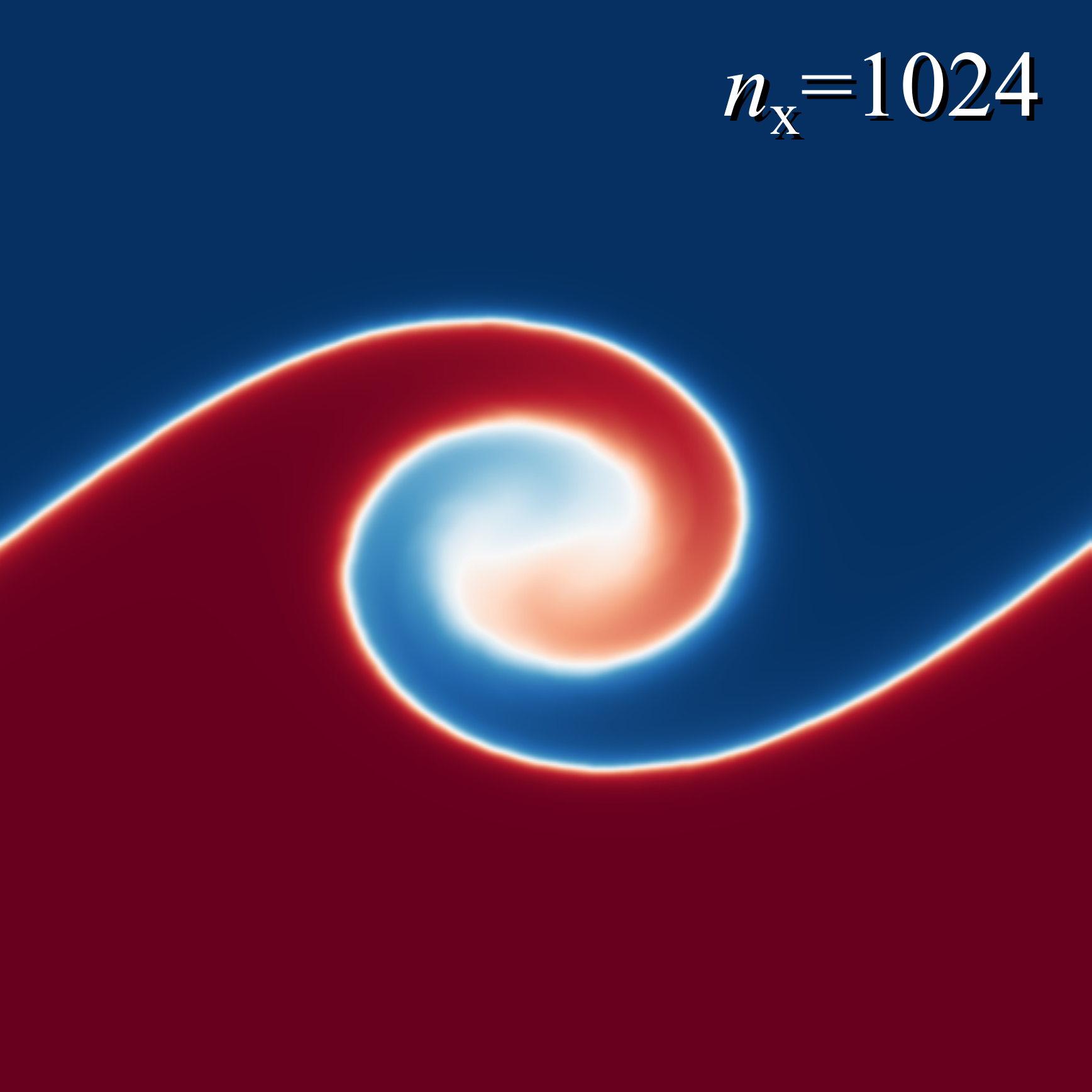} \hspace{-1.85mm}
\includegraphics[width=0.245\linewidth]{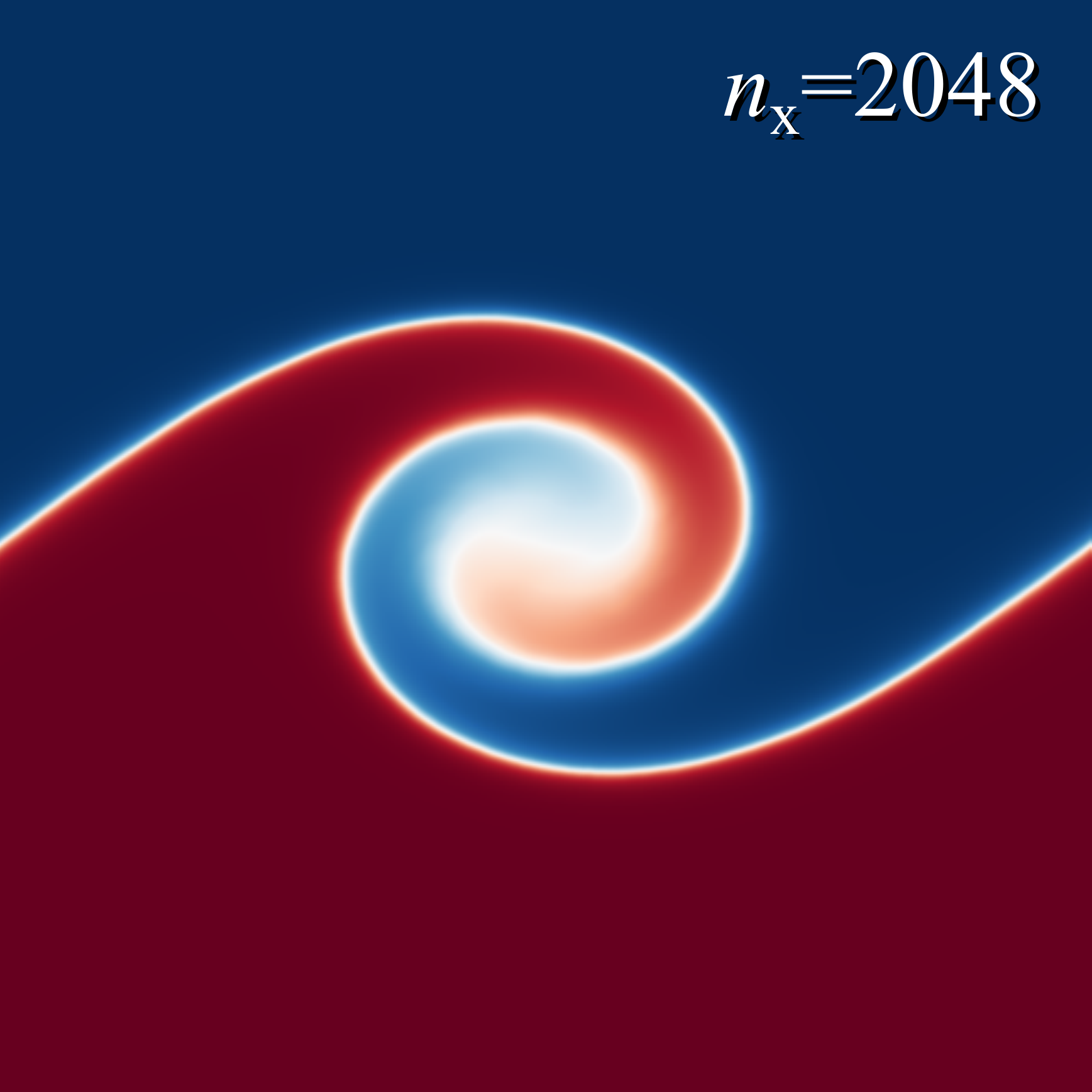}  \hspace{-1.85mm}
\includegraphics[width=0.245\linewidth]{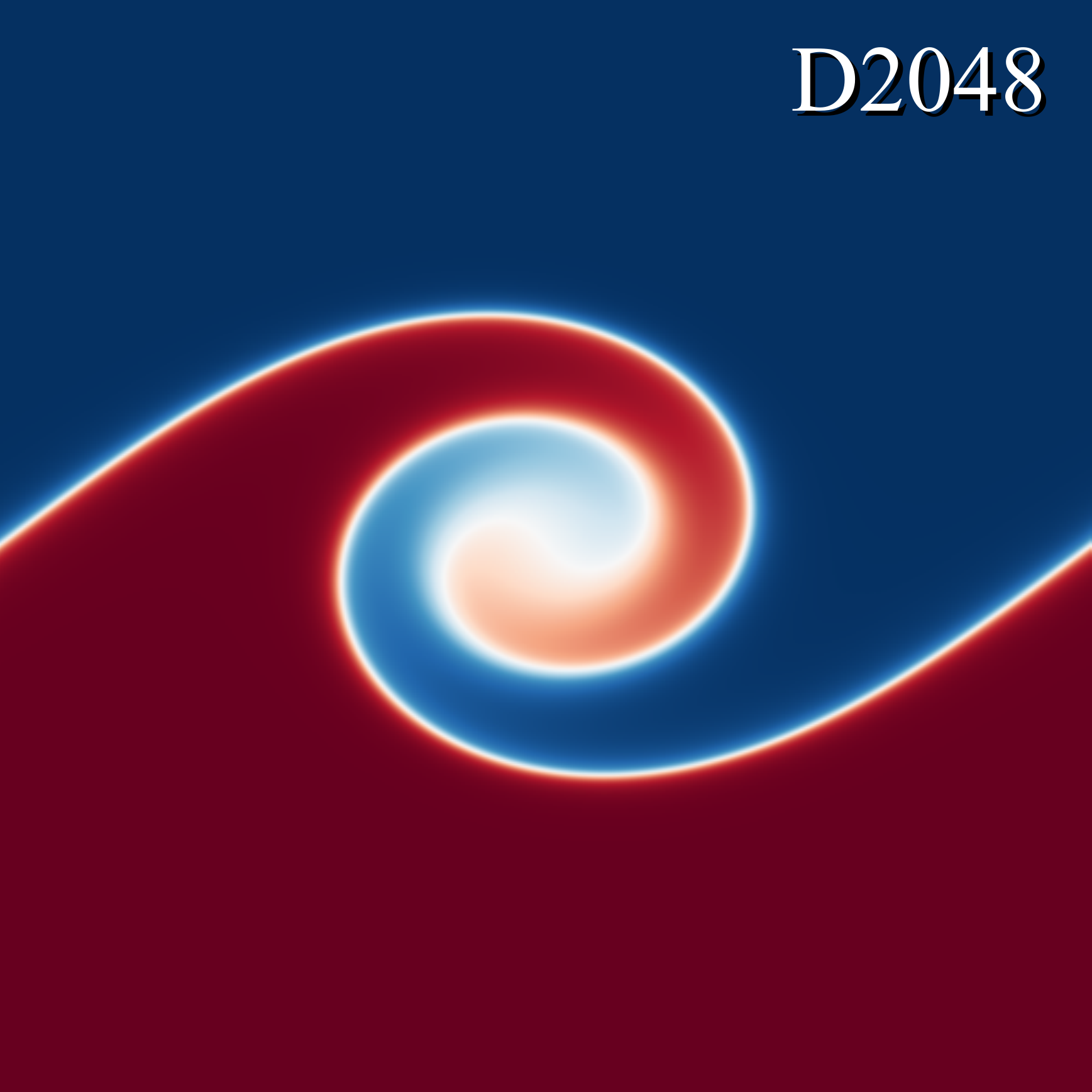} 
\includegraphics[width=0.983\linewidth]{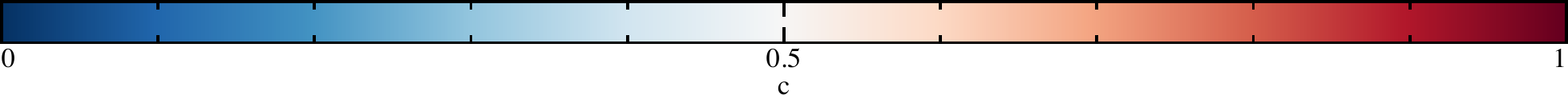} 
\caption{The colour field in the region $x,y \in [0,1]$ for the $n_{\rm x} = 512$, 1024, and 2048 particle SPH calculations (three leftmost panels) with the D2048 {\sc Dedalus} calculation (far right panel) at $t =2$. In all cases, the SPH calculations have grown the seeded mode to the correct shape. For the $n_{\rm x}=512$ particle calculation, there is some noise in the inner region of the curl arising from particle motion. This noise is absent from the $n_{\rm x}=2048$ particle calculation, which reproduces the D2048 reference solution in all aspects.}
\label{fig:t2}
\end{figure*}

Figure~\ref{fig:t2} shows the colour field at $t=2$ for the SPH calculations at resolutions of $n_{\rm x} = 512$, 1024, and 2048 particles, along with the D2048 reference solution (far right panel). Only the bottom half ($y < 1$) of the calculations are shown since the top and bottom mirror each other. At this time, the calculation is at the end of the linear evolution phase of the Kelvin-Helmholtz instability. 

In all cases, the SPH calculations form a single large curl corresponding to the seeded $k = 2 \pi$ mode, in agreement with the reference solution. For the $n_{\rm x}=512$ particle calculation, the inner tip of the curl is not wound as tightly as the reference solution. There is some additional noise in the inner region, due to particles breaking off lattice. These differences disappear for the $n_{\rm x}=2048$ particle calculation. Indeed, the visual appearance of the curl for the highest resolution SPH calculation ($n_{\rm x}=2048$) closely reproduces the reference solution.

The initial velocity perturbation grows exponentially in the linear regime of the Kelvin-Helmholtz instability, occurring at a rate $\propto \exp(\gamma t)$. The linear growth rate for an incompressible, infinite domain fluid is given by \citep{chandrasekhar61}
\begin{equation}
\gamma = \frac{(\rho_1 \rho_2)^{1/2}}{\rho_1 + \rho_2} \vert v_1 - v_2 \vert k . 
\label{eq:linear}
\end{equation}
For the initial conditions in Section~\ref{sec:ics}, the corresponding quantities are $k = 2 \pi$, $\rho_1 = \rho_2 = 1$, $v_1 = 1$ and $v_2 = -1$. This yields a growth rate of $\gamma = 2 \pi$. 

However, the growth rate in these calculations is expected to differ since the interfaces are smoothed. \citet*{wyl10} derive analytic estimates of the growth rate for incompressible fluids with smoothed density and/or velocity interfaces. Their key conclusion is that smoothing the velocity interface slows the growth rate. Using their analytic estimate (c.f. their Equation~18) with the initial conditions employed here yields $\gamma \approx 5.09$, about a $20\%$ reduction from the estimate using Equation~\ref{eq:linear}. 

Our calculations additionally smooth the velocity perturbation, and is for a compressible fluid in a periodic domain with peak density fluctuation of $\approx 10\%$. Therefore, it is expected that the growth rate will be further reduced. It is likely that an estimate using the approach of \citet{wyl10} merely represents an upper bound.

\citet{bp19} derive the linear theory for the Kelvin-Helmholtz instability with smooth initial velocity and density profiles (as well as the magnetised Kelvin-Helmholtz instability). Using their code, {\sc psecas}, a numerical solution was obtained for the linear growth rate for the initial conditions of \citet{lecoanetetal16}, with the growth rate of the $k = 2\pi$ mode found to be $\gamma \approx 3.227$.

To measure the growth rate of the $k = 2 \pi$ mode in our calculations, we use a calculation based on a discrete convolution, as described by \citet{mlp12}. A weighting term for each particle, $i$, is defined according to
\begin{equation}
d_i = 
\begin{cases}
h_i^2 \exp(- \vert y_i - \tfrac{1}{2} \vert / \sigma^2) & y < 1 , \\
h_i^2 \exp(- \vert (2 - y_i) - \tfrac{1}{2} \vert / \sigma^2) & y \geq 1,
\end{cases}
\end{equation}
where $h$ is the smoothing length and the weighting has characteristic size of the $v_y$ perturbation as given in the initial conditions (Equation~\ref{eq:icsvy}). The wave components for the $k = 2 \pi$ mode are given by
\begin{gather}
s_i = v_{y,i} \sin(2 \pi x_i), \\
c_i = v_{y,i} \cos(2 \pi x_i).
\end{gather}
The mode amplitude, $M$, may then be computed by the weighted summations of the components of the mode according to
\begin{equation}
M = 2 \left[ \left( \frac{\sum_{i=1}^N s_i d_i}{\sum_{i=1}^N d_i} \right)^2 + \left( \frac{\sum_{i=1}^N c_i d_i}{\sum_{i=1}^N d_i} \right)^2 \right]^{1/2} ,
\label{eq:mode}
\end{equation}
where the summation of $d_i$ in the denominator acts as a normalisation condition. The summations are over all particles. This approach avoids the need to interpolate to a fixed grid since the $v_y$ values are taken directly from the SPH particles.

\begin{figure}
\includegraphics[width=\linewidth]{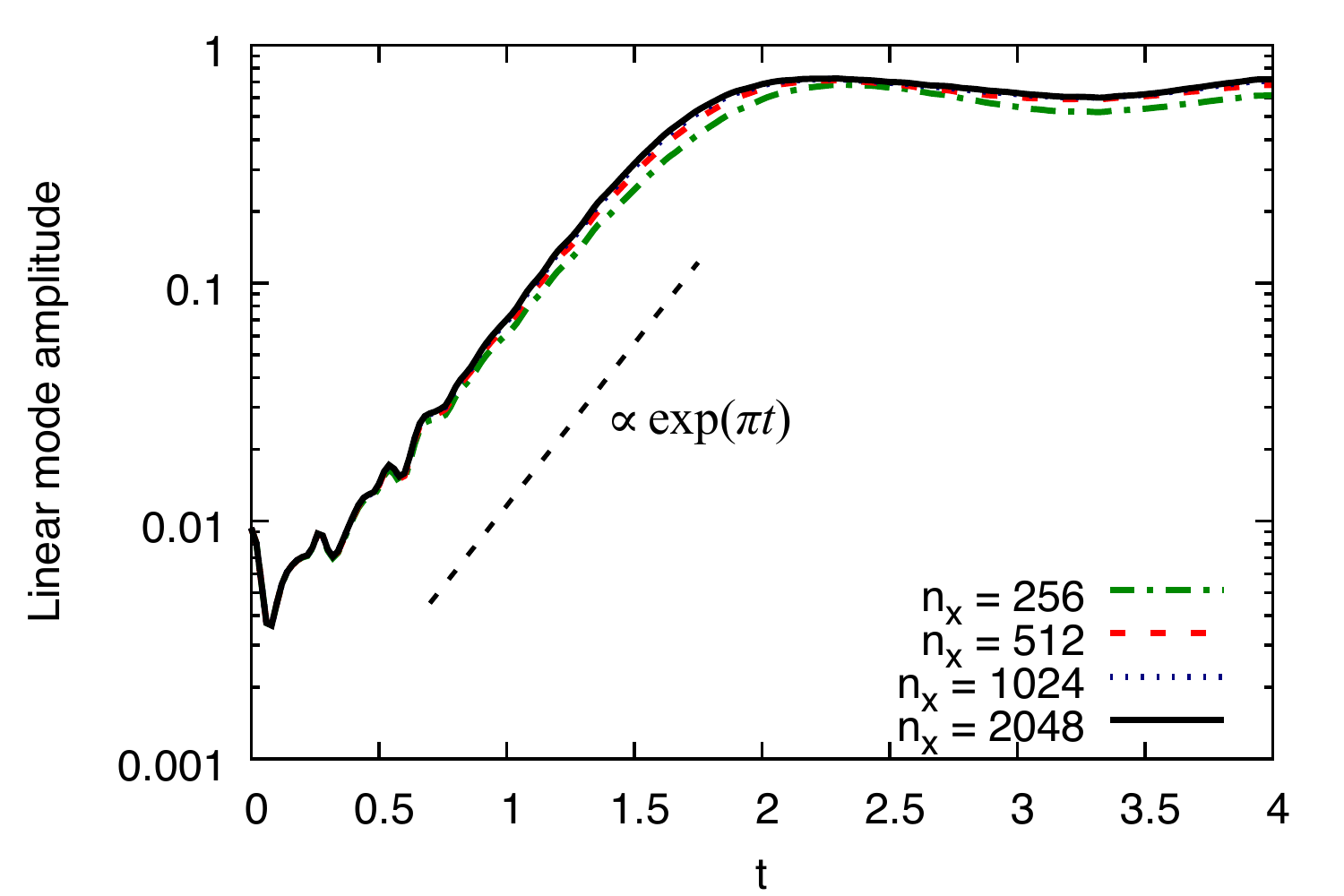}
\caption{Amplitude of the $k = 2\pi$ seeded mode for the $n_{\rm x}= 256$, 512, 1024 and 2048 SPH calculations. The growth is exponential in the linear regime, as expected for the Kelvin-Helmholtz instability. The $n_{\rm x} = 2048$ calculation has growth rate $\Gamma \propto \exp(3.139 t)$.}
\label{fig:modeamp}
\end{figure}

\begin{table}
\caption{Measured growth rates, $\Gamma$, of the linear phase.}
\label{tbl:growthrates}
\begin{tabular}{cc}
\hline \\[-6pt]
$n_{\rm x}$ & $\Gamma$ \\[2pt]
\hline \\[-6pt]
256 & 2.92 \\
512 & 3.055\\
1024 & 3.113 \\
2048 & 3.139 \\
\hline 
\end{tabular}
\end{table}

Figure~\ref{fig:modeamp} shows the amplitude of the $k = 2 \pi$ mode in the linear regime. For all calculations, it undergoes exponential amplification of the $y$-velocity, as expected for the linear regime of the Kelvin-Helmholtz instability. The mode amplitude reaches a maximum at $t\approx2$, establishing that the non-linear regime begins at this time. 

The measured growth rates per resolution are summarised in Table~\ref{tbl:growthrates}. The $n_{\rm x}=2048$ calculation has a growth rate $\gamma \approx 3.139$, which is in close agreement to the expected value of $\gamma \approx 3.227$ obtained using the {\sc psecas} code \citep{bp19}, with the growth rate converging linearly in resolution to the expected value.

\subsection{Non-linear regime; $t > 2$}
\label{sec:nonlinear}

\begin{figure*}
\centering
\hspace{-1.85mm}
\includegraphics[width=0.245\linewidth]{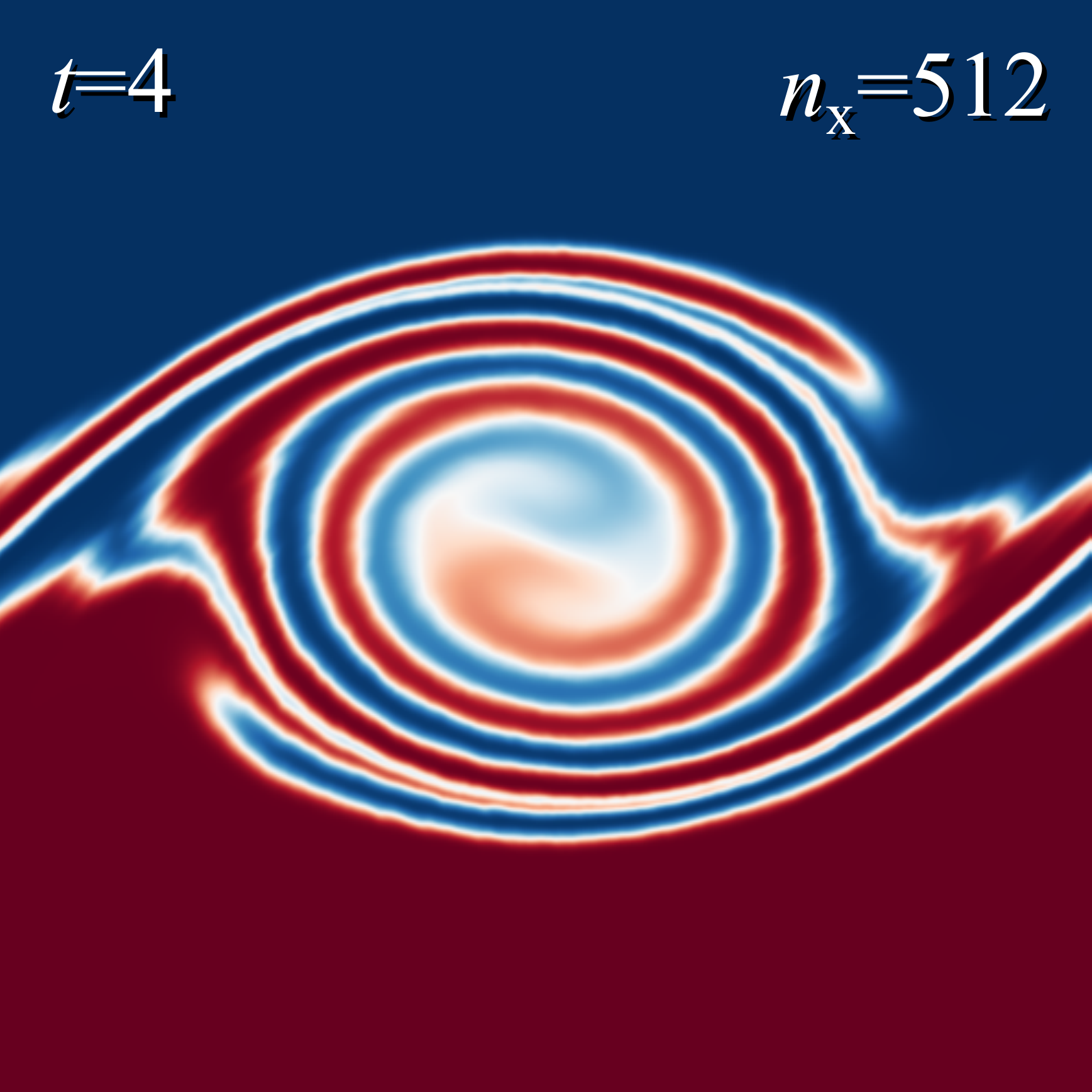}   \hspace{-1.85mm}
\includegraphics[width=0.245\linewidth]{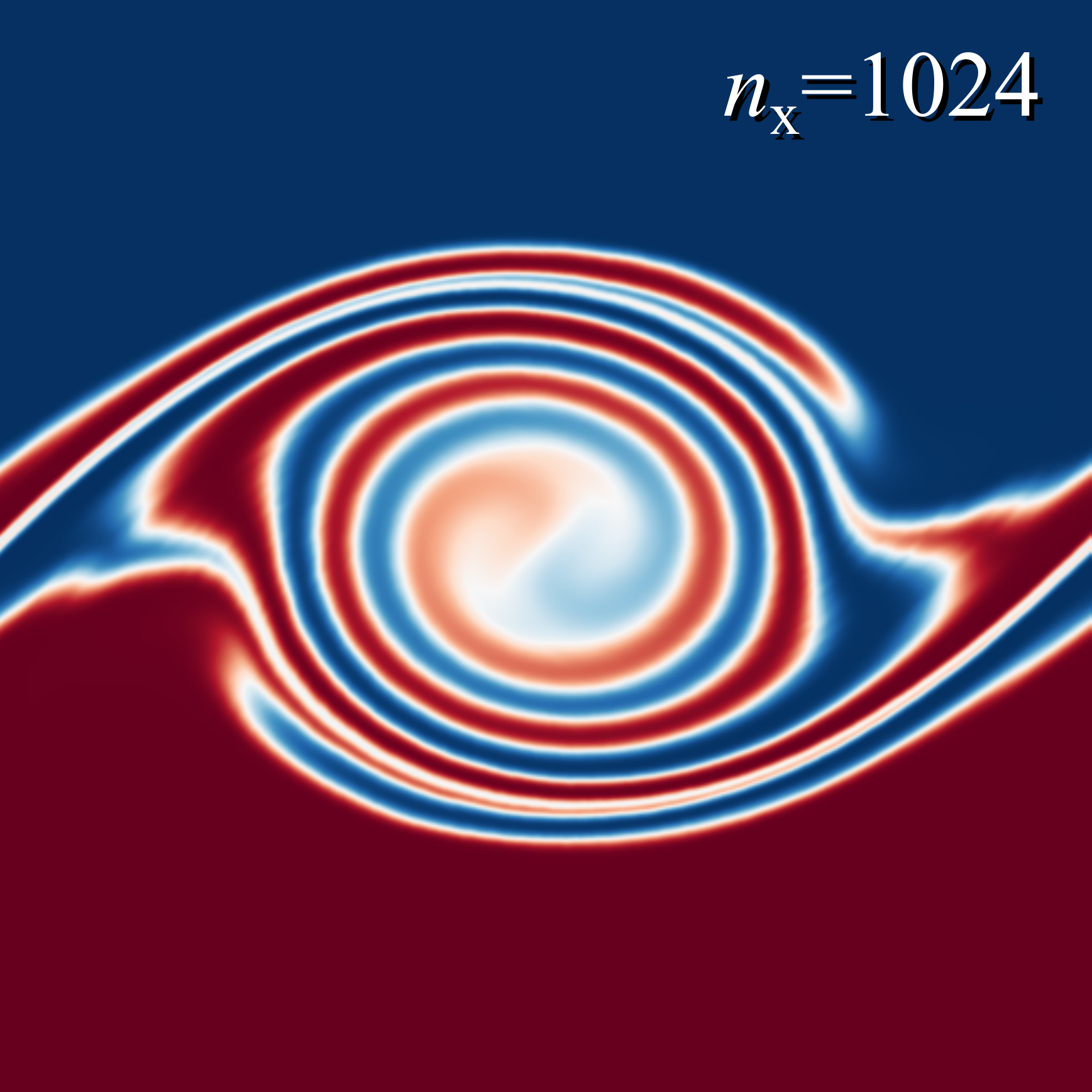} \hspace{-1.85mm}
\includegraphics[width=0.245\linewidth]{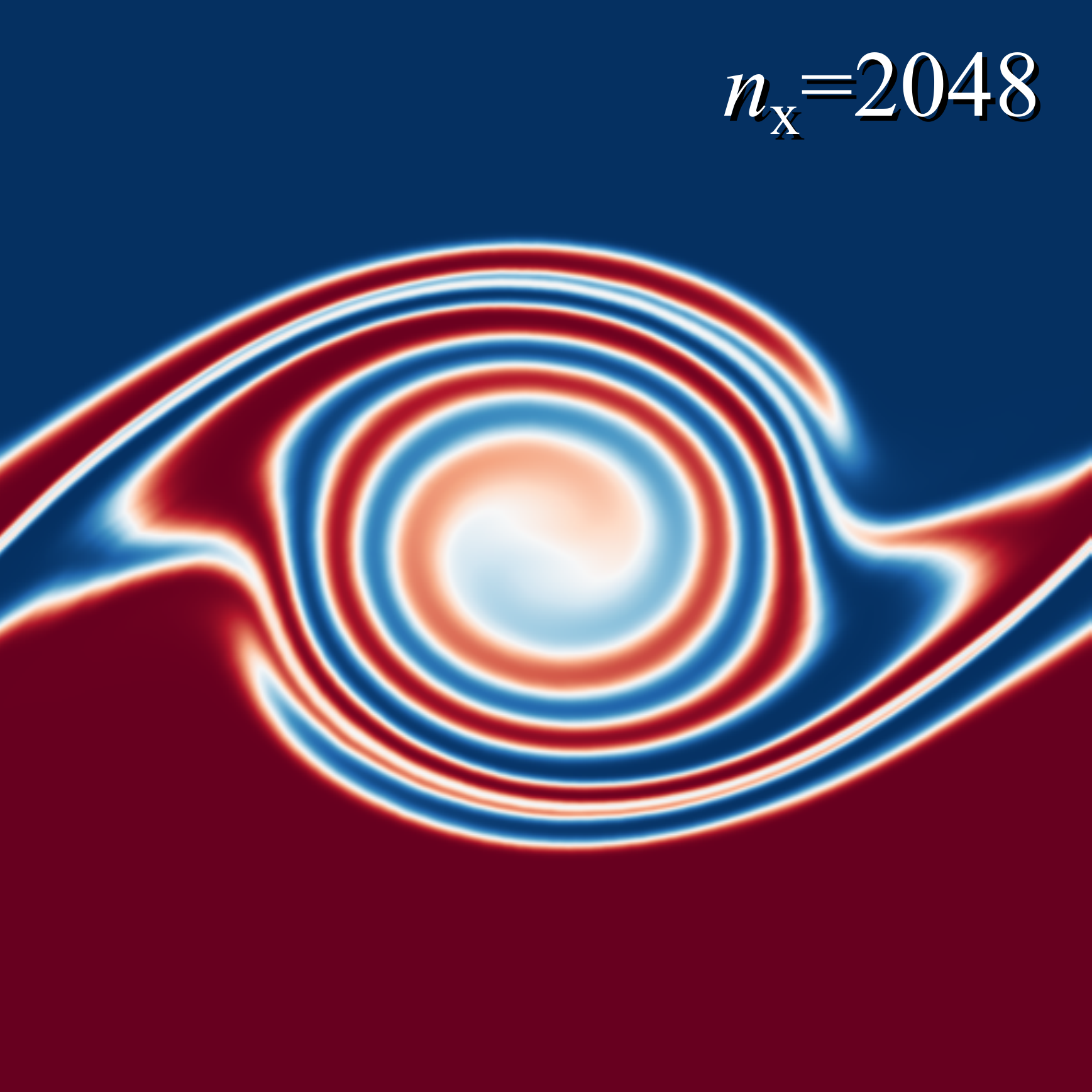} \hspace{-1.85mm}
\includegraphics[width=0.245\linewidth]{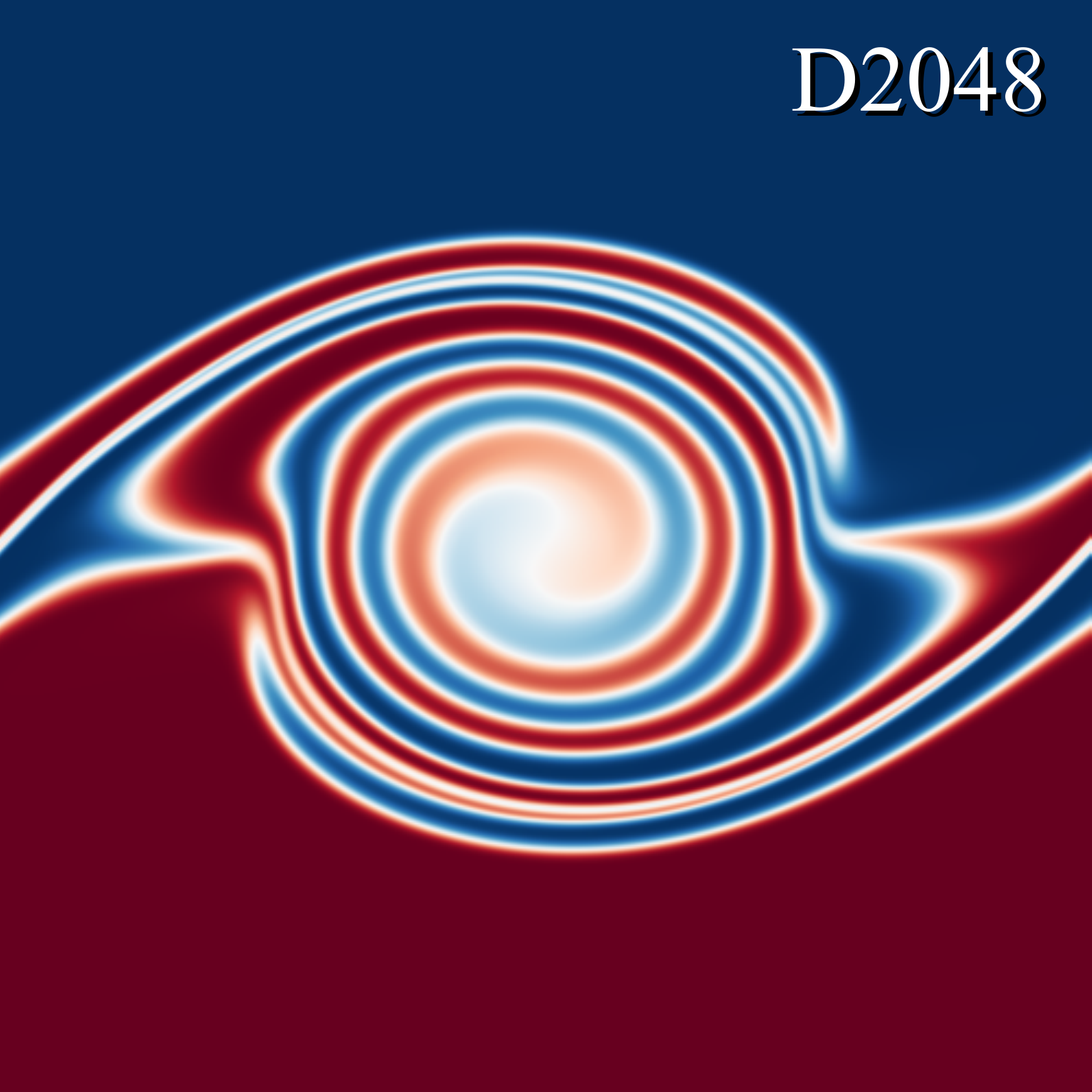} \\
\hspace{-1.85mm}
\includegraphics[width=0.245\linewidth]{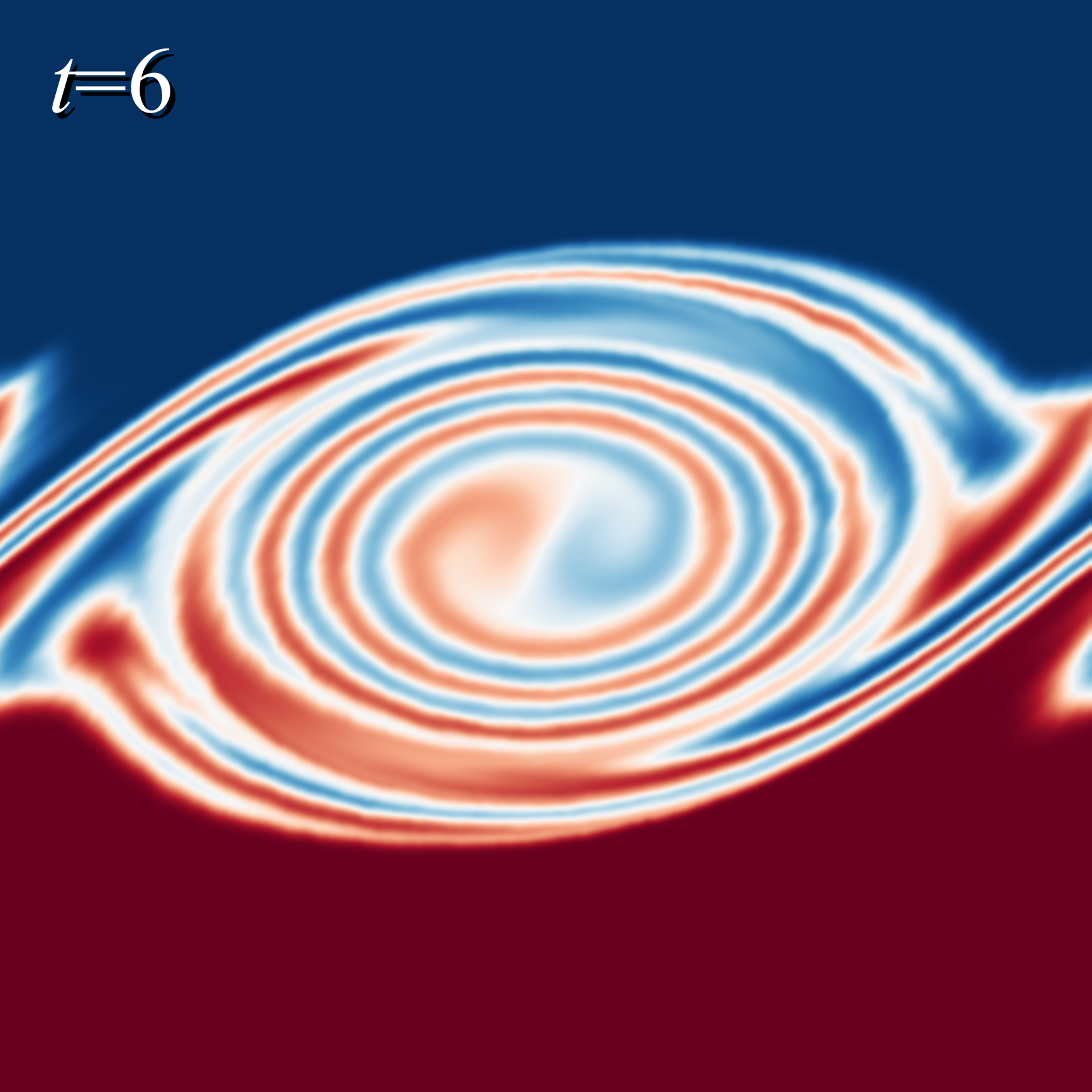}   \hspace{-1.85mm}
\includegraphics[width=0.245\linewidth]{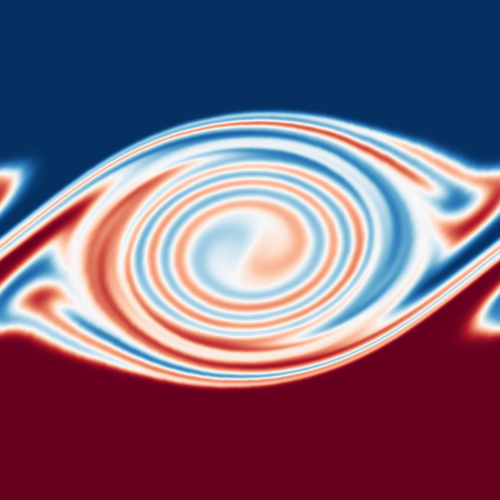} \hspace{-1.85mm}
\includegraphics[width=0.245\linewidth]{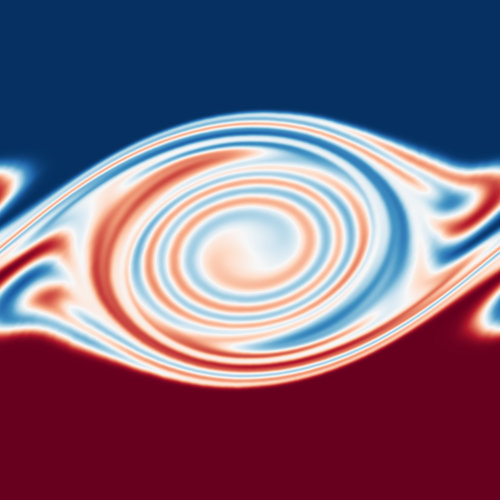} \hspace{-1.85mm}
\includegraphics[width=0.245\linewidth]{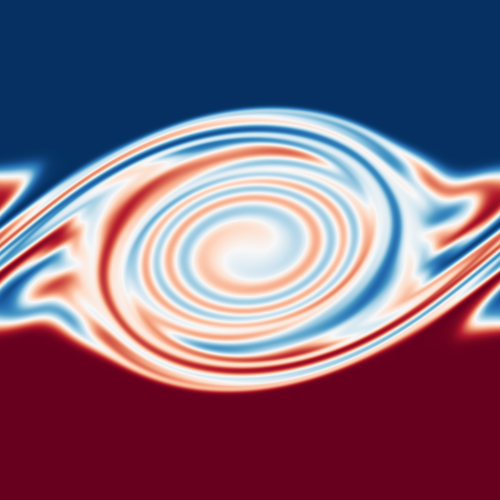} \\
\hspace{-1.85mm}
\includegraphics[width=0.245\linewidth]{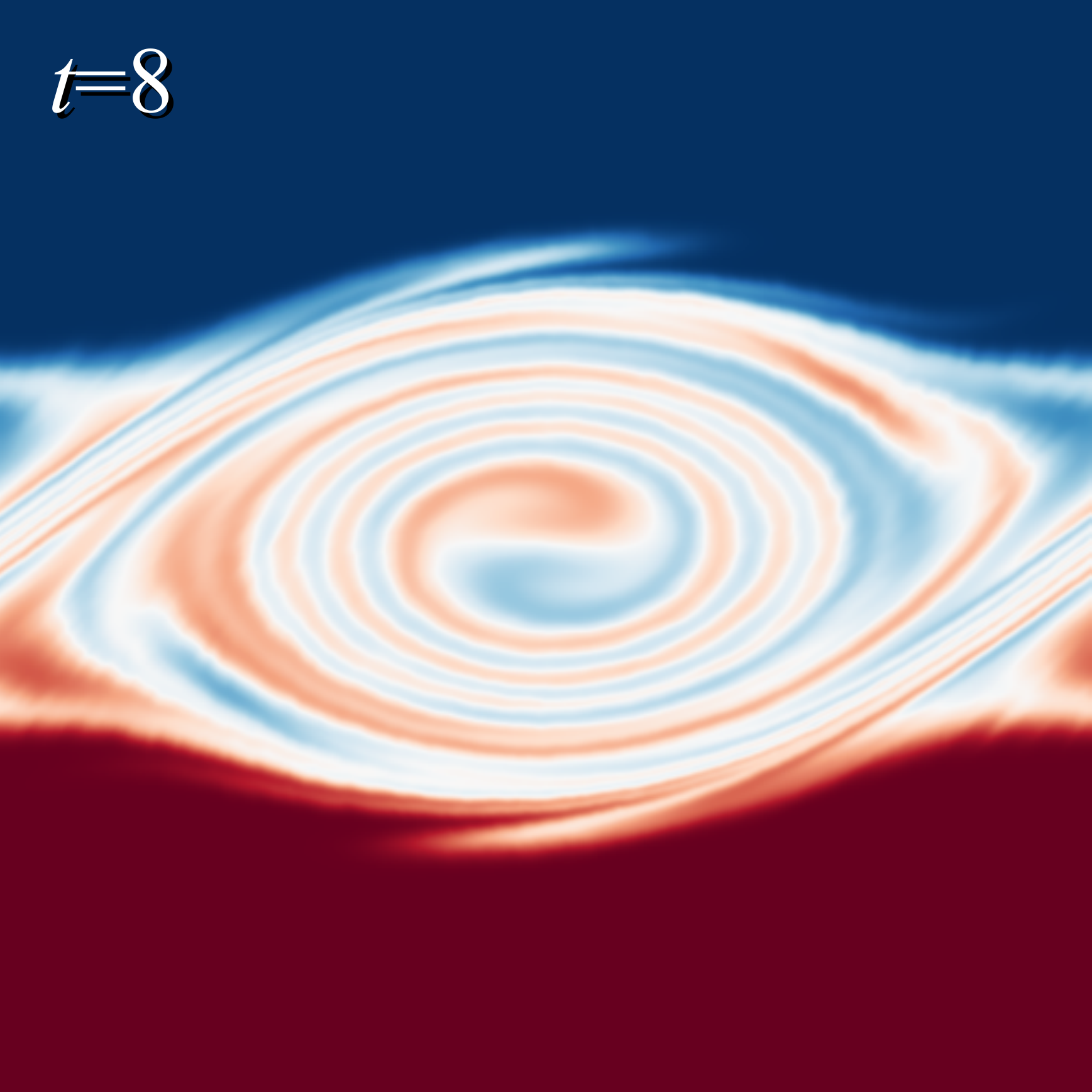}   \hspace{-1.85mm}
\includegraphics[width=0.245\linewidth]{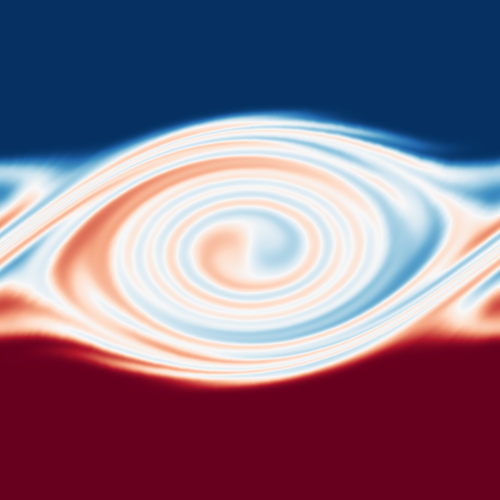} \hspace{-1.85mm}
\includegraphics[width=0.245\linewidth]{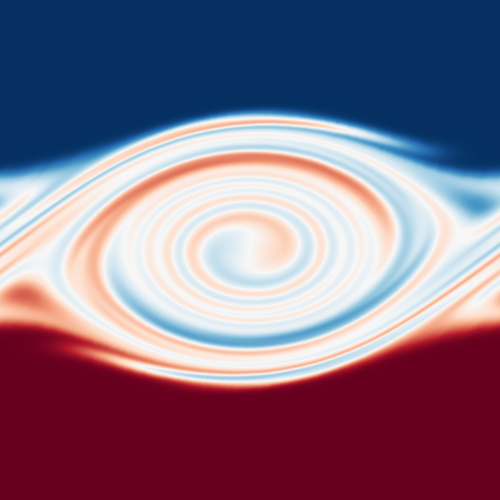} \hspace{-1.85mm}
\includegraphics[width=0.245\linewidth]{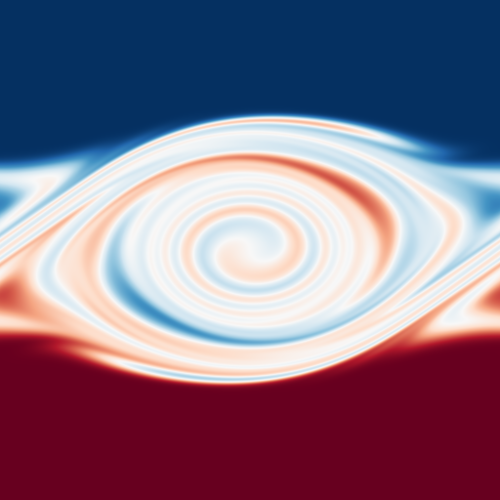} \\
\includegraphics[width=0.983\linewidth]{cbar-wide.pdf}
\caption{The colour field in the region $x,y \in [0,1]$ in the strongly non-linear regime for the $n_{\rm x} = 512$, 1024, and 2048 SPH calculations (three leftmost columns) with the D2048 {\sc Dedalus} calculation (rightmost column) at $t = 4$, 6 and 8 (top, middle and bottom rows, respectively). The SPH calculations reproduce the large-scale morphology of the {\sc Dedalus} calculation at all times. Differences in small-scale features are seen to be converging away as the resolution increases.}
\label{fig:nonlinear}
\end{figure*}

Figure~\ref{fig:nonlinear} shows the colour field at $t=4$, 6 and 8 for the $n_{\rm x}=512$, 1024 and 2048 particle SPH calculations with comparison to the D2048 {\sc Dedalus} calculation. Only the bottom half ($y < 1$) of the calculation is shown since the top and and bottom mirror each other. At these times, the Kelvin-Helmholtz instability is in the strongly non-linear regime. The evolution of the instability in the non-linear regime is dictated by the dissipation -- both numerical and physical. The curl produced in the linear regime ($t=2$; Figure~\ref{fig:t2}) continues to wind tighter as time progresses ($t=4$), producing multiple features as the curl turns into itself ($t>6$). At $t=8$, the physical dissipation applied to the colour field has blended many of the smaller features together, producing a well-mixed colour field along the interface.

The SPH calculations qualitatively reproduce the evolution and morphology of the reference solution. For the $n_{\rm x}=512$ particle calculation, there is noise in the colour field due to particle motion, evident by the `jaggedness' of the filaments comprising the curl. The origin of the noise is attributed to the high Reynolds number (Re=$10^5$) of the calculations, which is challenging to meet at low resolutions. This noise is absent in the $n_{\rm x}=2048$ particle calculation. 

Differences between the solutions, as seen in multiple local features, are reducing as the resolution improves. For all time snapshots, the innermost tip of the red curl is more tightly wound with increasing resolution. At $t=4$, the outer features of the curl become more sharply defined, and, at $t=8$, the top and bottom spurs are lengthening with increasing resolution. In all features, it is clear that the SPH calculations are converging to the reference solution.

\subsection{Colour entropy}

The volume-integrated colour entropy is measured to quantify the mixing of the two regions. Following \citet{lecoanetetal16}, the specific colour entropy is defined to be
\begin{equation}
s \equiv - c \ln(c) ,
\end{equation}
with the total colour entropy given by the volume integral
\begin{equation}
S = \int \rho s {\rm d}V.
\end{equation}
In SPH, this integral can be computed by summation according to
\begin{equation}
S = \sum_a m_a s_a ,
\end{equation}
such that $\rho {\rm d}V$ is equivalent to the mass element ${\rm d}m$. This quantity has the property that it monotonically increases in time when $\nu_{\rm c} > 0$.

\begin{figure}
\includegraphics[width=\linewidth]{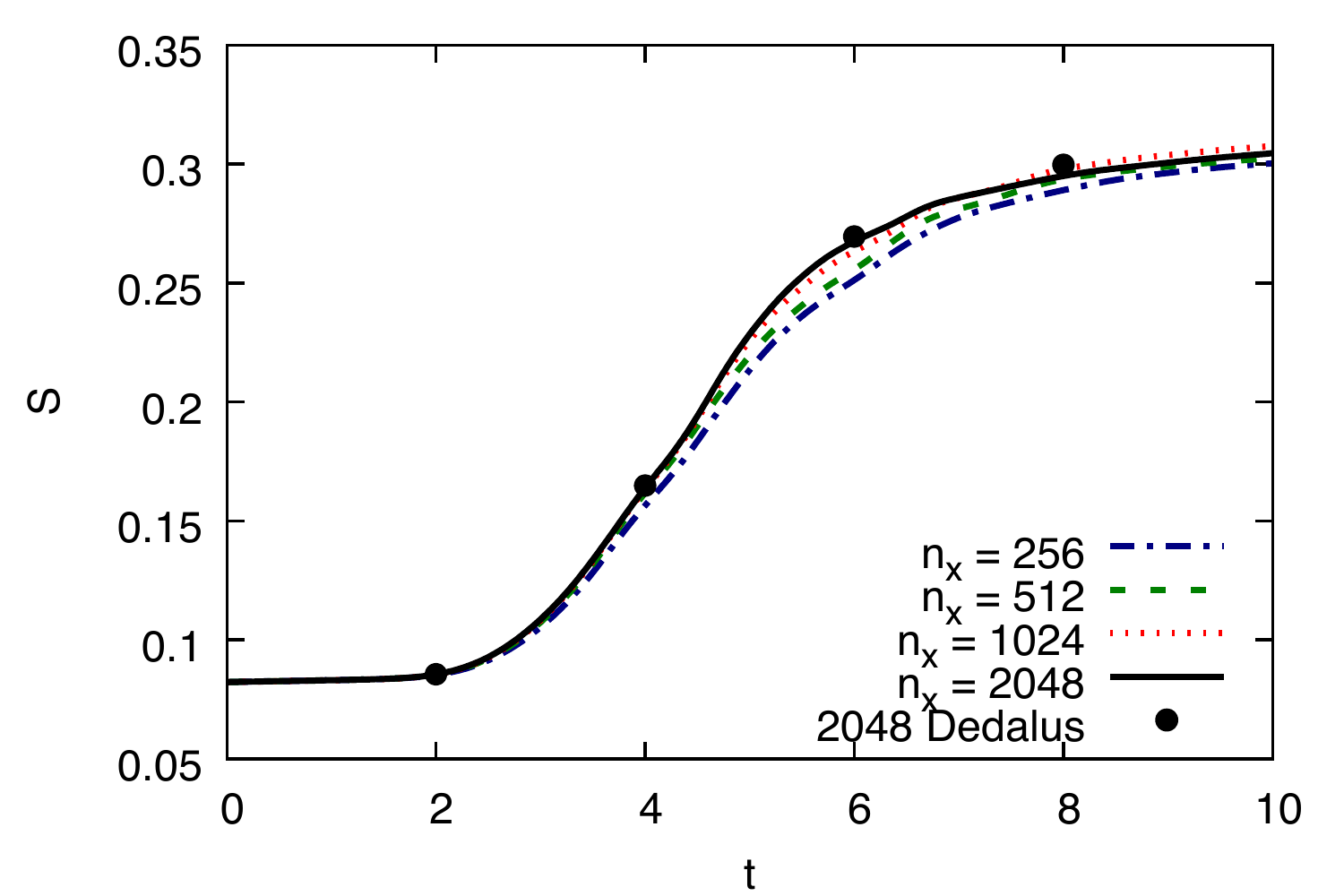}
\caption{Total colour entropy for the $n_{\rm x}= 256$, 512, 1024 and 2048 particle SPH calculations, with the black dots the values from the D2048 solution. The shape and magnitude of the total colour entropy curve agrees with the reference data points for all resolutions.}
\label{fig:colourentropy}
\end{figure}

Figure~\ref{fig:colourentropy} shows the total colour entropy as a function of time for the SPH calculations. The black dots at $t=2$, 4, 6, and 8 are the values from the reference solution. The total colour entropy of the SPH calculations reproduces the behaviour of the reference solution (compare also to Figure~3 in \citealt{lecoanetetal16}). In the nonlinear phase ($t>2$), the difference in total colour entropy between the SPH calculations and the reference data points decreases as the resolution increases. That is, the SPH calculations are converging to the reference solution. At late times, the lower resolution SPH calculations have lower total colour entropy than the reference solution, in contrast to Re=$10^6$ {\sc Athena} results by \citet{lecoanetetal16} which had increased colour entropy for lower resolutions. This difference in behaviour is caused by advection in grid-based codes introducing numerical diffusion, thereby artificially increasing the rate of mixing. Advection is exact in SPH.

Deviations of the total colour entropy from the reference solution are minor even for our lowest resolution of $n_{\rm x}=256$ particles,  and the shape and magnitude of the total colour entropy curve is well represented. From a practical perspective, SPH, even at modest resolutions, not only reproduces the correct qualitative features of the Kelvin-Helmholtz instability (c.f.~Figure~\ref{fig:nonlinear}), but also the correct amount of mixing.

\subsection{$\mathcal{L}_2$ error}

The $\mathcal{L}_2$ error is computed to quantify the convergence of the SPH calculations towards the D2048 calculation. To compute the error between the two calculations, the SPH particles are interpolated to a $2048 \times 4096$ grid so that the two solutions can be directly compared. The colour at each grid point is computed according to
\begin{equation}
c({\bm r}) = \frac{\sum_b m_b c_b W_{ab}(h_b)}{\sum_b m_b W_{ab}(h_b	)} ,
\end{equation}
where the summation in the denominator is the normalisation condition. The $\mathcal{L}_2$ error between the SPH interpolated grid and D2048 grid is computed as
\begin{equation}
\mathcal{L}_2 = \left[ {\rm d}V \sum_a \left| c_a^{\rm SPH} - c_a^{\rm D2048} \right|^2 \right]^{1/2} ,
\end{equation}
where ${\rm d}V$ is the volume of each grid cell (i.e., $1 / 2048^2$), and $c^{\rm SPH}$ and $c^{\rm D2048}$ are the colour fields of the SPH interpolated grid and the D2048 grid, respectively.

\begin{figure}
\includegraphics[width=\linewidth]{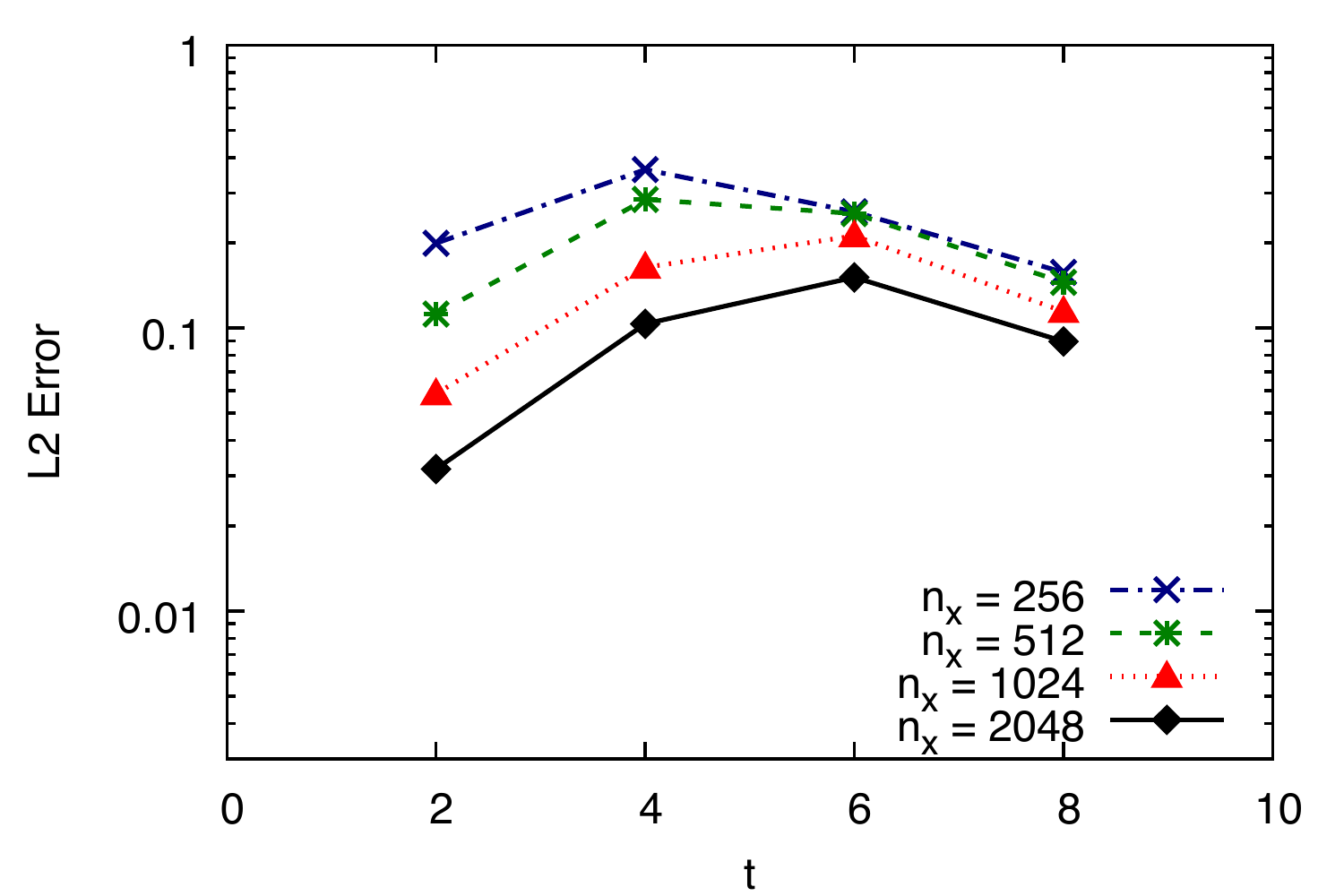} 
\caption{$\mathcal{L}_2$ error of the SPH calculations with respect to the D2048 solution at $t=2$, 4, 6, and 8. The connecting lines are merely for visual aid. The convergence rate is linear at $t=2$. In the late non-linear regime ($t>6$), the error only decreases once the Navier-Stokes viscosity begins to be resolved. The SPH calculations are converging to the reference solution at all times, with the $\mathcal{L}_2$ error of the $n_{\rm x}=2048$ particle calculation lower than $n_{\rm x}=1024$ particle calculation, which is lower than the $n_{\rm x}=512$ particle calculation.}
\label{fig:l2err}
\end{figure}

\begin{table}
\caption{Convergence rates of the $\mathcal{L}_2$ error fit to $\propto n_{\rm x}^{-\sigma}$}
\label{tbl:convergence}
\begin{tabular}{ccc}
\hline \\[-6pt]
t & $\sigma$ (artificial viscosity switch) & $\sigma$ ($\alpha$=0) \\[2pt]
\hline \\[-6pt]
2.0 &  0.875 & 1.108 \\
4.0 & 0.623 & 0.754 \\
6.0 &  0.257 & 0.381 \\
8.0 & 0.278 & 0.376 \\
\hline 
\end{tabular}
\end{table}

Figure~\ref{fig:l2err} shows the $\mathcal{L}_2$ error between the SPH calculations and the D2048 {\sc Dedalus} calculation. The error is computed at $t=2$, 4, 6 and 8, with the lines between points present only for visual aid. The convergence rates at each time are listed in Table~\ref{tbl:convergence}. The convergence rate at $t=2$ is close to linear, with each doubling of resolution reducing the $\mathcal{L}_2$ error by half. The solution at $t=4$ is converging towards the reference solution, though at a rate slower than linear. In the late non-linear regime of the instability ($t=6$ and $t=8$), the SPH solution only begins to demonstrate convergence for resolutions above $n_{\rm x}=512$ particles. The error between the $n_{\rm x}=256$ and $n_{\rm x}=512$ is similar at these times, a consequence of the artificial viscosity still dominant over the Navier-Stokes viscosity. The long term behaviour, for the resolutions presented, is governed still by the numerical dissipation. Only when the physical dissipation begins to be resolved, that is in the $n_{\rm x}=1024$ and $n_{\rm x}=2048$ particle calculations, does the $\mathcal{L}_2$ error decrease with resolution, though at a rate slower than linear. The effect of numerical dissipation is investigated in detail in Section~\ref{sec:numdiss}.

Overall, the SPH calculations are converging towards the D2048 reference solution. At all times, the $\mathcal{L}_2$ errors of the $n_{\rm x}=2048$ particle calculation are lower than for the $n_{\rm x}=1024$ particle calculation, which are lower than the $n_{\rm x}=512$ particle calculation.

\subsection{Convergence of the dissipation rate}
\label{sec:numdiss}

One question is whether the dissipation of kinetic energy is dominated by the physical Navier-Stokes viscosity or the numerical dissipation of the artificial viscosity. This may be estimated by simple analytic arguments and through measurements directly from the calculations.

As discussed by \citet{al94}, \citet{murray96}, and \citet{lp10}, the $\alpha$ term of the artificial viscosity can be equated to a shear and bulk Navier-Stokes viscosity, with shear component
\begin{equation}
\nu^{\rm AV} = \frac{1}{10} \alpha v_{\rm sig} h ,
\end{equation}
and bulk component that is $\zeta^{\rm AV} = 5/3 \times \nu^{\rm AV}$. Since the fluid is subsonic, the \citet{mm97} artificial viscosity switch is effective at reducing $\alpha$ for all particles in all calculations to $\alpha \approx \alpha_0$ where $\alpha_0=0.1$. For the $n_{\rm x}=256$ particle calculation, this would imply that $\nu^{\rm AV} \approx 18 \nu$, that is, the numerical dissipation is eighteen times greater than the Navier-Stokes viscosity. For our highest resolution calculation ($n_{\rm x}=2048$), it would thus be expected that $\nu^{\rm AV}$ be roughly twice as large as $\nu$, with the Navier-Stokes viscosity becoming dominant over the numerical dissipation only for resolutions $n_{\rm x} > 4096$ particles.

\begin{figure}
\includegraphics[width=\linewidth]{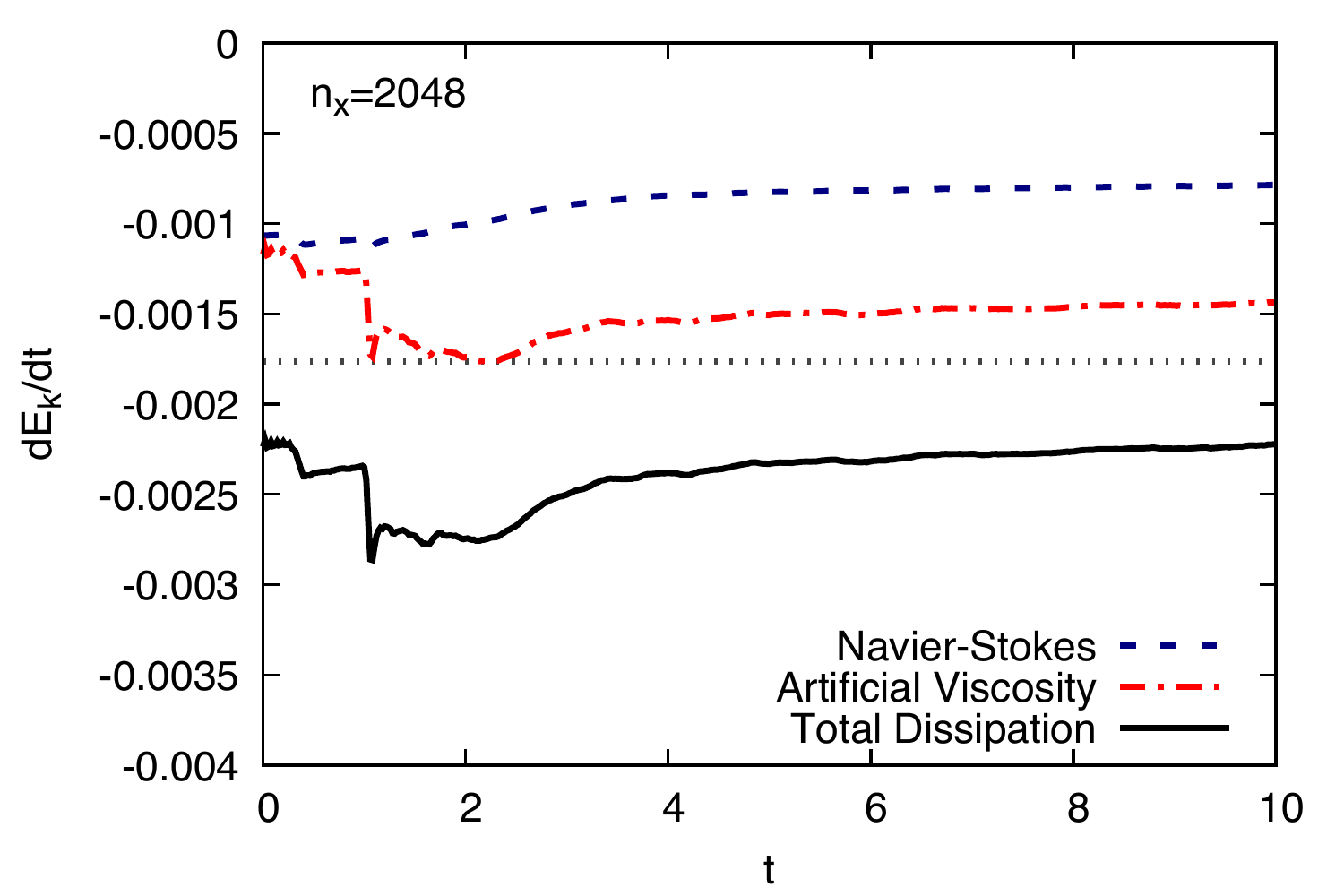}
\caption{Rate of change of kinetic energy (${\rm d}E_k/{\rm d}t$) from sources of dissipation, as directly measured from the $n_{\rm x}=2048$ particle calculation. The grey dotted line represents twice the average Navier-Stokes dissipation. In agreement with analytic expectation, the artificial viscosity is about twice as dissipative as the explicit Navier-Stokes viscosity at this resolution. The increase of dissipation by the artificial viscosity at $t \approx 1$ corresponds to when the particles first break off the initial lattice arrangement.}
\label{fig:diss}
\end{figure}

The rate at which kinetic energy is dissipated by each source of dissipation may be directly measured from the calculations. Both the artificial viscosity and Navier-Stokes viscosity are explicitly added sources of dissipation. The rate of change of kinetic energy can be measured by computing
\begin{equation}
\frac{{\rm d}E_k}{{\rm d}t} = \sum_a m_a {\bm v}_a \cdot \frac{{\rm d}{\bm v}_a}{{\rm d}t} ,
\end{equation}
and substituting in ${\rm d}{\bm v}/{\rm d}t$ from the artificial viscosity (from Equation~\ref{eq:sphmom}) and Navier-Stokes viscosity (Equation~\ref{eq:sphNS}). 

Figure~\ref{fig:diss} shows ${\rm d}E_k/{\rm d}t$ for the artificial viscosity, Navier-Stokes viscosity, and the sum of the two for the $n_{\rm x}=2048$ calculation. This measurement confirms the analytic estimate that the rate at which kinetic energy is dissipated by the artificial viscosity is approximately twice that of the Navier-Stokes viscosity. The $\beta$ viscosity must be an insignificant contribution to the overall numerical dissipation in this case. The sharp increase of dissipation by the artificial viscosity at $t\approx 1$ occurs from the particles breaking off the initial lattice arrangement. 

These analyses suggest that the non-linear evolution of the SPH calculations is influenced considerably by numerical dissipation, with a ratio of 2:1 of numerical to physical dissipation for our highest resolution calculation. The ratio of kinetic energy dissipation is expected to be equal for $n_{\rm x}=4096$ particles, with the explicit Navier-Stokes viscosity becomining the dominant source of dissipation only at resolutions greater than that. 

It is worth noting that the calculations in this work have used the \citet{mm97} switch with $\alpha_0=0.1$. More sophisticated artificial viscosity limiters have been developed in recent years that are better at reducing numerical dissipation \citep{cd10, rh12, gasoline2}, in particular permitting the use of $\alpha_0=0$ in some instances. Using one of these more advanced switches should improve the convergence properties of SPH on this Kelvin-Helmholtz test. This is explored further in Section~\ref{sec:alpha0}.

\subsection{Reducing the artificial viscosity ($\alpha=0$)}
\label{sec:alpha0}

To investigate the convergence properties when the Navier-Stokes viscosity is the dominant source of kinetic energy dissipation, we have repeated our highest resolution calculation ($n_{\rm x}=2048$) using a fixed $\alpha=0$ for all particles. By reducing the numerical dissipation, this result should be indicative of higher resolutions calculations that use typical artificial viscosity parameters and switches.

It is important to be careful about setting $\alpha=0$ since the primary role of artificial viscosity in these calculations is to maintain particle regularity. As discussed in Section~\ref{sec:numdiss}, the shear viscosity introduced by the artificial viscosity for the $n_{\rm x}=2048$ particle calculation is about twice that of the explicitly added Navier-Stokes viscosity. It is reasonable to expect that the Navier-Stokes viscosity may be sufficient by itself to keep the particles regular for this resolution. Note that the $\beta$ term of the artificial viscosity is retained for these calculations since it represents a von Neumann-Richtmyer type viscosity. Its contribution to the overall dissipation of kinetic energy should be small.

\begin{figure}
\centering
\hspace{-1.85mm}
\includegraphics[width=0.49\linewidth]{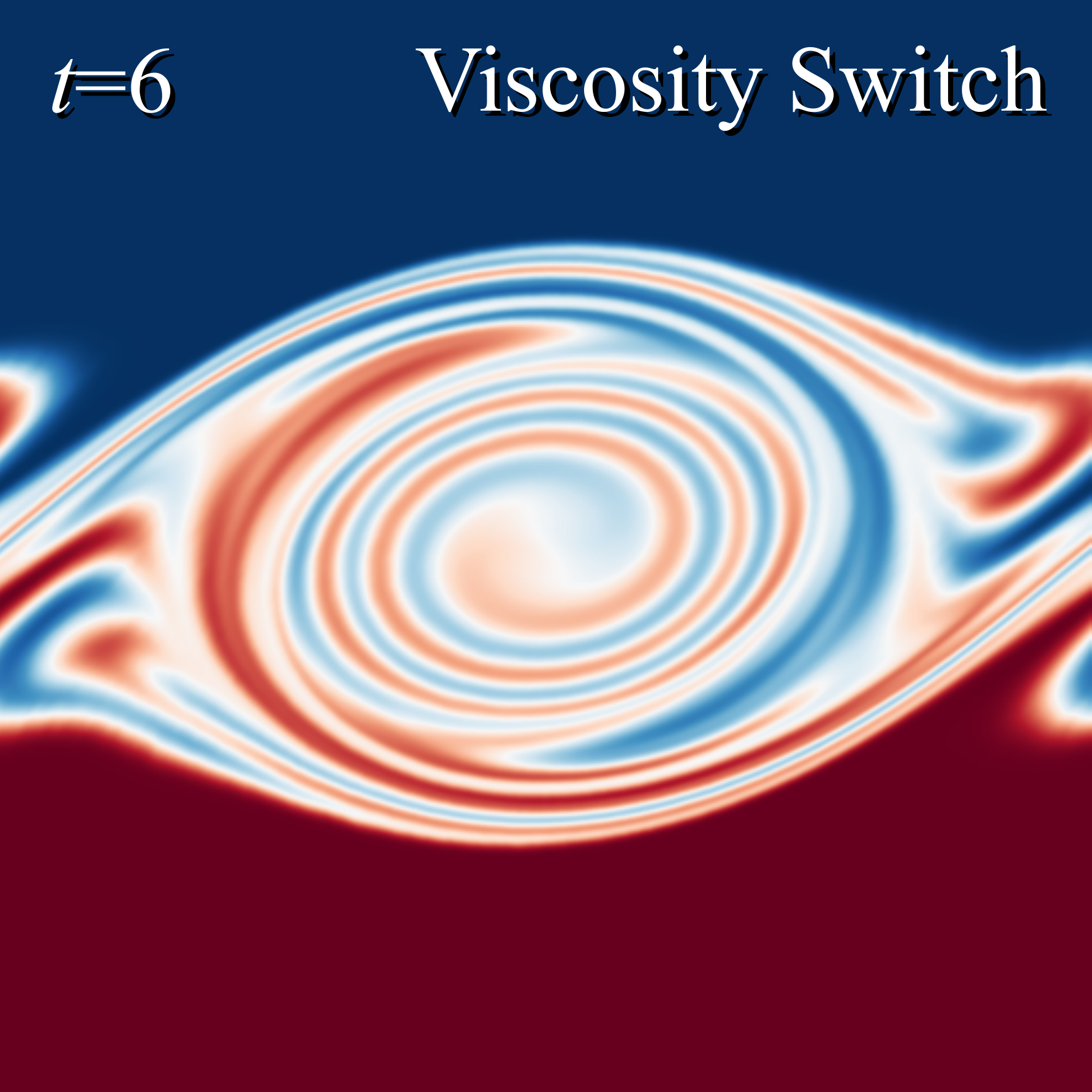}   \hspace{-1.85mm}
\includegraphics[width=0.49\linewidth]{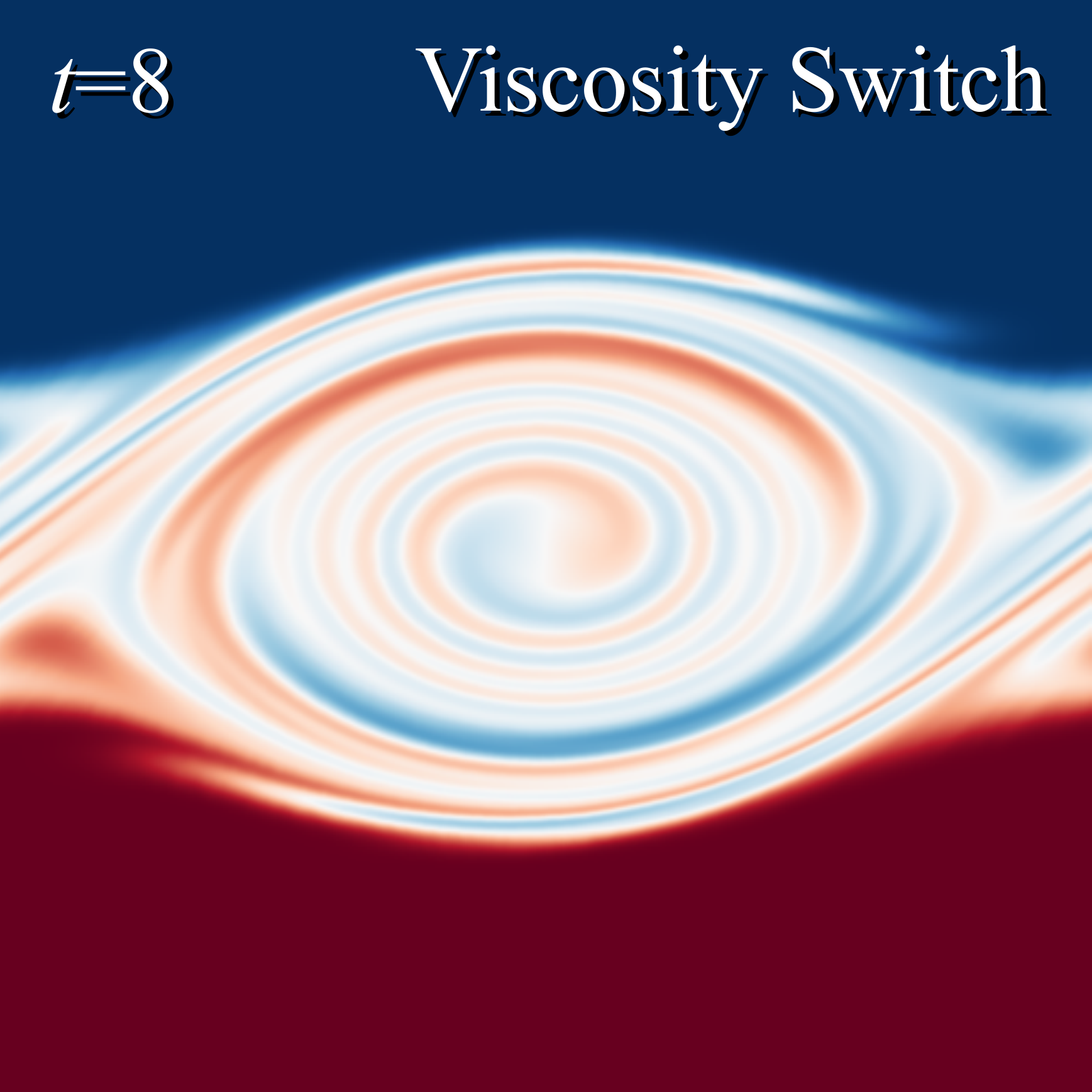} \\
\hspace{-1.85mm}
\includegraphics[width=0.49\linewidth]{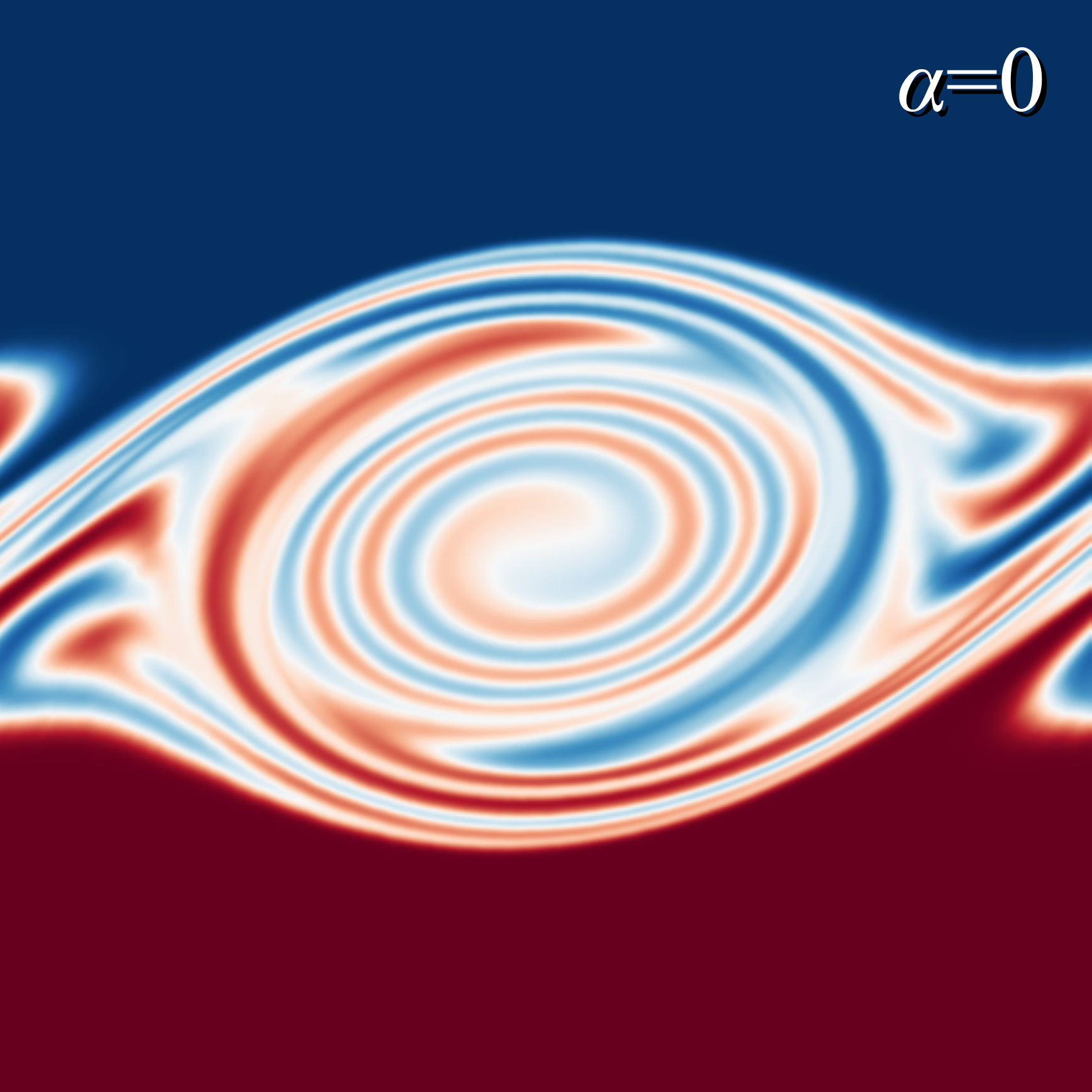}   \hspace{-1.85mm}
\includegraphics[width=0.49\linewidth]{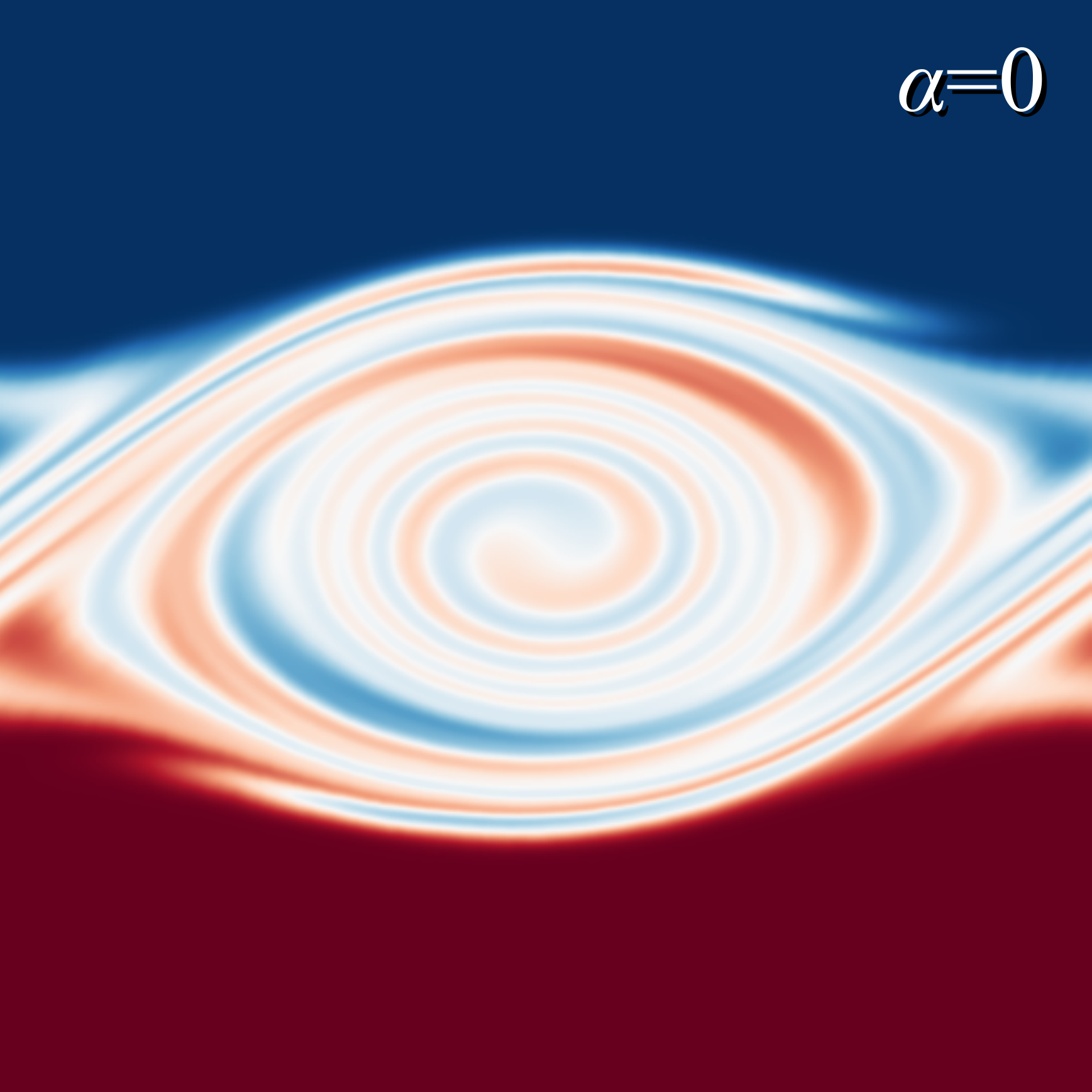} \\
\hspace{-1.85mm}
\includegraphics[width=0.49\linewidth]{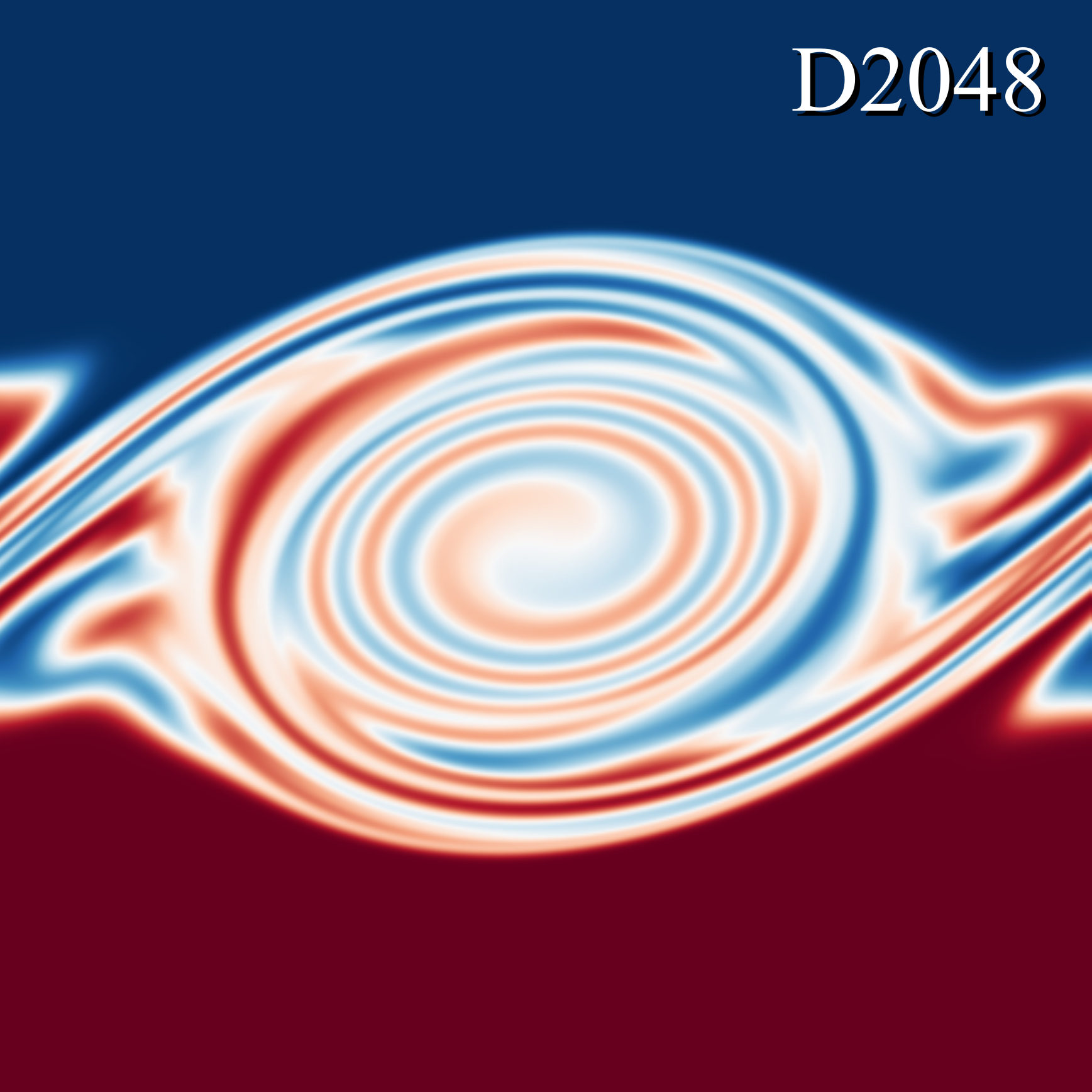} \hspace{-1.85mm}
\includegraphics[width=0.49\linewidth]{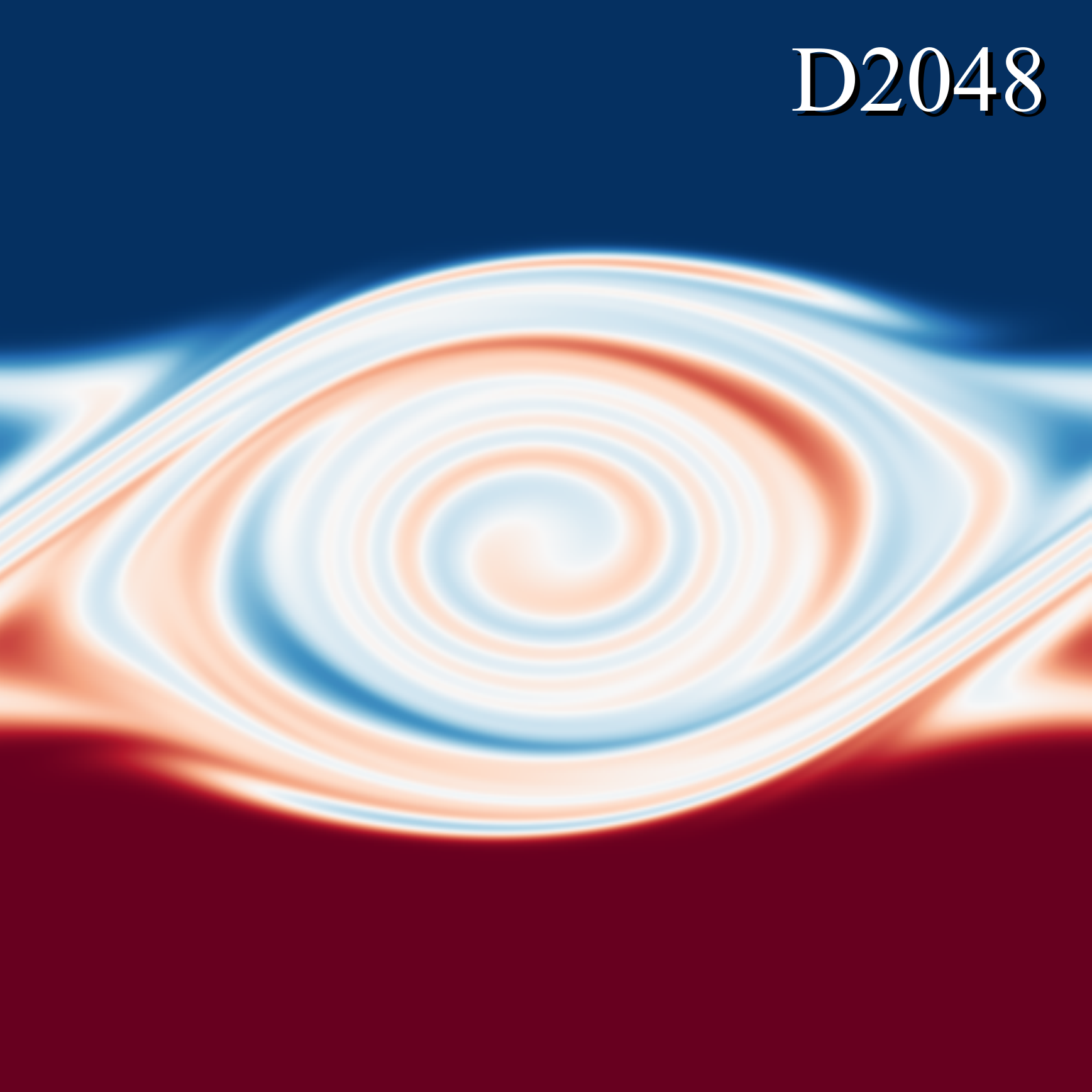} 
\includegraphics[width=0.983\linewidth]{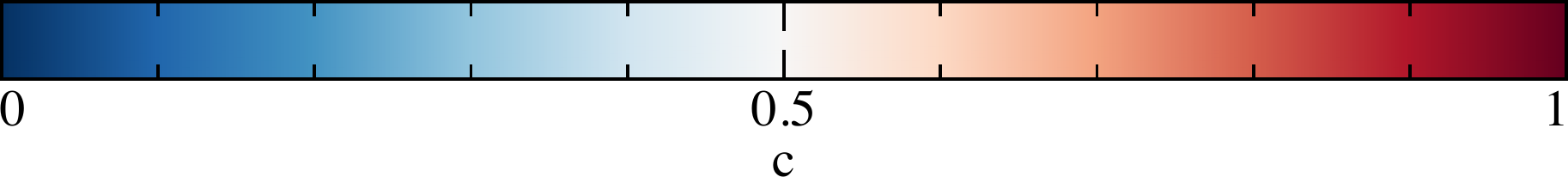}
\caption{The colour field in the region $x,y \in [0,1]$ at $t=6$ (left column) and $t=8$ (right column) for the standard $n_{\rm x}=2048$ particle SPH calculation using the \citet{mm97} viscosity switch (first row), using a fixed $\alpha=0$ to reduce artificial viscosity (middle row), and the D2048 reference solution for comparison (bottom row). At this resolution, the explicit Navier-Stokes viscosity is sufficient to keep the particles regular. By reducing the numerical dissipation ($\alpha = 0$), the qualitative agreement between the SPH and D2048 solutions is dramatically improved in comparison to calculations that employ the \citet{mm97} artificial viscosity switch. }
\label{fig:noav}
\end{figure}

Figure~\ref{fig:noav} shows the colour field at $t=6$ and $t=8$ for fixed $\alpha=0$. The qualitative agreement between the SPH calculation and the reference solution is significantly improved in comparison to the standard calculation (c.f.~Figure~\ref{fig:nonlinear}). Most noticeable is the inner tip of the curl, which is wound as tightly as the reference solution in both time snapshots. The thick red filament in the upper region of the curl at $t=8$ is concentrated towards the upper right, as in the reference solution, whereas this feature in the standard calculation is diffuse across the top of the curl. Overall, numerous features can be identified by eye as resembling the reference solution more closely in this calculation than for the standard calculation.

\begin{figure}
\includegraphics[width=\linewidth]{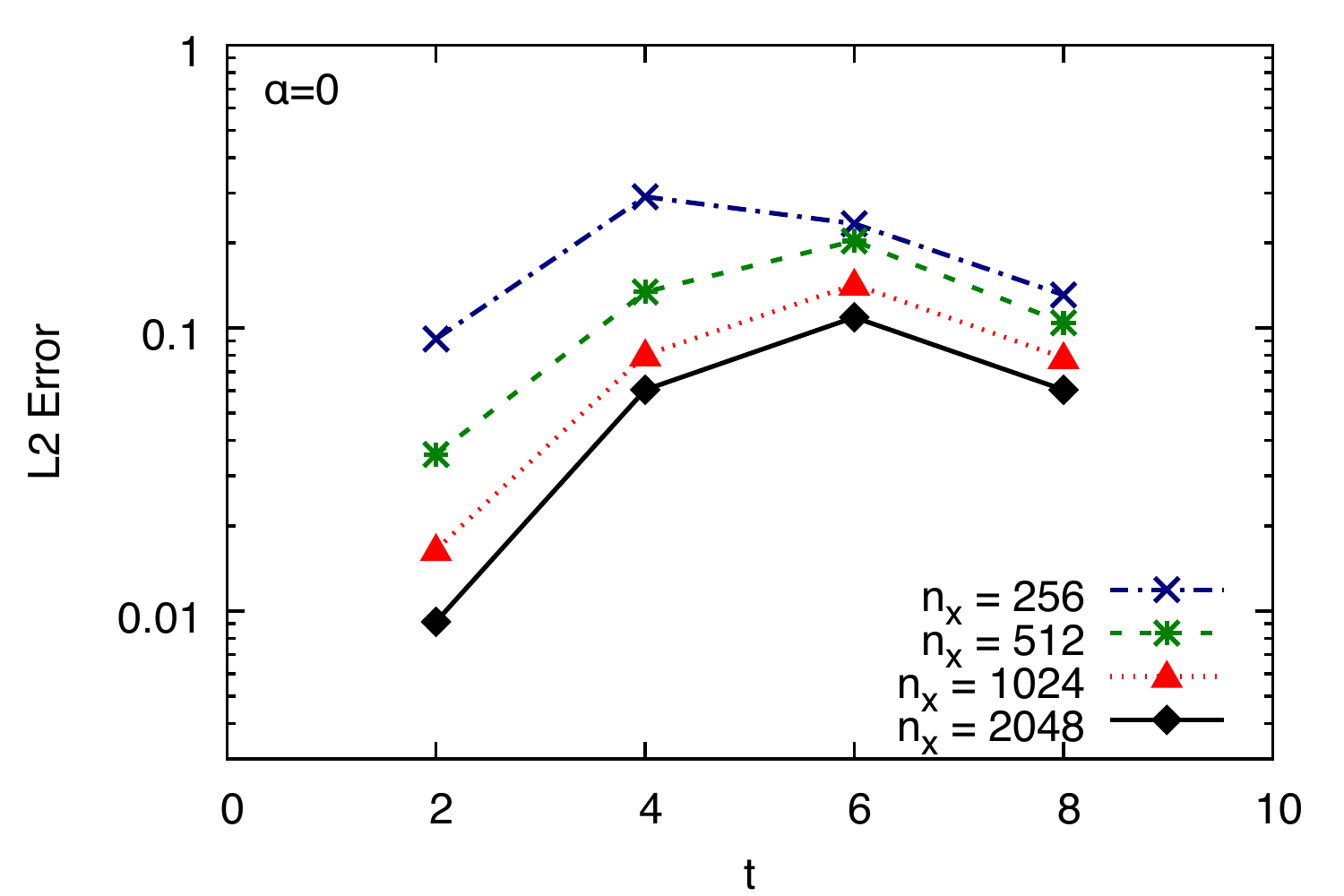} 
\caption{$\mathcal{L}_2$ error of SPH calculations using fixed $\alpha=0$ in the artificial viscosity with respect to the D2048 solution at $t=2$, 4, 6, and 8. The connecting lines are for visual aid. The solution of the $n_{\rm x}=2048$ particle calculation with reduced numerical dissipation is in better agreement to the D2048 reference solution than the standard SPH calculation (c.f.~Figure~\ref{fig:l2err}).}
\label{fig:l2err-noav}
\end{figure}

Figure~\ref{fig:l2err-noav} shows the $\mathcal{L}_2$ error for fixed $\alpha=0$ calculations for resolutions of $n_{\rm x}=256$ to $n_{\rm x}=2048$ particles. The convergence rates are given in Table~\ref{tbl:convergence}. At low resolutions, substantial noise occurs in the particle arrangement due to the absence of sufficient viscosity (either numerical or physical). The results at low resolution are not significantly improved in the non-linear regime, except for at $t=2$ since the particles have yet to re-arrange significantly. However, at high resolution ($n_{\rm x} = 2048$), the $\mathcal{L}_2$ error is uniformly lower than the standard calculation (c.f.~Figure~\ref{fig:l2err}), by approximately $20\%$ at $t=2$ and $30\%$ in the late non-linear regime ($t\ge4$). The solution obtained at this resolution is more accurate than the standard calculation, suggesting that SPH with typical artificial viscosity parameters will indeed converge to the reference solution as the resolution increases due to decreasing numerical dissipation. This is substantiated by the improved rate of convergence when the artificial viscosity is reduced (c.f. Table~\ref{tbl:convergence}), although the rate of convergence still remains sub-linear in the non-linear regime of the instability.

It is expected that the results of this calculation would be similar to a calculation which used a higher order artificial viscosity limiter, such as the approaches by \citet{cd10}, \citet{rh12} or \citet{gasoline2}. The \citet{cd10} artificial viscosity switch, for example, uses the time derivative of the divergence of the velocity, ${\rm d}(\nabla \cdot {\bm v})/{\rm d}t$, promising more accurate shock detection than the \citet{mm97} switch. In practice, this permits the use of $\alpha_0=0$ (e.g.,~Equation~\ref{eq:mm97}), granting significant reduction in numerical dissipation. Therefore, it is reasonable to expect that more accurate switches would allow for $\alpha$ to be near zero for this problem, and that the result obtained here are indicative of results that use a higher order artificial viscosity switch.

\subsection{Kernel bias}
\label{sec:kernelbias}

The calculations performed in this work used the M8 septic spline. It is important to note that the use of a high-order kernel is not an intrinsic requirement to capture the Kelvin-Helmholtz instability. Rather, a high-order B-spline has been used to minimise velocity noise caused by errors in the pressure gradient. This is examined in greater detail below.

\begin{figure*}
\hspace{-1.85mm}
\includegraphics[width=0.195\linewidth]{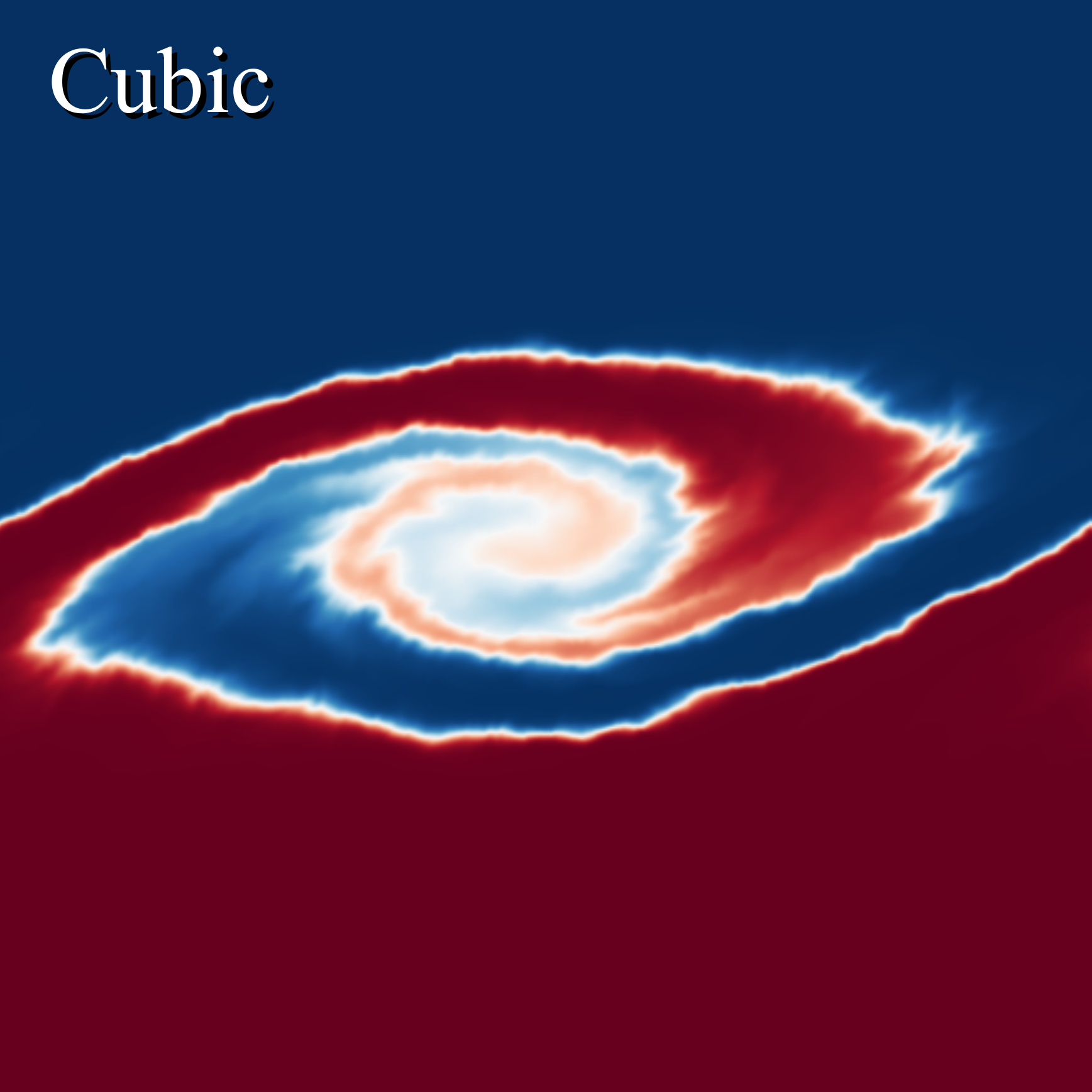}  \hspace{-1.85mm}
\includegraphics[width=0.195\linewidth]{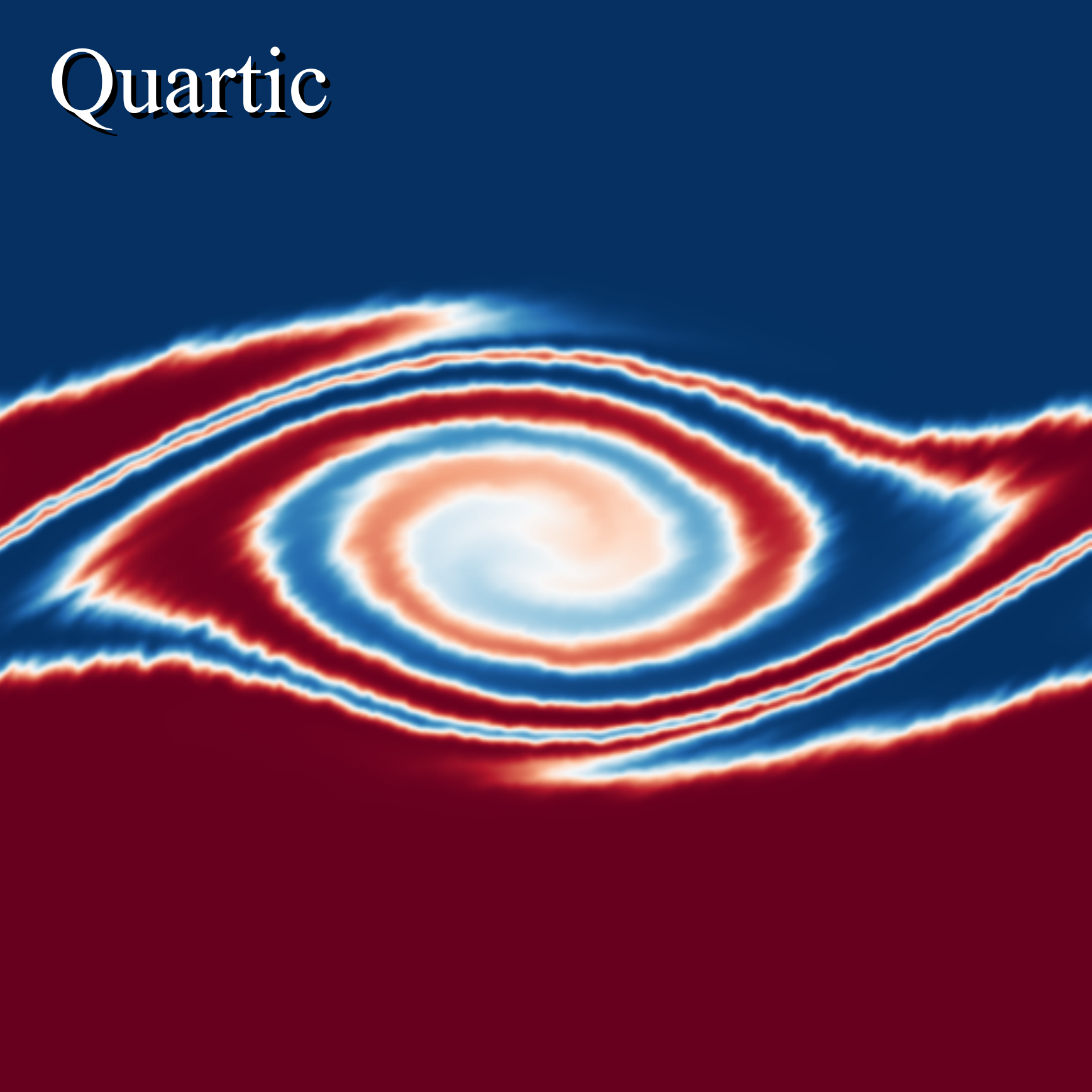} \hspace{-1.85mm}
\includegraphics[width=0.195\linewidth]{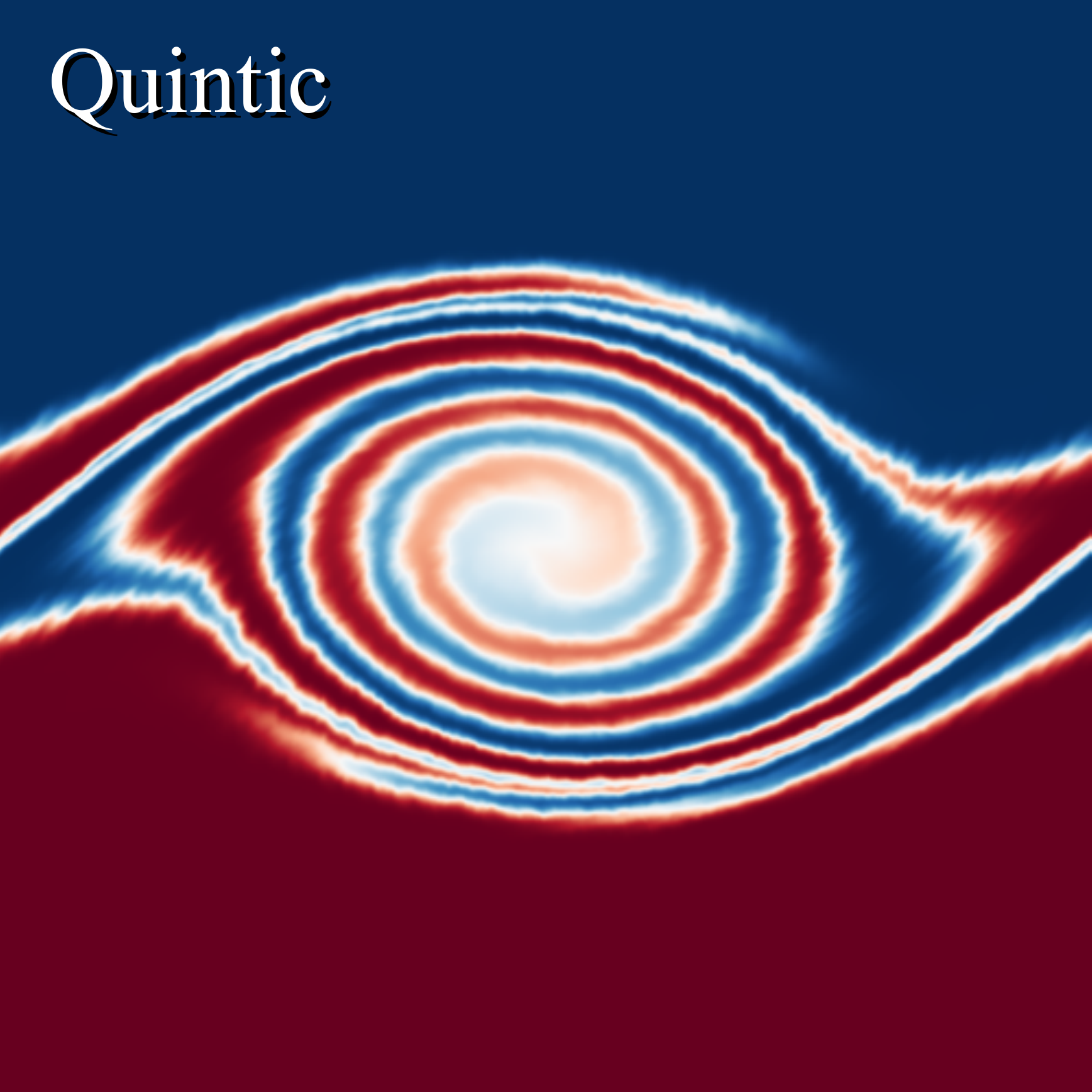} \hspace{-1.85mm}
\includegraphics[width=0.195\linewidth]{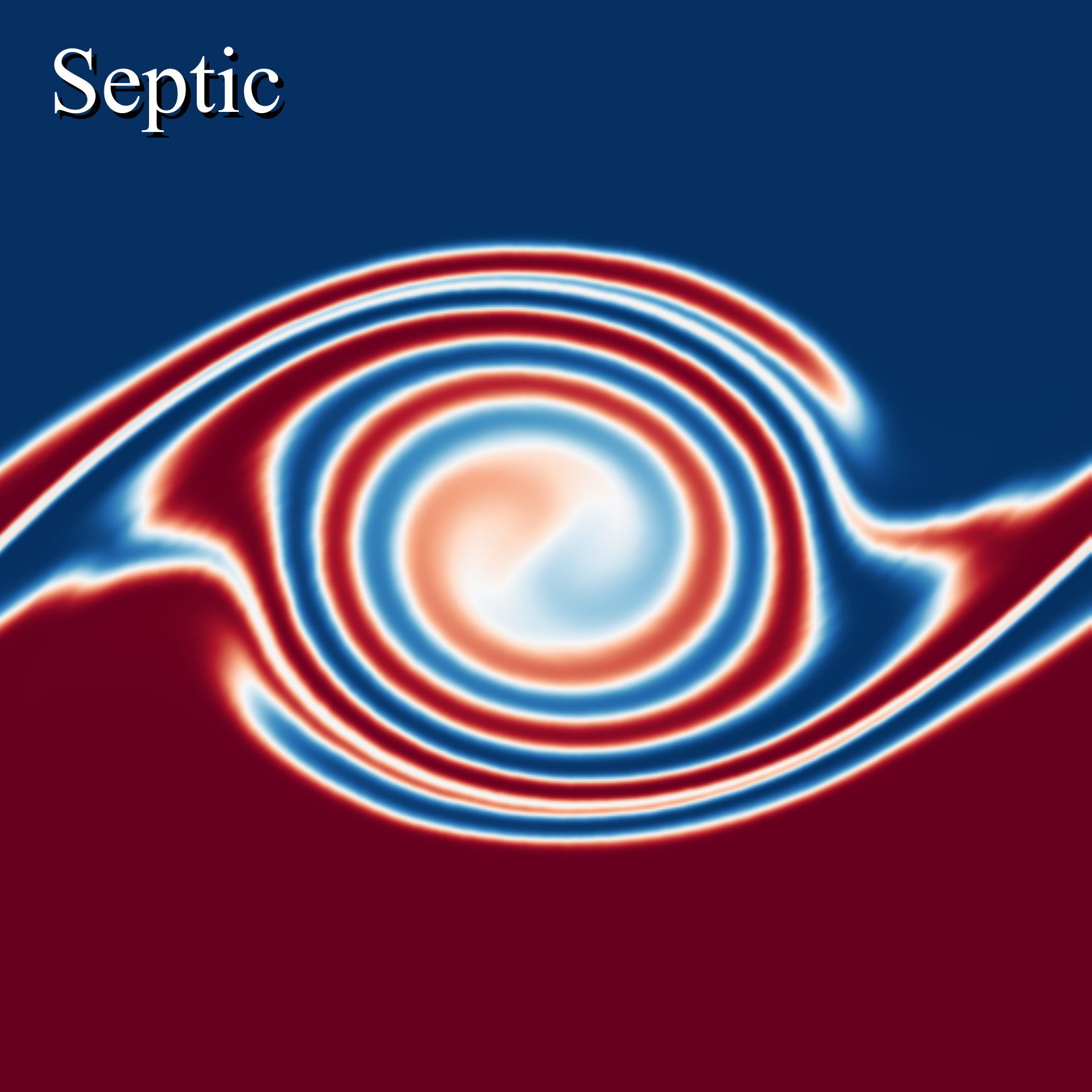} \hspace{-1.85mm}
\includegraphics[width=0.195\linewidth]{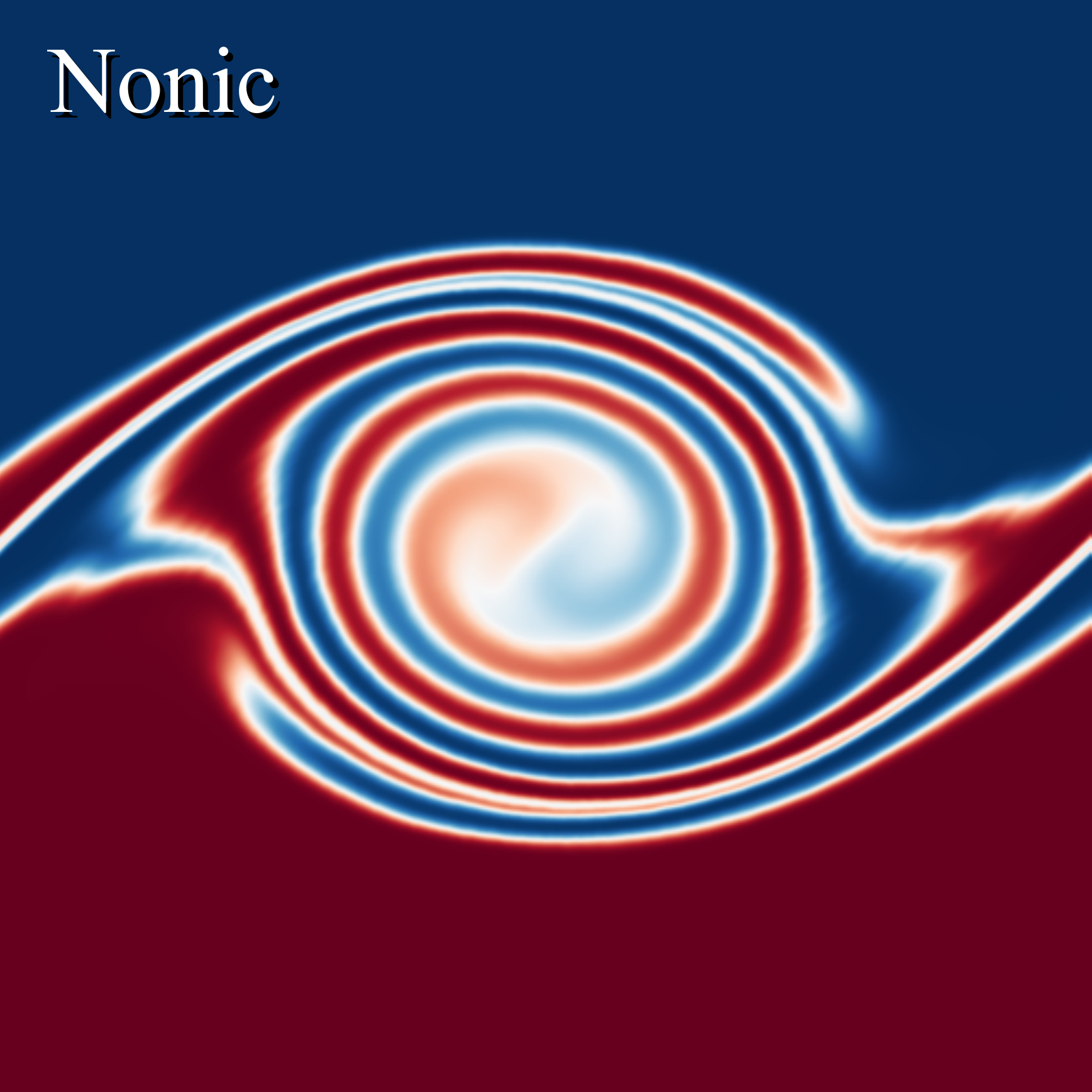} \\
\includegraphics[width=0.978\linewidth]{cbar-wide.pdf}
\caption{The colour field in the region $x,y \in [0,1]$ for $n_{\rm x}=1024$ particle SPH calculations at $t=4$ using the cubic, quartic, quintic, septic, and nonic splines (left to right). All produce a dominant curl of the correct mode frequency, but the cubic spline, and to a lesser degree the quartic spline, evolve incorrectly in comparison to the reference solution (c.f.~Figure~\ref{fig:nonlinear}). The quintic spline shows substantial noise along the spiral arm edges, arising due to particle motion. The colour field of the septic and nonic spline results are smooth and virtually indistinguishable, demonstrating that the kernel bias does not affect the evolution for these high-order splines.}
\label{fig:kernels}
\end{figure*}

Figure~\ref{fig:kernels} show the colour field at $t=4$ for $n_{\rm x}=1024$ particle calculations using the B-splines between cubic (M4) and nonic (M10). In all cases, a single dominant curl forms, mixing the two fluids along the interface. However, the shape of this vortex changes distinctly with respect to the quality of the smoothing kernel. The cubic spline does not resemble the reference solution except that the correct mode has developed. It lacks the interior structure present in the reference solution. The calculation with the quartic spline improves upon this. It resembles the reference solution better than the cubic spline result, though the curl is not wound as tightly as it should, and the filaments are broader and stunted. It is also narrower in the $y$-direction. The quintic spline improves upon this further. The overall shape and structure of the vortex is in close proximity the reference solution. However, the interface along the spirals show substantial noise, noticeable by the `jaggedness' of the spiral edges. 

The best results are obtained when the septic or nonic spline is used. In both cases, a smooth spiral structure develops that closely resembles the reference solution, as discussed in Section~\ref{sec:nonlinear}. The noise present in the quintic spline result is gone. The nonic spline solution is virtually identical to the septic spline result, demonstrating that the kernel bias is no longer the dominant source of error.

\begin{figure}
\includegraphics[width=\linewidth]{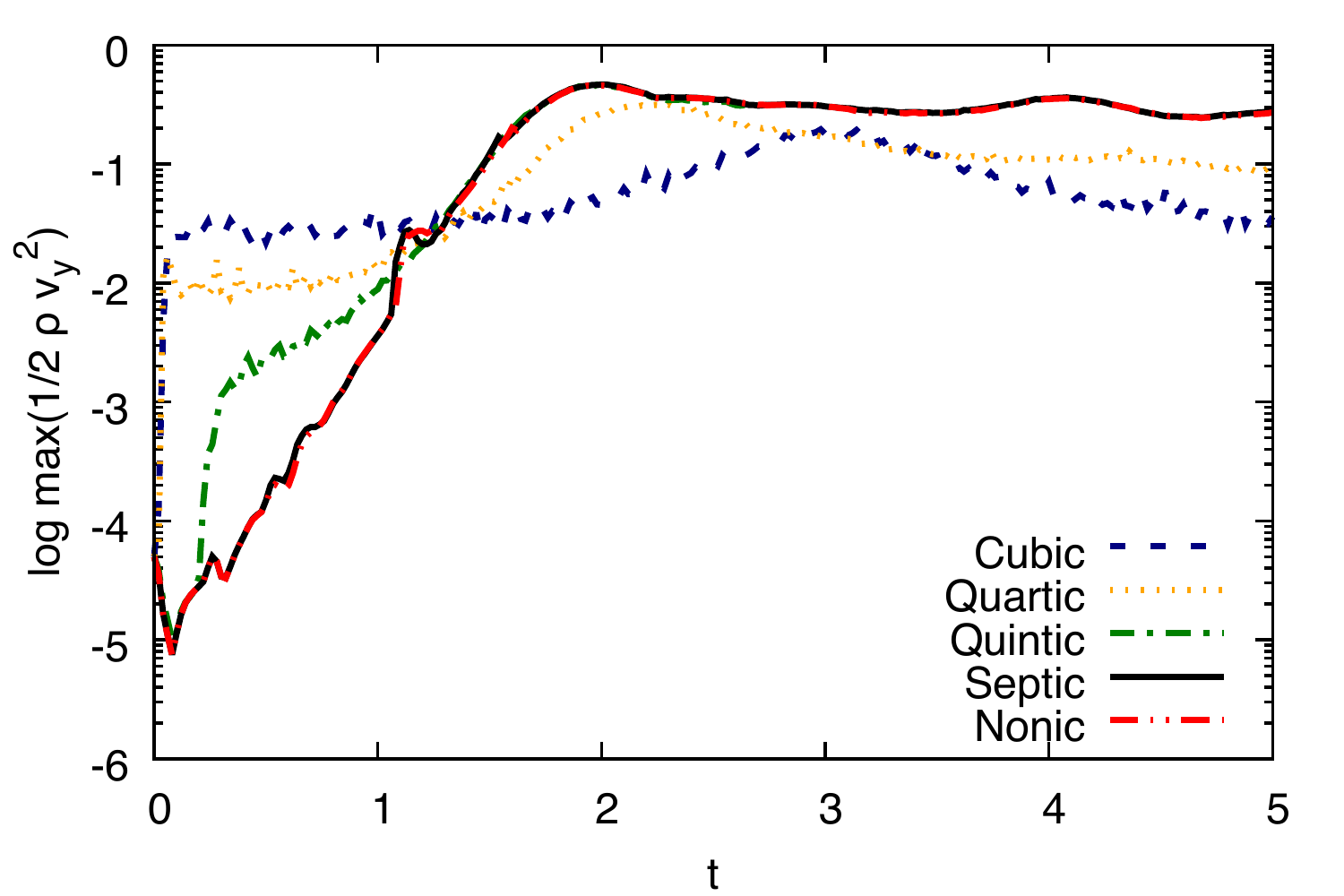}
\caption{The maximum $y$-kinetic energy, $\tfrac{1}{2} \rho v_y^2$, over all particles for calculations using the cubic to nonic spline kernels. The results using the cubic and quartic splines undergo sharp initial jumps, occurring due to errors in the pressure gradient leading to particle rearrangement. The quintic spline result shows similar behaviour. In contrast, calculations using the septic spline show steady exponential growth, as expected of the Kelvin-Helmholtz instability. The nonic spline result matches the septic spline result, showing that the kernel bias no longer affects the evolution of the instability.}
\label{fig:yke}
\end{figure}

To better understand the effect of the kernel bias on the growth of the Kelvin-Helmholtz instability, we consider the maximum kinetic energy in the $y$-direction. Figure~\ref{fig:yke} shows the maximum of $\tfrac{1}{2} \rho v_y^2$ over all particles. This maximum is initially $\sim 10^{-4}$, as defined by the initial conditions. 

For the septic and nonic spline calculations, the maximum $y$-kinetic energy undergoes steady exponential amplification over four orders of magnitude between $0 < t < 2$. This is the expected behaviour in the linear regime of the Kelvin-Helmholtz instability, and further establishes that the non-linear regime begins at $t > 2$. Additionally, the results between the septic and nonic spline are virtually indistinguishable at all times, showcasing further that the kernel bias has no effect on the results obtained in the standard calculations.

On the other hand, the cubic and quartic spline results show an immediate discontinuous jump of two to three orders of magnitude in $y$-kinetic energy. The maximum $y$-velocities are $\mathcal{M}\approx 0.05$ and $\mathcal{M}\approx 0.03$, respectively, significantly larger than the $\mathcal{M}\approx 0.0025$ amplitude of the initial velocity perturbation. Similar behaviour was noted by \citet{mlp12}. Despite the differing growth of $y$-kinetic energy, the peak energy for the cubic and quartic spline calculations is of similar magnitude to the high-order splines. The peaks are delayed, however, occurring at later times. Overall, the initial jump and delayed peak of $y$-kinetic energy demonstrate that the evolution of the instability is strongly affected by the kernel bias for these low-order kernels (see also Figure~\ref{fig:kernels}).

The quintic spline result is between these two extremes. It exhibits a sharp jump in $y$-kinetic energy, as with the low-order splines, but at $t \approx 0.25$ and not as large in magnitude, with a corresponding maximum $y$-velocity of $\mathcal{M} \approx 0.01$. However, between $1 < t < 2$, the $y$-kinetic energy undergoes steady exponential amplification along the track of the high-order splines. This is reflected in the colour field (Figure~\ref{fig:kernels}), which qualitatively contains the features of the reference solution, but with noise introduced by the early sharp increase in energy.

The sharp increase in $y$-kinetic energy for the low-order splines is due to errors in the pressure gradient which generates spurious velocity noise. The error in the discretisation of the pressure gradient scales as $\mathcal{O}(1)$ with respect to resolution. That is, the errors are not reduced by increasing the resolution of the calculation. However, the pressure gradient in the standard formulation of SPH is derived from the discretised Lagrangian using the density summation \citep{price12}, such that the SPH equations of motion are exactly the Lagrangian equations of motion for the particle system. The errors in the pressure gradient act to regularise the particles, pushing them towards a better arrangement which in turn leads to better interpolation properties. This means that maintaining particle regularity is inherently built into SPH, though at the cost of a small degree of velocity noise. Using a more accurate discretisation for the pressure gradient is problematic because it requires the introduction of a particle regularisation scheme. Instead, the primary approach to reduce pressure gradient errors is to use a better smoothing kernel. This has led to investigation of new kernels, such as the Wendland family of kernels \citep{da12}, but the same effect can be obtained by switching to high-order splines.

Figure~\ref{fig:yke} demonstrates that the magnitude of the velocity noise decreases as the quality of the smoothing kernel improves. It is clear that the amplitude of the velocity perturbation used to seed the Kelvin-Helmholtz instability must be greater than the velocity noise introduced by particle arrangement in order to correctly model the linear regime. For these calculations, the lowest order spline kernel which meets this criterion is the septic spline. This may not be the sole requirement. It would additionally be expected that other modes not seeded in the initial conditions will be excited by any introduced velocity noise, as if they had initial amplitudes similar to the threshold specific to each kernel. The growth of these spurious modes would have the potential to affect the solution.

\begin{figure}
\includegraphics[width=0.49\linewidth]{kh2048-septic-t4.pdf}   \hspace{-1.85mm}
\includegraphics[width=0.49\linewidth]{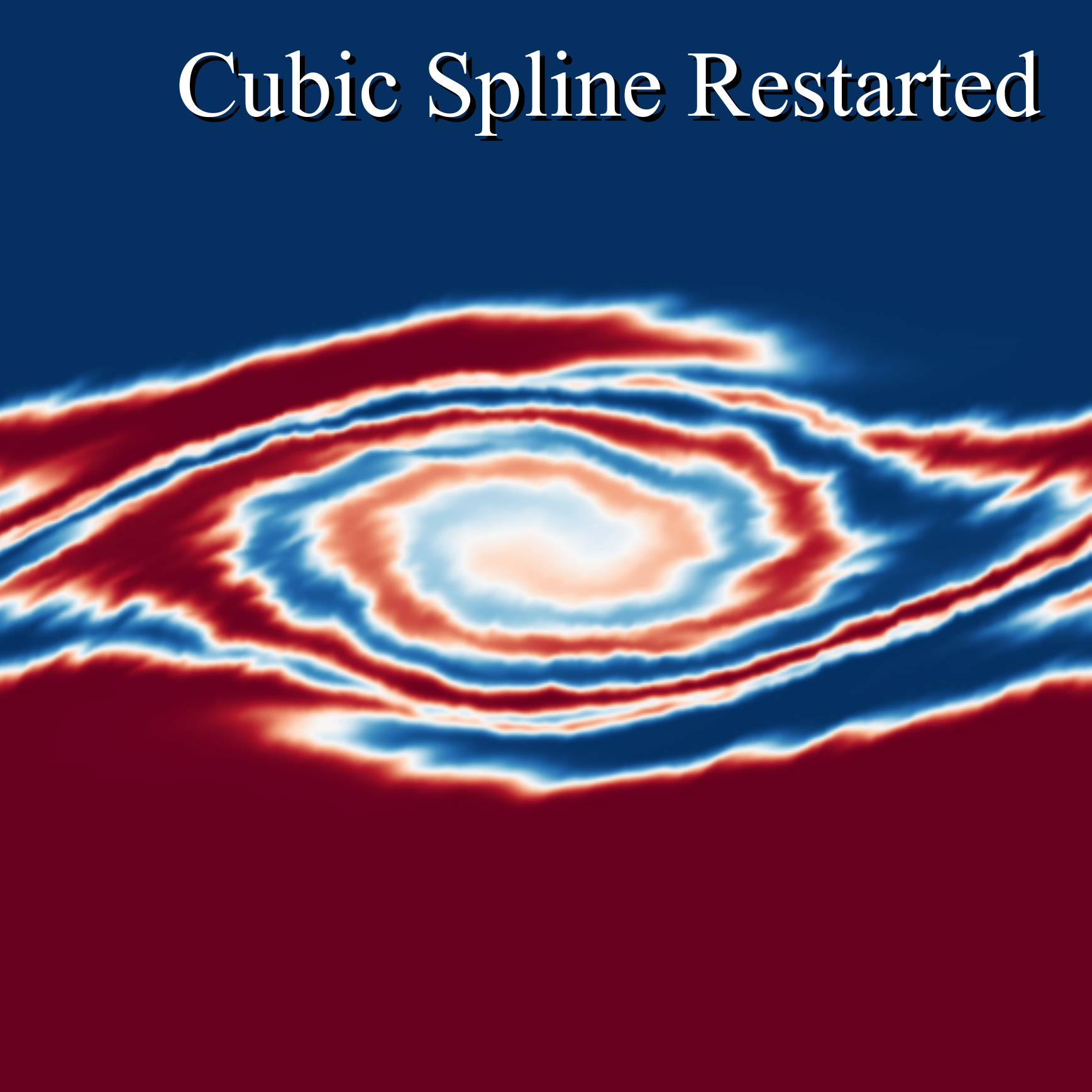} \\
\includegraphics[width=0.983\linewidth]{cbar-onecol.pdf}
\caption{The $n_{\rm x}$=2048 calculation at $t=4$ (left panel) alongside a calculation where the smoothing kernel was changed from the septic spline to the cubic spline at $t=2$ (right panel). Switching to the low order kernel partway through the calculation introduces substantial noise into the solution.}
\label{fig:restart}
\end{figure}

This conjecture was tested by restarting the $n_{\rm x}$=2048 calculation at $t=2$ with the smoothing kernel switched from the septic spline to the cubic spline. In this way, the linear growth phase will be accurately modeled using a high-order kernel before switching to a low-order kernel. Figure~\ref{fig:restart} shows the colour field of the cubic spline restarted calculation at $t=4$, alongside the septic spline calculation for comparison. A substantial degree of noise is present in the vortex structure of the cubic spline calculation, occurring due to velocity noise introduced by errors in the pressure gradient. This demonstrates that a high-quality smoothing kernel is necessary for both the linear and non-linear phases.

\subsection{Calculations without physical dissipation}
\label{sec:artdiss}

Astrophysical simulations often rely solely on numerical dissipation, that is, without explicit physical dissipation. Here we present the results of calculations without Navier-Stokes viscosity or thermal conductivity. The artificial viscosity is as described in Section~\ref{sec:sph}. An artificial thermal conductivity is applied according to
\begin{equation}
\frac{{\rm d}u_a}{{\rm d}t} = \frac{1}{10} \sum_b \frac{m_b}{\overline{\rho}_{ab}} \sqrt{\frac{\vert P_a - P_b \vert}{\overline{\rho}_{ab}}} (u_a - u_b) \hat{\bf{r}}_{ab} \cdot \nabla_a W_{ab}(h_a) ,
\end{equation}
making use of the \citet{price08} switch.  Since advection is dissipationless in SPH, there is no numerical dissipation associated with the colour field in the absence of explicitly prescribed terms. In such a case, the total colour entropy would remain constant over time as the colour of each particle would remain fixed. Rather than creating a contrived `artificial colour diffusion' term for the colour field for the purposes of diffusing colour between particles, we simply retain the `physical' dissipation term given by Equation~\ref{eq:sphcolourdiffusion}.

\begin{figure}
\centering
\hspace{-1.85mm}
\includegraphics[width=0.49\linewidth]{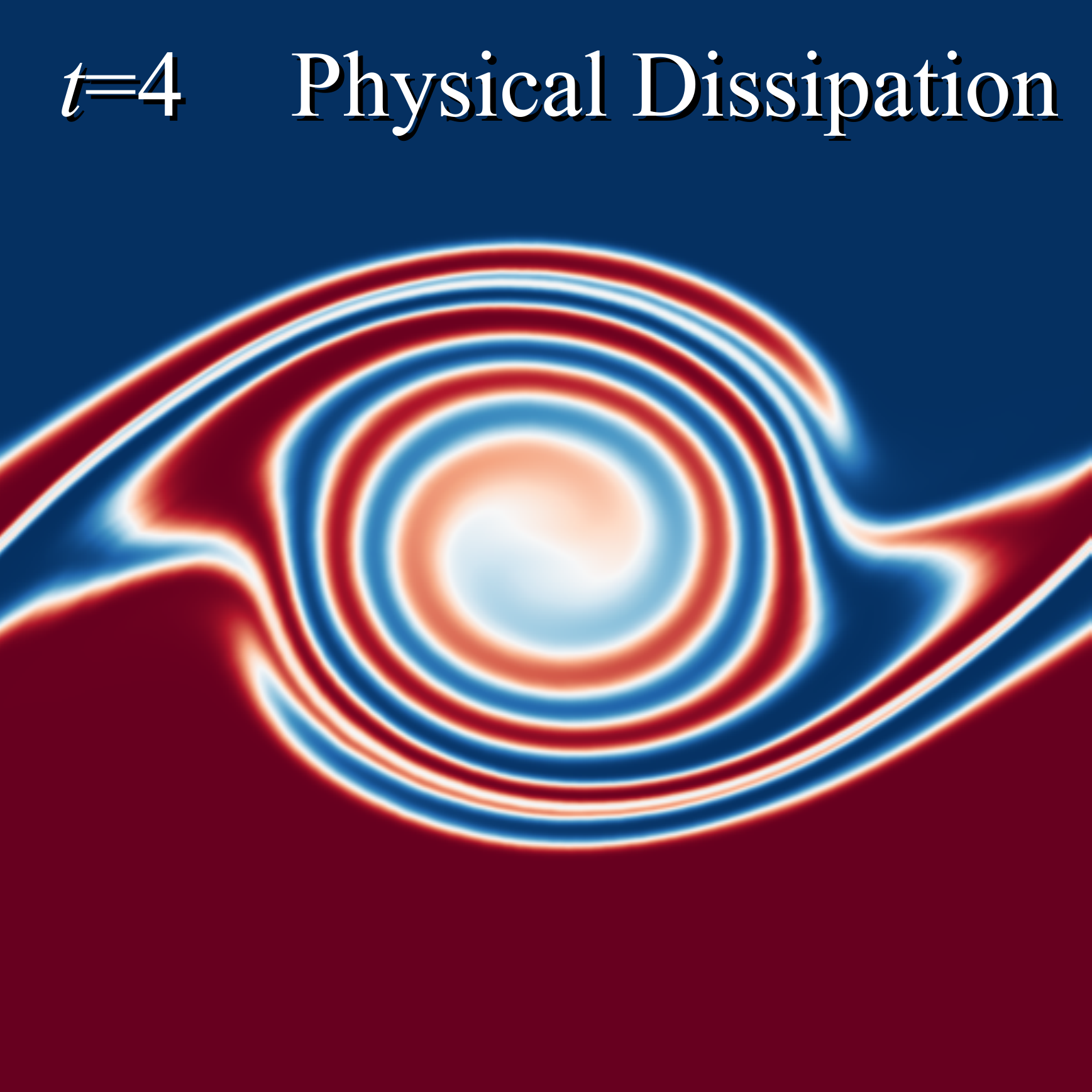} \hspace{-1.85mm}
\includegraphics[width=0.49\linewidth]{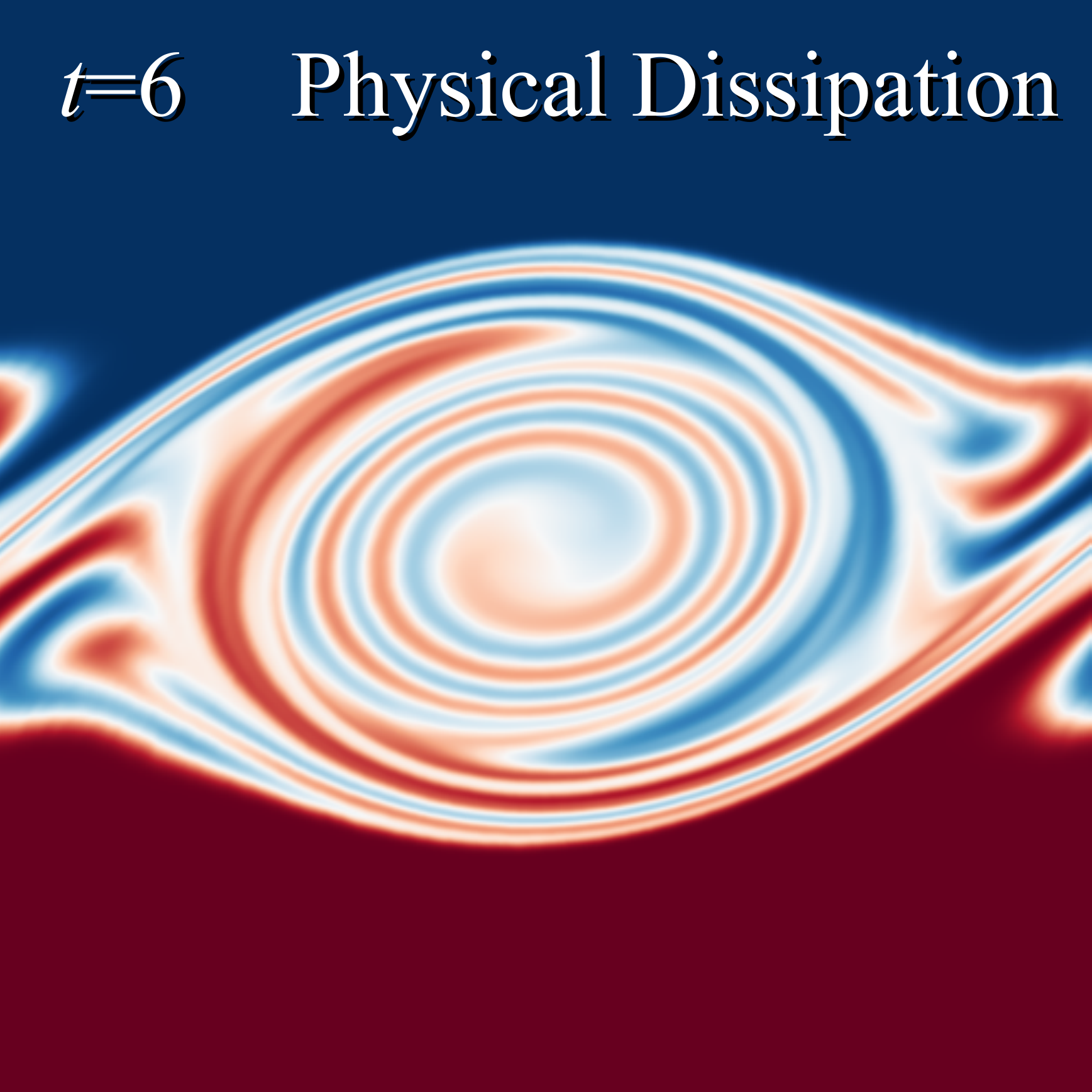} \hspace{-1.85mm} \\
\hspace{-1.85mm}
\includegraphics[width=0.49\linewidth]{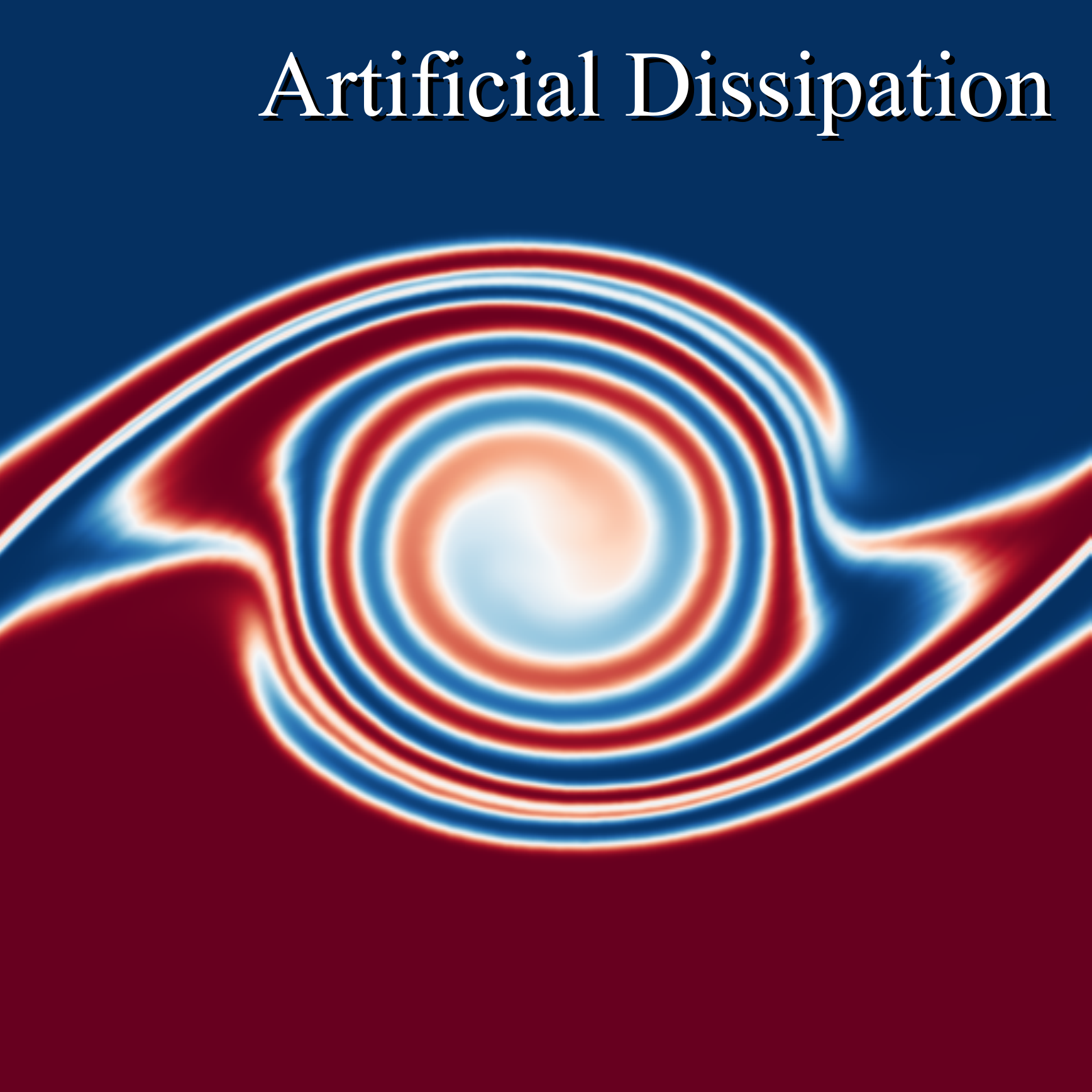} \hspace{-1.85mm}
\includegraphics[width=0.49\linewidth]{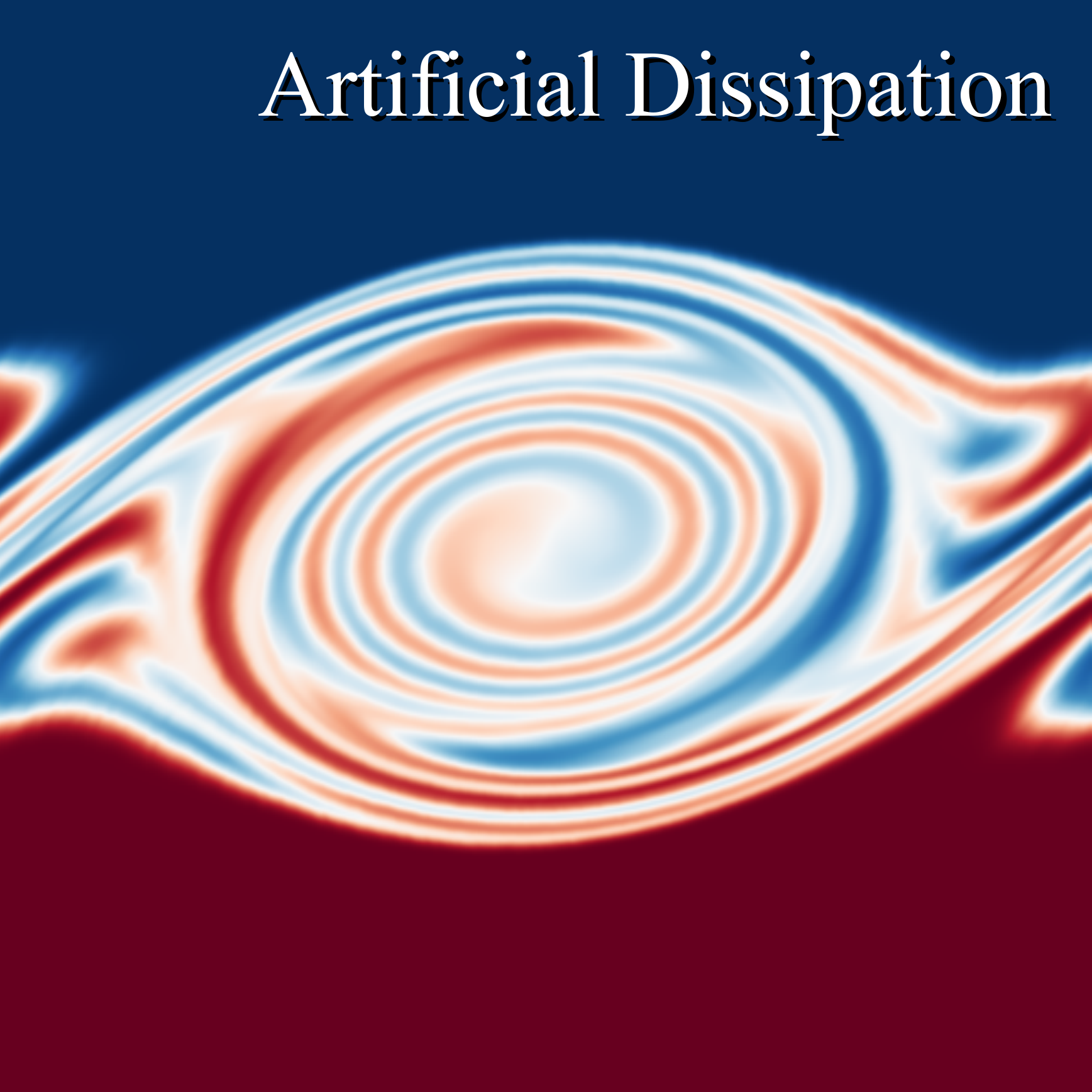} \hspace{-1.85mm} \\
\hspace{-1.85mm}
\includegraphics[width=0.49\linewidth]{c-t4.pdf} \hspace{-1.85mm}
\includegraphics[width=0.49\linewidth]{c-t6.pdf} \hspace{-1.85mm} \\
\hspace{-1.85mm} 
\includegraphics[width=0.983\linewidth]{cbar-onecol.pdf}
\caption{Calculations relying solely on artificial dissipation terms for the velocity and internal energy fields, contrasted with the standard SPH calculations with physical dissipation terms, and the {\sc Dedalus} reference solution. At this resolution, $n_{\rm x}=2048$ particles, there is no significant difference in the vortex structure as the artificial dissipation is the dominant source of dissipation even when physical dissipation is present.}
\label{fig:artdiss}
\end{figure}

Figure~\ref{fig:artdiss} shows the colour field for these calculations at $t=4$ and 6, alongside the standard calculations with physical dissipation terms. Slight differences are evident to the calculation with physical dissipation, however, as the artificial dissipation is the dominant source of kinetic dissipation for both calculations (c.f. Section~\ref{sec:alpha0}), the differences are minimal. Since the artificial viscosity can be equated to a Navier-Stokes shear and bulk viscosity, it is expected that a more complex vortex structure would develop only when the artificial dissipation is lower than the physical dissipation. That is, when the Reynolds number exceeds $10^5$, occurring for resolutions $n_{\rm x} > 4096$ particles.

\subsection{Calculations with a density contrast}
\label{sec:densitycontrast}

\citet{lecoanetetal16} studied the Kelvin-Helmholtz for both initial conditions that have uniform density and a difference in density between the two flow regions. Including a density contrast introduces a significant degree of complexity to the structure and evolution of the instability. In this case, the computational demand to achieve convergence is strict, with \citet{lecoanetetal16} only achieving convergence between {\sc athena} and {\sc dedalus} when $16~384 \times 32~768$ grid cells were used.

We study the behaviour of SPH on the Kelvin-Helmholtz instability problem of \citet{lecoanetetal16} that has an initial density contrast. The initial density profile is specified as in Equation~\ref{eq:ic-rho} with $\Delta \rho / \rho_0 = 1$. The particles are arranged using the stretch mapping technique \citep{phantom} to achieve the correct density profile. The calculations are performed for $n_{\rm x} = 512$, 1024 and 2048 particles. The D4096 solution from \citet{lecoanetetal16} is used as the reference solution. 

\begin{figure*}
\centering
\hspace{-1.85mm}
\includegraphics[width=0.245\linewidth]{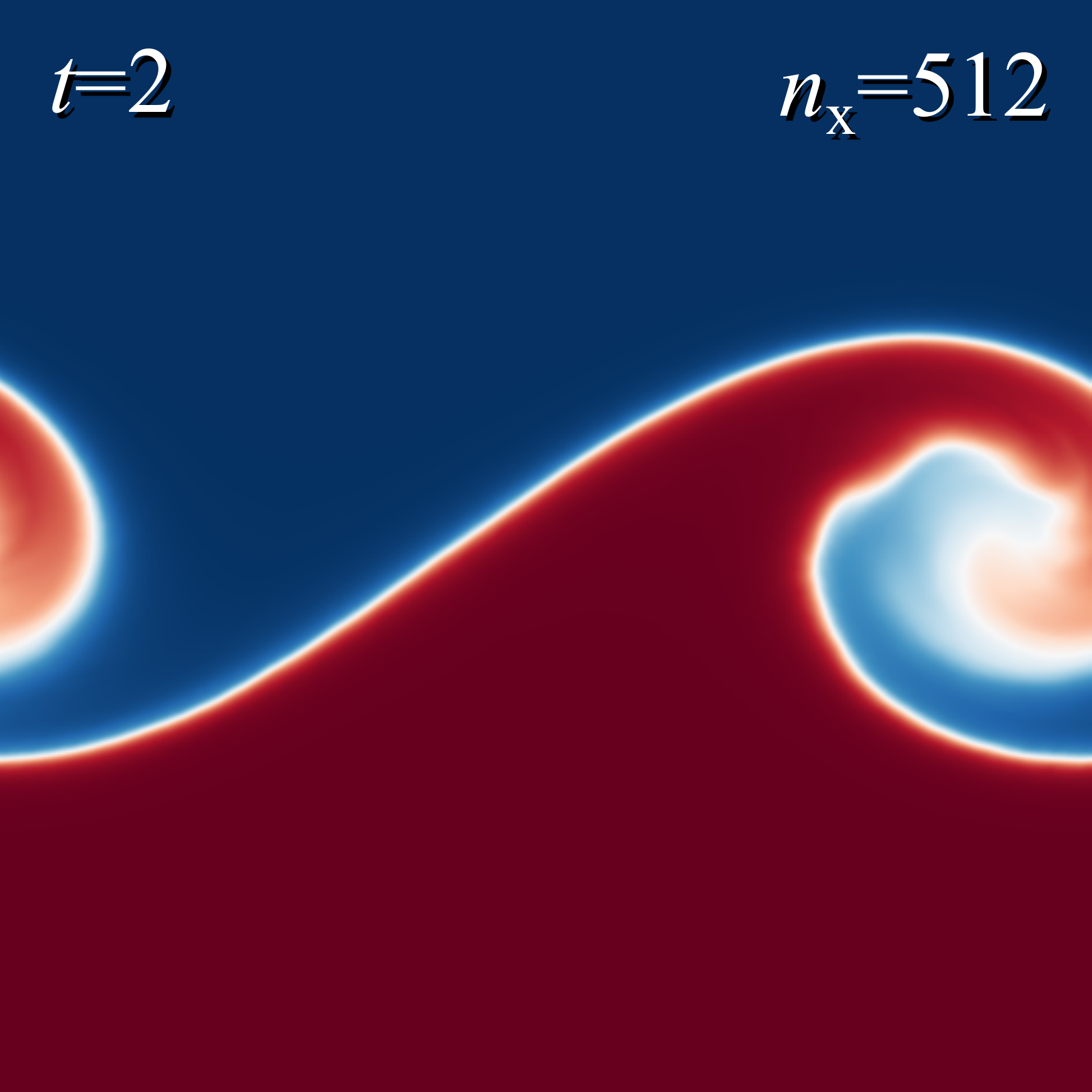}   \hspace{-1.85mm}
\includegraphics[width=0.245\linewidth]{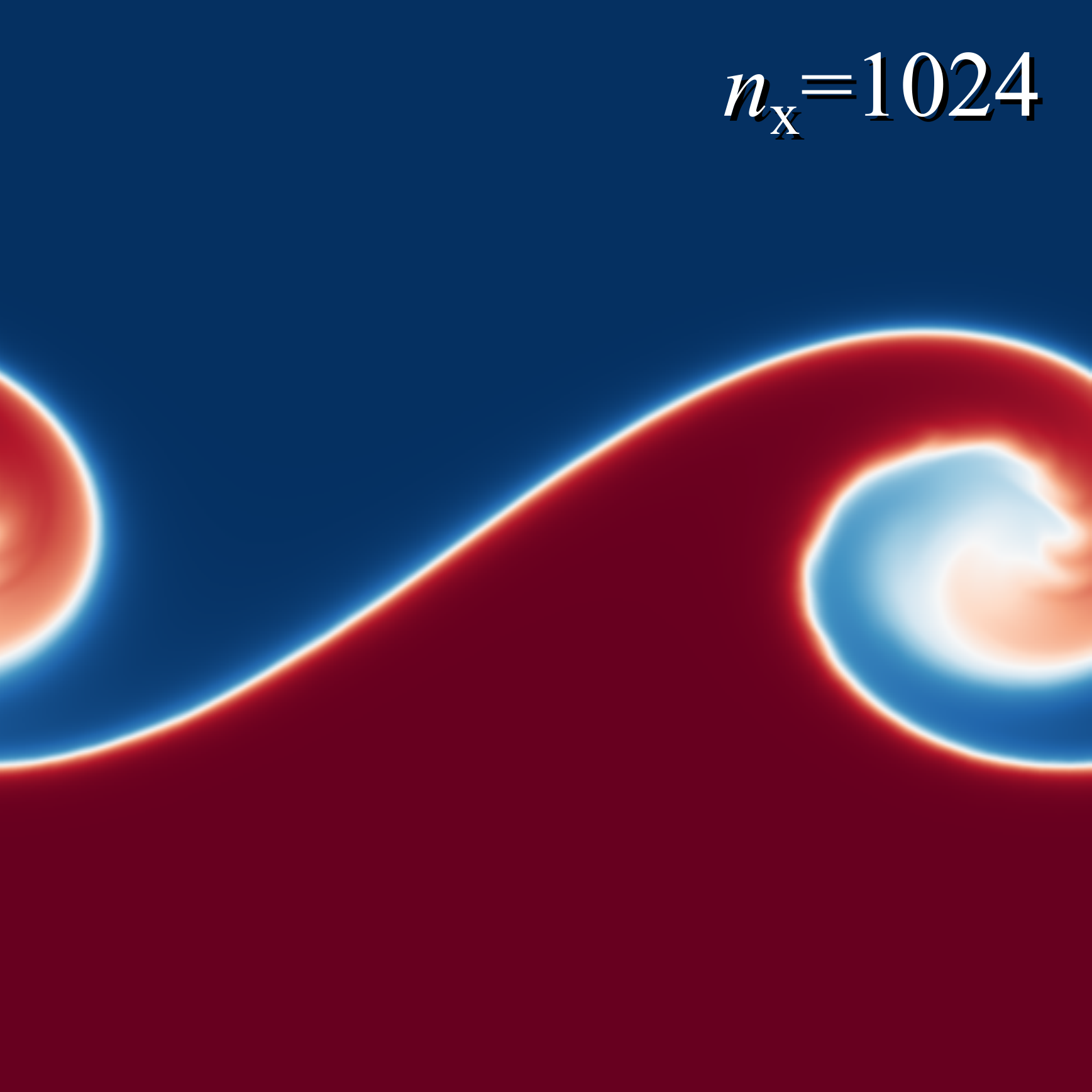} \hspace{-1.85mm}
\includegraphics[width=0.245\linewidth]{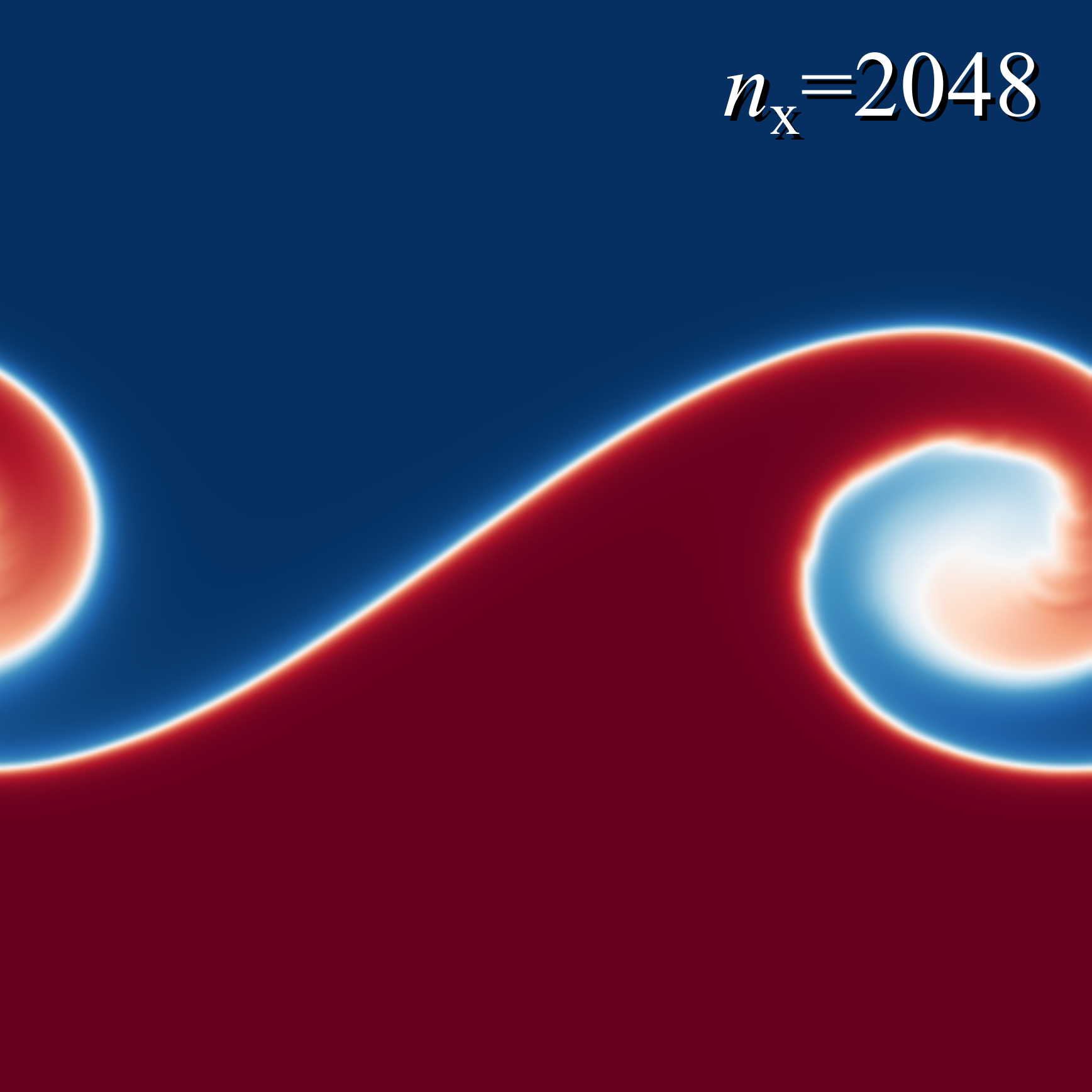} \hspace{-1.85mm}
\includegraphics[width=0.245\linewidth]{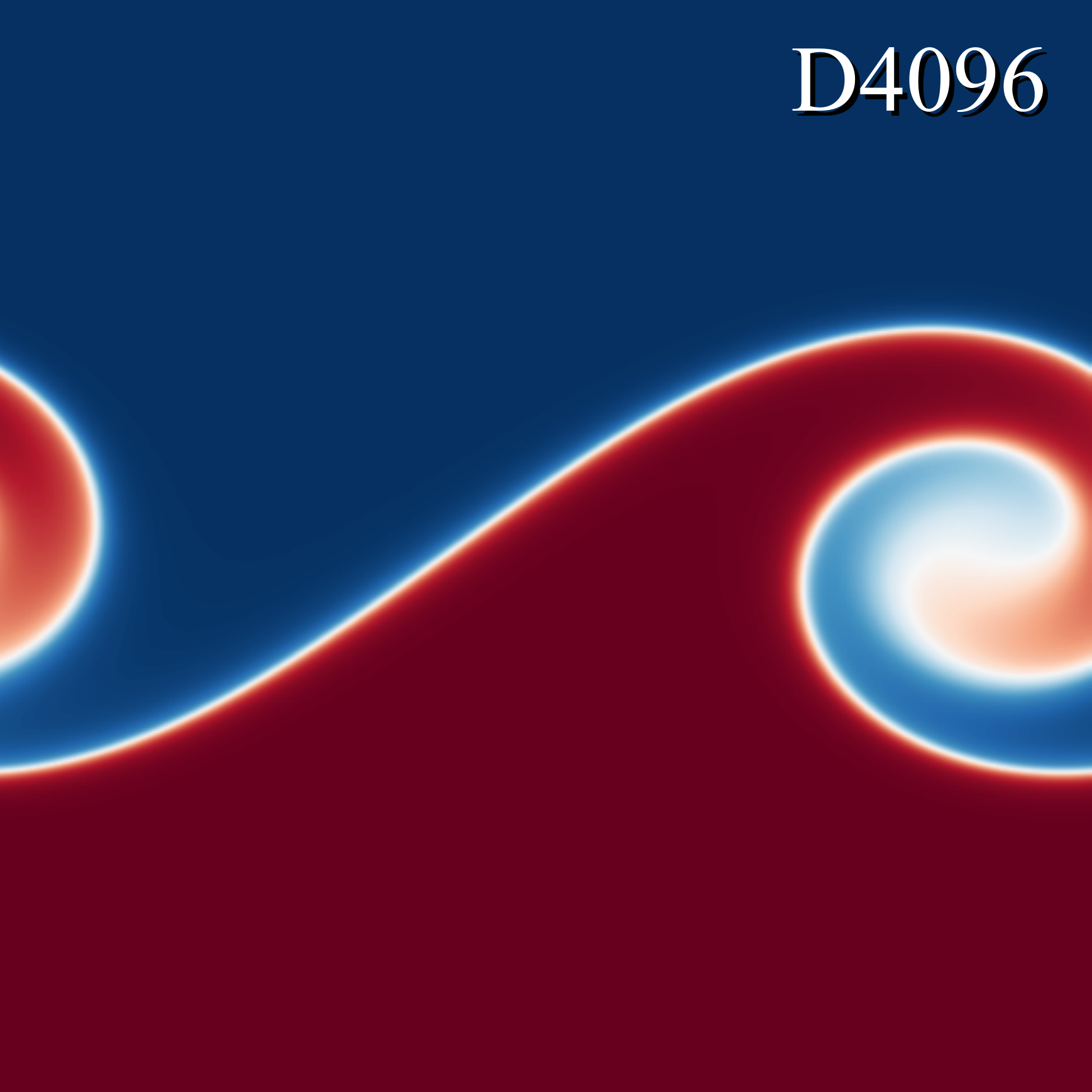} \\
\hspace{-1.85mm}
\includegraphics[width=0.245\linewidth]{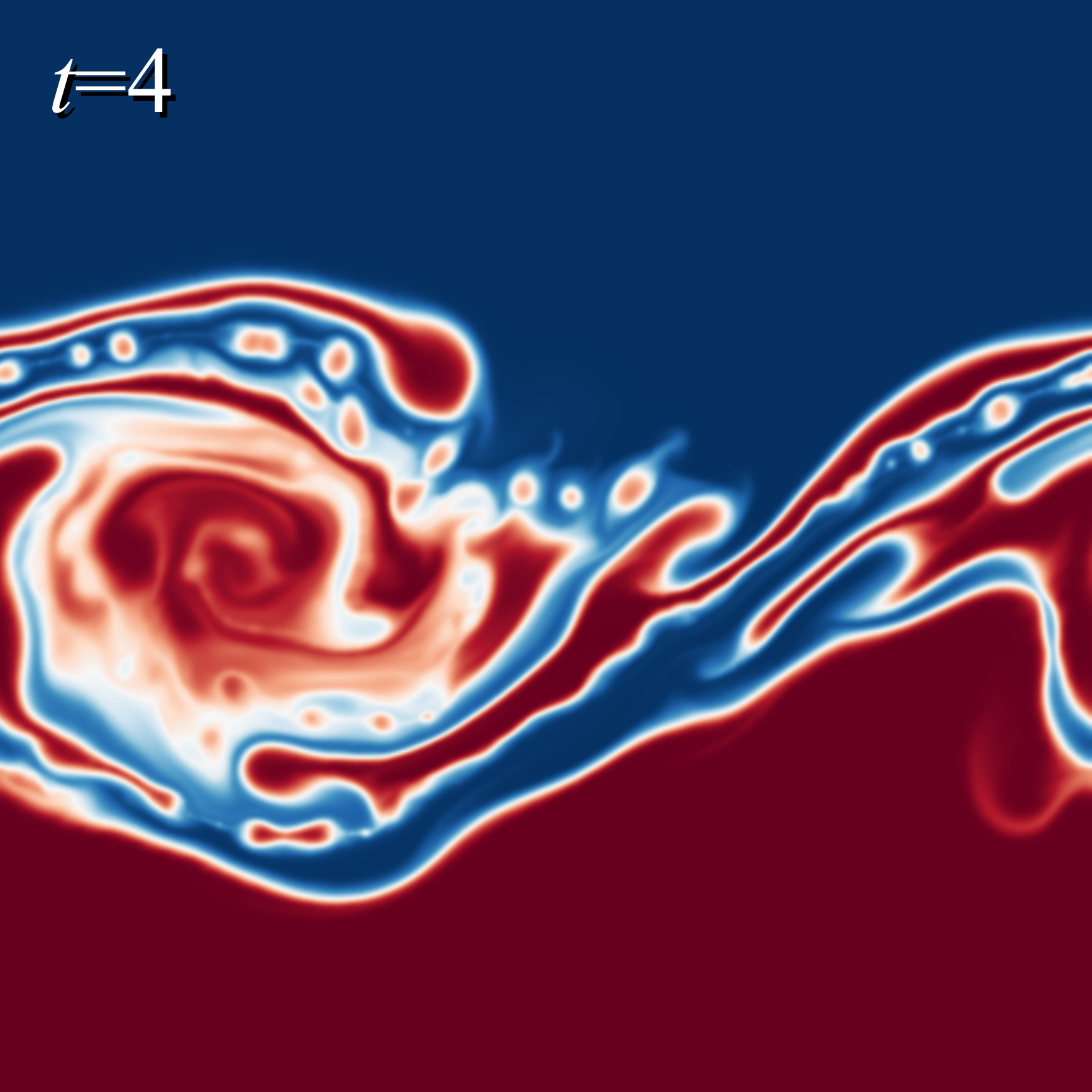}   \hspace{-1.85mm}
\includegraphics[width=0.245\linewidth]{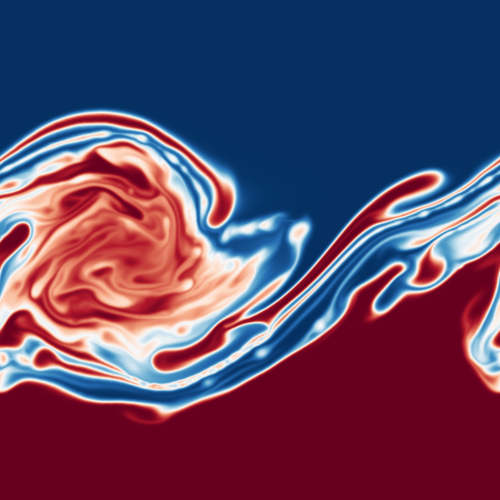} \hspace{-1.85mm}
\includegraphics[width=0.245\linewidth]{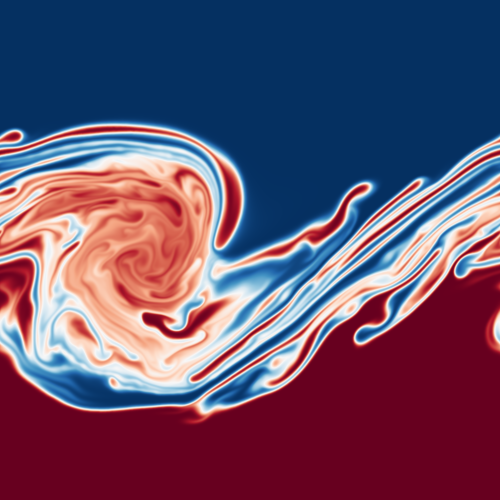} \hspace{-1.85mm}
\includegraphics[width=0.245\linewidth]{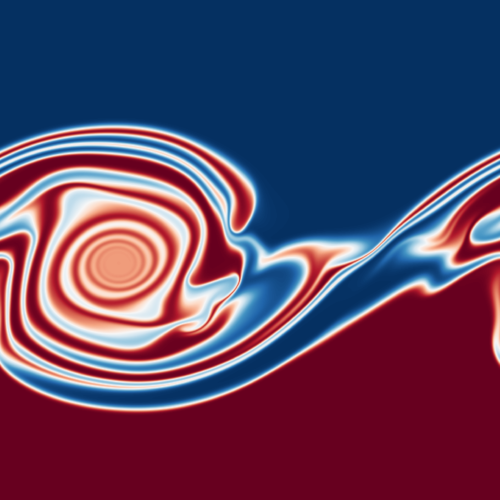} \\
\hspace{-1.85mm}
\includegraphics[width=0.245\linewidth]{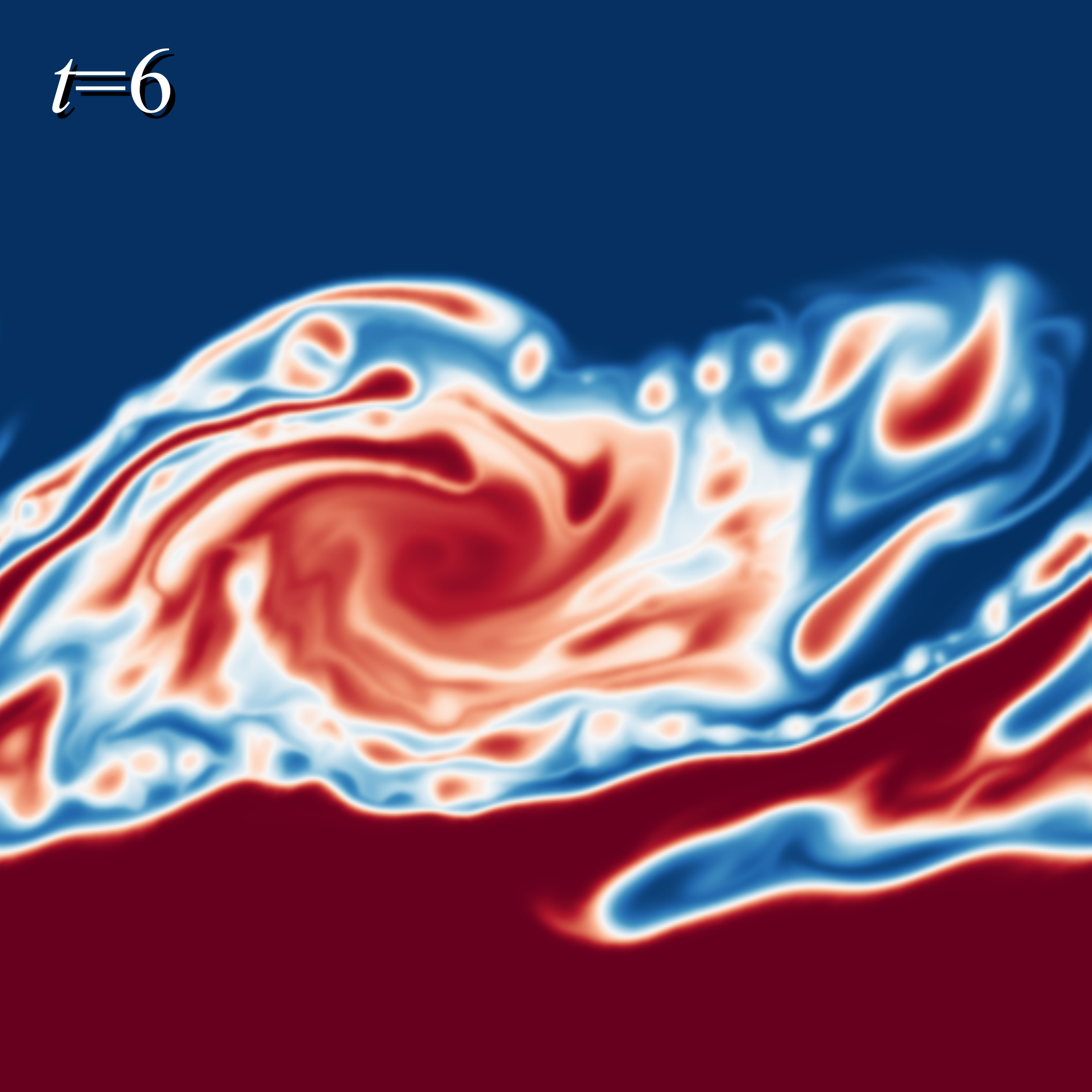}   \hspace{-1.85mm}
\includegraphics[width=0.245\linewidth]{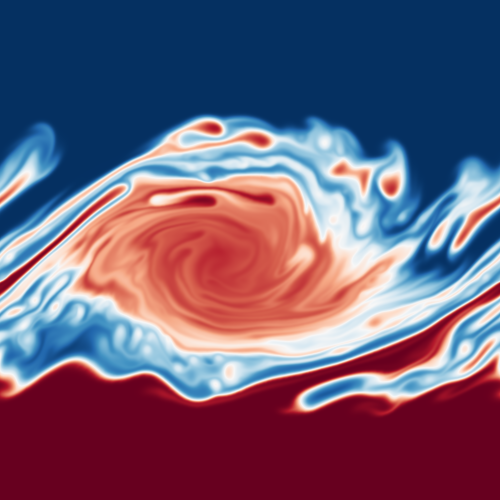} \hspace{-1.85mm}
\includegraphics[width=0.245\linewidth]{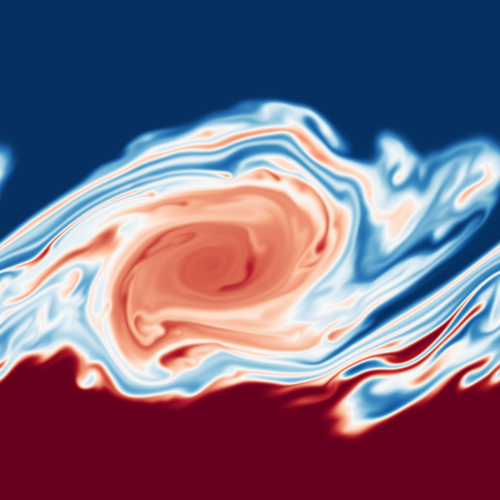} \hspace{-1.85mm}
\includegraphics[width=0.245\linewidth]{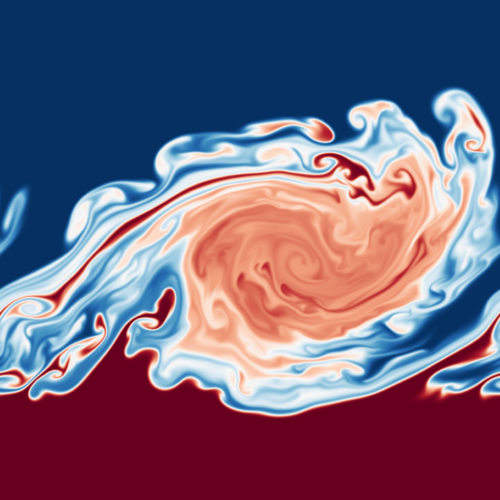} \\
\includegraphics[width=0.983\linewidth]{cbar-wide.pdf}
\caption{The colour field with a density contrast in the initial conditions ($\Delta \rho / \rho_0 = 1$) for SPH calculations of $n_{\rm x} = 512$, 1024 and 2048 particles (first to third columns), with the {\sc dedalus} reference solution for comparison (D4096) at $t=2, 4$ and 6 (top to bottom rows). The instability is triggered in SPH even at modest resolutions. The vortex structure is trending towards the reference solution as the resolution increases, but would require significantly higher resolution to match the reference solution for $t \ge 4$. }
\label{fig:dens}
\end{figure*}

Figure~\ref{fig:dens} shows the colour field at $t = 2$, 4 and 6 for the SPH calculations for resolutions up to $n_{\rm x}=2048$ particles. The instability is triggered in all cases, with a curled vortex forming by $t=2$ that is similar in shape to the reference solution. The similarity improves as the resolution increases.

At $t=4$, the physical thermal conductivity is insufficient to properly mix the two flow regions for the $n_{\rm x}=512$ particle calculation, resulting in a clumpy structure. This issue is alleviated at higher resolution ($n_{\rm x}=2048$). Including an artificial thermal conductivity would help prevent this issue from occurring at low resolution \citep{price08}. 

The non-linear evolution of the vortex structure for the $n_{\rm x} = 1024$ and 2048 particle SPH calculations exhibits a greater degree of mixing in comparison to the reference solution. Similar behaviour was found for {\sc athena} calculations \citep{lecoanetetal16} at lower resolution, thus this outcome is not surprising given that the highest resolution {\sc athena} calculation has length resolution one-eighth that of the $n_{\rm x}=2048$ particle SPH calculation. Overall, the qualitative behaviour of the SPH calculations is as expected for the resolutions simulated.

\section{Conclusion}
\label{sec:summary}

We have performed calculations of the Kelvin-Helmholtz instability using SPH. Our tests have used the unstratified, uniform density test of \citet{lecoanetetal16} with a Reynolds number of Re=$10^5$. The calculations have included a passive scalar `colour' field to quantify the degree of mixing by the instability. Physical dissipation terms have been included in order to control the evolution of the instability in the non-linear regime. 

We have compared our results qualitatively and quantitatively to the D2048 results obtained using the pseudo-spectral code {\sc Dedalus}, as presented by \citet{lecoanetetal16}. Results have been examined in both the linear and non-linear regime of the instability. Our key observations are:

\begin{itemize}
\item The SPH calculations qualitatively agreed with the reference solution on the linear and non-linear evolution of the Kelvin-Helmholtz instability. A single dominant curl formed in the linear regime, which continued to wind producing substantial substructure in the non-linear regime. At late times ($t>8$), the instability, in combination with the physical dissipation terms, produced a well-mixed interface. The internal structure of the vortices resembled the reference solution for the times compared, with the visual agreement improving as the resolution was increased.

\item No modifications to SPH were required to activate or correctly evolve the instability. The Kelvin-Helmholtz instability can be captured using the standard density formulation of SPH with artificial dissipation terms, of the kind that has been used for decades. 

\item The exponential growth rate of the seeded mode in the linear regime was in agreement with the expected growth rate of $\propto \exp(3.227 t)$, as obtained numerically using the {\sc psecas} code \citep{bp19}. This is half the analytic estimate for an incompressible fluid with sharp interfaces. The SPH growth rate was linearly converging to the expected rate, and was accurate even at our lowest resolution of $n_{\rm x}=256$ particles.

\item The Kelvin-Helmholtz instability can be activated in SPH even for modest resolutions. For our lowest resolution calculation of $n_{\rm x}=256$ particles, the total entropy of the colour field increased monotonically along a similar curve to the reference solution. Even though small-scale features in the vortices were not as well represented at low resolutions, the overall degree of mixing experienced, with respect to the reference solution, was similar. 

\item The rate of convergence of $\mathcal{L}_2$ error was linear in the linear regime of the Kelvin-Helmholtz instability. That is, the $\mathcal{L}_2$ error decreased by a factor of two for each factor of two increase in resolution in each spatial direction. In the non-linear regime, the $\mathcal{L}_2$ errors were decreasing, but the convergence rate was sub-linear.

\item Numerical dissipation was the primary factor in convergence rate. Performing calculations with reduced artificial viscosity, using $\alpha=0$, yielded results that qualitatively and quantitatively better agreed with the reference solution. The filaments and spiral structure of the vortices more closely resembled the reference solution, and the $\mathcal{L}_2$ errors were lower at all times sampled. That the calculations could be performed with $\alpha=0$ was permissible only because, at high resolution, the physical Navier-Stokes viscosity was as dissipative as the numerical dissipation and was sufficient to maintain particle regularity.

\item It was found that the septic spline was the minimum order of B-spline required for these Kelvin-Helmholtz calculations in order to accurately model both the linear and non-linear regimes. High-order kernels are needed to minimize the generation of spurious velocity noise from errors in the pressure gradient. This background velocity noise must be less than the initial velocity perturbation in order to correctly model the linear regime. Velocity noise can additionally excite other modes, disrupting the non-linear phase even when the linear phase is correctly modeled, as demonstrated using the cubic spline.

\item Calculations employing an initial density contrast qualitatively agree with the reference solution. A single curl forms during the linear regime, which continues to wind producing substantial substructure in the non-linear regime. The interior structure of the vortex is over-mixed compared to the reference solution, but similar behaviour was found for {\sc athena} calculations at similar resolutions \citep{lecoanetetal16}.

\end{itemize}

The true test of any numerical method is whether it converges to an agreed upon solution, and, in this work, it has been demonstrated that SPH is converging towards the Kelvin-Helmholtz solution of \citet{lecoanetetal16}. It would be of interest to test the convergence properties of voronoi moving mesh or other particle-based schemes. Our core conclusion is that SPH can correctly model the Kelvin-Helmholtz instability.

\section{Acknowledgements}
Thank you to Daniel Lecoanet for providing the data from the {\sc Dedalus} simulations, Daniel Price for useful discussions on physical dissipation terms in SPH, and the anonymous referee whose comments helped improve the quality of this work. TST was supported by a CITA Postdoctoral Research Fellowship.

\appendix

\section{B-spline kernels}
\label{sec:kernels}

\begin{table*}
\caption{List of bell-shaped spline kernels, with the functional form, normalisation ($\sigma$), and radial extent ($R$). The functional form, $w(q)$, is for the piece $R-1 \le q < R$, from which the full piecewise function can be constructed.}
\label{tbl:kernels}
\begin{tabular}{llcccc}
\hline \\[-6pt]
Name & $w(q)$ & $\sigma$ (1D) & $\sigma$ (2D) & $\sigma$ (3D) & $R$ ($q$) \\[2pt]
\hline \\[-6pt]
Cubic 	& $(2 - q)^3 - 4 (1 - q)^3$ & $3!$ & $5/(14\pi)$ & $6/(4!\pi)$ & 2 \\[2pt]
Quartic 	& $(\tfrac{5}{2} - q)^4 - 5 (\tfrac{3}{2} - q)^4 + 10 (\tfrac{1}{2} - q)^4$ & $4!$ & $96/(1~199\pi)$ & $6/(5!\pi)$ & 2.5 \\[2pt]
Quintic 	& $(3 - q)^5 - 6 (2 - q)^5 + 15 (1 - q)^5$ & $5!$ & $7/(478\pi)$ & $6/(6!\pi)$ & 3 \\[2pt]
Sextic 	& $(\tfrac{7}{2} - q)^6 - 7 (\tfrac{5}{2} - q)^6 + 21 (\tfrac{3}{2} - q)^6 - 35 (\tfrac{1}{2} - q)^6$ & $6!$ & $256/(113~149\pi)$& $6/(7!\pi)$  & 3.5 \\[2pt]
Septic 	& $(4 - q)^7 - 8 (3 - q)^7 + 28 (2 - q)^7 - 56 (1 - q)^7$ & $7!$ & $9/(29~749\pi)$ & $6/(8!\pi)$ & 4 \\[2pt]
Octic 	& $(\tfrac{9}{2} - q)^8 - 9 (\tfrac{7}{2} - q)^8 + 36 (\tfrac{5}{2} - q)^8 - 84 (\tfrac{3}{2} - q)^8 + 126 (\tfrac{1}{2} - q)^8$ & $8!$ & $512/(14~345~663\pi)$ & $6/(9!\pi)$ & 4.5 \\[2pt]
Nonic 	& $(5 - q)^9 - 10 (4 - q)^9 + 45 (3 - q)^9 - 120 (2 - q)^9 + 210 (1- q)^9$ & $9!$ & $11/(2~922~230\pi)$ & $6/(10!\pi)$  & 5 \\[2pt]
\hline 
\end{tabular}
\end{table*}

The family of B-spline kernels \citep{schoenberg46} can be written in the functional form
\begin{equation}
W_{ab}(h_a) = \frac{\sigma}{h^p} w(q) ,
\end{equation}
where $\sigma$ is the normalisation, $p$ is the number of dimensions and $q = r_{ab}/h_a$. Each spline is a piecewise polynomial. Table~\ref{tbl:kernels} lists the M4 (cubic) to M10 (nonic) polynomial spline kernels. In the table, the kernel function, $w(q)$, is for the range $R-1 \le q < R$, where $R$ is the radial extent of the kernel. The full piecewise function can be constructed from this.

For clarity, we also explicitly write the cubic spline,
\begin{equation}
w(q) =
\begin{cases}
(2 - q)^3 - 4 (1 - q)^3 & q < 1, \\
(2 - q)^3  & 1 \leq q < 2 , \\
0 & q \geq 2.
\end{cases}
\end{equation}
quintic spline, 
\begin{equation}
w(q) = 
\begin{cases}
(3 - q)^5 - 6 (2 - q)^5 + 15 (1 - q)^5 & q < 1, \\
(3 - q)^5 - 6 (2 - q)^5 & 1 \leq q < 2 , \\
(3 - q)^5  & 2 \leq q < 3 , \\
0 & q \geq 3 .
\end{cases}
\end{equation}
and septic spline,
\begin{equation}
w(q) = 
\begin{cases}
 \!\begin{aligned}%[b]
       &(4 - q)^7 - 8 (3 - q)^7 + 28 (2 - q)^7 \\
       &\hspace{1mm} - 56 (1 - q)^7
    \end{aligned} & q < 1,\\
(4 - q)^7 - 8 (3 - q)^7 + 28 (2 - q)^7 & 1 \leq q < 2, \\
(4 - q)^7 - 8 (3 - q)^7 & 2 \leq q < 3, \\
(4 - q)^7  & 3 \leq q < 4, \\
0 & q \geq 4 .
\end{cases}
\end{equation}

The increased computational cost for higher order kernels compared to the cubic spline can estimated by the increased number of particles within the radial extent of the kernel. In two dimensions, such as the calculations in this paper, it would imply that the increased computational cost of switching to the quintic spline over the cubic spline would be $2.25\times$. For the septic spline over the cubic spline, this would be $4\times$. There is a further floating point operations cost incurred to compute the more complex splines.

\section{Derivation of the Navier-Stokes heating term}
\label{sec:NSheating}

The heat gained from the two first derivatives implementation of Navier-Stokes viscosity in SPH can be derived as follows. The total energy is the sum of kinetic and thermal energy from all particles, 
\begin{equation}
E = \sum_a m_a \left( \tfrac{1}{2}{\bm v}_a^2 + u_a \right) .
\end{equation}
Conservation of energy requires the total energy to remain constant, thus
\begin{equation}
\frac{{\rm d}E}{{\rm d}t} = \sum_a m_a \left( {\bm v}_a \cdot \frac{{\rm d}{\bm v}_a}{{\rm d}t} + \frac{{\rm d}u_a}{{\rm d}t} \right) = 0.
\end{equation}
Re-arranging yields
\begin{equation}
\sum_a m_a \frac{{\rm d}u_a}{{\rm d}t} = - \sum_a m_a {\bm v}_a \cdot \frac{{\rm d}{\bm v}_a}{{\rm d}t} .
\label{eq:appndx1}
\end{equation}
Substituting in the Navier-Stokes term from the SPH momentum equation (Equation~\ref{eq:sphNS}) yields, for the RHS,
\begin{equation}
\sum_a \sum_b m_a m_b v_a^i \Bigg[ \frac{\Pi^{ij}_a}{\Omega_a \rho_a^2} \nabla_a^j W_{ab}(h_a) + \frac{\Pi^{ij}_b}{\Omega_b \rho_b^2} \nabla_a^j W_{ab}(h_b) \Bigg] .
\end{equation}
Substituting in the definition the Navier-Stokes stress tensor (Equation~\ref{eq:stresstensor}) yields
\begin{align}
& \sum_a \sum_b m_a m_b \nu \Bigg[ v_a^i \left( \frac{\partial v_a^i}{\partial x_a^j} + \frac{\partial v_a^j}{\partial x_a^i} - \frac{2}{3} \frac{\partial v_a^k}{\partial x_a^k} \delta^{ij} \right)  \frac{\nabla_a^j W_{ab}(h_a)}{\Omega_a \rho_a} \nonumber \\
& \hspace{3mm} + v_a^i \left( \frac{\partial v_b^i}{\partial x_b^j} + \frac{\partial v_b^j}{\partial x_b^i} - \frac{2}{3} \frac{\partial v_b^k}{\partial x_b^k} \delta^{ij} \right)  \frac{\nabla_a^j W_{ab}(h_b)}{\Omega_b \rho_b} \Bigg] .
\end{align}
Swapping the summation indices on the second term produces
\begin{align}
& \sum_a \sum_b m_a m_b \nu \Bigg[ v_a^i \left( \frac{\partial v_a^i}{\partial x_a^j} + \frac{\partial v_a^j}{\partial x_a^i} - \frac{2}{3} \frac{\partial v_a^k}{\partial x_a^k} \delta^{ij} \right)  \frac{\nabla_a^j W_{ab}(h_a)}{\Omega_a \rho_a} \Bigg] \nonumber \\
& + \sum_a \sum_b m_a m_b \nu \Bigg[ v_b^i \left( \frac{\partial v_a^i}{\partial x_a^j} + \frac{\partial v_a^j}{\partial x_a^i} - \frac{2}{3} \frac{\partial v_a^k}{\partial x_a^k} \delta^{ij} \right)  \frac{\nabla_b^j W_{ab}(h_a)}{\Omega_a \rho_a} \Bigg] .
\end{align}
The kernel derivative in the second term can be substituted using the identity $\nabla_b W_{ab} \equiv - \nabla_a W_{ab}$, allowing the two terms to be combined to produce
\begin{equation}
\sum_a \sum_b m_a m_b \nu \Bigg[ \left( \frac{\partial v_a^i}{\partial x_a^j} + \frac{\partial v_a^j}{\partial x_a^i} - \frac{2}{3} \frac{\partial v_a^k}{\partial x_a^k} \delta^{ij} \right) v_{ab}^i  \frac{\nabla_b^j W_{ab}(h_a)}{\Omega_a \rho_a} \Bigg] .
\end{equation}
By utilising the definition of the difference derivative estimate (Equation~\ref{eq:sphdiffderiv}), one can arrive, after simplification, to
\begin{equation}
\sum_a m_a \nu \left[ \left( \frac{\partial v_a^i}{\partial x_a^j} + \frac{\partial v_a^j}{\partial x_a^i} \right) \left( \frac{\partial v_a^i}{\partial x_a^j} \right) - \frac{2}{3} \left( \frac{\partial v_a^k}{\partial x_a^k} \right)^2 \right] .
\end{equation}
The second term represents the divergence of the velocity, $\nabla \cdot {\bm v}$. The first term may be simplified by splitting the sum into two equal halves and swapping the $i$, $j$ indices on the second half, 
\begin{align}
&\sum_a m_a \nu \Bigg[ \frac{1}{2} \left( \frac{\partial v_a^i}{\partial x_a^j} + \frac{\partial v_a^j}{\partial x_a^i} \right) \left( \frac{\partial v_a^i}{\partial x_a^j} \right) - \frac{1}{3} \left( \frac{\partial v_a^k}{\partial x_a^k} \right)^2 \Bigg]  \nonumber \\
&+ \frac{1}{2} \left( \frac{\partial v_a^j}{\partial x_a^i} + \frac{\partial v_a^i}{\partial x_a^j} \right) \left( \frac{\partial v_a^j}{\partial x_a^i} \right)  - \frac{1}{3} \left( \frac{\partial v_a^k}{\partial x_a^k} \right)^2 \Bigg] 
\end{align}
which after combining yields
\begin{equation}
\sum_a m_a \frac{\nu}{2} \left[ \left( \frac{\partial v_a^i}{\partial x_a^j} + \frac{\partial v_a^j}{\partial x_a^i} \right) \left( \frac{\partial v_a^i}{\partial x_a^j} + \frac{\partial v_a^j}{\partial v_a^i} \right) - \frac{2}{3} \left( \frac{\partial v_a^k}{\partial x_a^k} \right)^2 \right] .
\end{equation}
The heat gain from the Navier-Stokes viscosity is obtained by substituting the preceding into Equation~\ref{eq:appndx1}, yielding
\begin{equation}
\frac{{\rm d}u_a}{{\rm d}t} = \frac{\nu}{2} \left[ \left(\frac{\partial v_a^i}{\partial x_a^j} + \frac{\partial v_a^j}{\partial x_a^i} \right)^2 - \frac{2}{3} \left( \frac{\partial v_a^k}{\partial x_a^k} \right)^2 \right] .
\end{equation}
This is guaranteed to yield positive definite increases in thermal energy, provided that the velocity derivatives are computed using difference derivative estimates.

\bibliographystyle{mnras}
\bibliography{bib}

\label{lastpage}
\end{document}